\documentclass[preprint,authoryear,11pt]{elsarticle}
\usepackage{amssymb,amsmath,latexsym}
\usepackage[utf8]{inputenc}
\usepackage[T1]{fontenc}
\usepackage{amsmath,amsthm,amsfonts,amssymb,dsfont,amscd}
\usepackage[english]{babel}
\usepackage{latexsym}
\usepackage{epsfig}
\usepackage{graphics}
\usepackage[autostyle]{csquotes}
\usepackage{enumerate}
\usepackage{multirow}

\usepackage{tabularx}
\usepackage[left=2.5cm,right=2cm,top=2cm,bottom=2cm]{geometry}
\usepackage{graphicx}
 \usepackage{natbib}
\usepackage{here}
\usepackage{bm}
\usepackage[linesnumbered]{algorithm2e}
\usepackage{booktabs} 
\usepackage{lipsum}
\usepackage{multicol}
\usepackage{changepage}
\usepackage{dsfont}

\usepackage{accents}
\usepackage{caption}
\usepackage{xcolor} 
\usepackage{graphicx}
\usepackage{subcaption}
\usepackage{float}

\usepackage[colorlinks=true, linkcolor=red, citecolor=red, urlcolor=blue]{hyperref}
\usepackage{cleveref}

\numberwithin{propo}{section}

\numberwithin{thm}{section}

\numberwithin{rem}{section}

\numberwithin{exemple}{section}

\numberwithin{lem}{section}

\numberwithin{cor}{section}
\newtheorem{l1}{Lemma}

\newtheorem{t1}{Theorem}
\newtheorem{r1}{Remark}
\newtheorem{pr1}{Proposition}

\newcommand{\Var}{\operatorname{Var}} 
\newcommand{\Cov}{\operatorname{Cov}}
\newcommand{\Vect}{\operatorname{vec}}

\newcommand{\nbR}{\mathbb{R}}
\newcommand{\nbN}{\mathbb{N}}

\newcommand{\nbC}{\mathbb{C}}

\newcommand{\nbP}{\mathbb{P}}
\newcommand{\nbZ}{\mathbb{Z}}

\newcommand{\calS}{\mathcal{S}}

\makeatletter

\@addtoreset{equation}{section}
\makeatother

\begin{document}


\begin{frontmatter}

\title{Asymptotic properties for weak Markov-switching AR$(p)$ models}

\author[uphf]{Yacouba Boubacar Ma\"{i}nassara \corref{cor1}} \ead{Yacouba.BoubacarMainassara@uphf.fr}
\cortext[cor1]{Corresponding author}
\address[uphf]{ Univ. Polytechnique Hauts-de-France, INSA Hauts-de-France, CERAMATHS - Laboratoire de
Mat\'{e}riaux C\'{e}ramiques et de Math\'{e}matiques, F-59313 Valenciennes, France}
\author[ufc]{Armel Bra}\ead{kja.bra@univ-fcomte.fr}
\address[ufc]{Universit\'{e} Marie et Louis Pasteur, Laboratoire de math\'{e}matiques de Besan\c{c}on,  UMR CNRS 6623, 16 route de Gray,  25030 Besan\c{c}on, France}

\author[ufc]{Landy Rabehasaina}\ead{landy.rabehasaina@univ-fcomte.fr}

\begin{abstract}
In this paper we study the asymptotic properties of an autoregressive model of order~\(p\) (AR for short), possibly non-centered,  whose parameters switch according to an unobserved Markov chain. Departing from the conventional assumption of independent innovations, we merely require the errors to be uncorrelated but not necessarily independent, thereby widening substantially the range of potential applications of Markov-switching AR models. We establish sufficient conditions for the consistency and asymptotic normality of the proposed moment estimators.  We also propose a consistent estimator of the asymptotic variance matrix of the moment estimators. The finite-sample performance of the proposed procedures is illustrated through Monte Carlo experiments and an empirical study based on hourly meteorological data, which together provide strong support for the theoretical findings.
\end{abstract}
\begin{keyword}
Weak AR models, Regime-switching models, Markov-switching models, Times series with changes in regime, Moment's method, Asymptotic normality. 
\end{keyword}

\end{frontmatter}

\section{Introduction}\label{Introduction}
\noindent  We consider a time series exhibiting changes in regime with random coefficients, where the associated noise is multiplicative. Such models can be interpreted as a Markovian mixture of dynamic systems. To capture regime-switching or Markov-switching dynamics, the seminal works of \cite{hamilton1988rational}, \cite{hamilton1989new}, \cite{mcculloch1994bayesian}, \cite{chib1996calculating} use latent Markov chains with time-invariant transition probability matrices. 

\noindent  Let \((\Delta_t)_{t\in\nbZ}\) be an \textit{unobserved} homogeneous, irreducible Markov chain with a stationary distribution \(\pi=(\pi_1,\ldots,\pi_K)\) taking values in a discrete finite set \(\mathcal{S} := \{1, \ldots, K\}\).
In this paper, we consider the real valued process \((X_t)_{t\in\nbZ}\) defined as
\begin{align}
\label{eq:1}
   X _t = \sum_{i=1}^p a_i(\Delta_t)X_{t-i}+f(\Delta_t)\eta_t, \  \ \forall t\in \mathbb{Z},
\end{align}
where the process $(\eta_t)_{t\in\nbZ}$ is a weak white noise satisfying $\mathbb{E}(\eta_t) = 0$, $\mathbb{E}(\eta_t \eta_{t'}) =\sigma^2_\eta \mathds{1}_{[t = t']}$, $f : \mathcal{S}\rightarrow \mathbb{R}\setminus\{0\}$, $a_i$ is a $\mathcal{S}\rightarrow \mathbb{R}$ function for all $1\leq i \leq p$ and $p_{i,j} := \mathbb{P}(\Delta_t = j \mid \Delta_{t-1} = i)$ denotes the transition probability from state $i\in\calS$ to state $j\in\calS$ of the Markov chain $(\Delta_t)_{t \in \mathbb{Z}}$. These models are called \emph{weak} Autoregressive Hidden Markov Chain (ARHMC, for short) and can be viewed as Markovian mixtures of dynamic systems, belonging to the class of Markov regime-switching models. More precisely, a Markov-switching model is a non-linear specification in which different states of the world affect the evolution of a time series (see, for examples, \cite{francq1997white, Hamilton1990, HS1994autoregressive}).

\noindent An example of weak white noise is the autoregressive conditional heteroscedastic (GARCH for short) models introduced by \cite{engle1982autoregressive} and extended by \cite{bollerslev1986generalized} (see \cite{BMKS2022} and see also \cite{francq2019garch}, for a reference book on GARCH models). Others examples of weak white noises can be found in \cite{romano1996inference}. It is customary to say that $(X_t)_{t\in \mathbb{Z}}$ is a strong ARHMC$(p)$ representation and we will do this henceforth if in (\ref{eq:1}) $(\eta_t)_{t\in \mathbb{Z}}$ is a strong white noise, namely an
independent and identically distributed (i.i.d. for short) sequence of random variables with mean 0 and common variance $\sigma^2_\eta$. A strong white noise is obviously a weak white noise because independence entails uncorrelatedness. Of course, the converse is not true. It is clear from these definitions that the following inclusion hold: 
\begin{align*}
    \{\text{Strong ARHMC(p)}\} \subset \{\text{Weak ARHMC(p)}\}.
\end{align*}
\noindent {As mentioned by \cite{fz05,francq1998estimating} in the case of ARMA models, many important classes of nonlinear processes admit  weak ARMA
representations in which the linear innovation is not a martingale difference. 
For instance, in financial econometrics, the returns are often assumed to be martingale increments (though they are not generally independent sequences). However, many works have shown that for some exchange rates, the martingale difference assumption is not satisfied (see for instance \cite{EscVel2006}).   There is no doubt that it is  important to have a soundness inference procedure for the parameter in the ARHMC model when the (possibly dependent) error is subject to unknown conditional  heteroscedasticity. Little is thus known when the martingale difference assumption is relaxed. Our aim in this paper is to consider a flexible ARHMC specification and to relax the independence assumption (and even the martingale difference assumption) in order to be able to cover weak ARHMC representations of general nonlinear models. 
 Weak ARHMC models are then especially relevant for time series whose innovations are uncorrelated but not independent. This situation occurs in many empirical applications where the first-order dynamics can be adequately described by a regime-switching autoregressive structure, while the innovation process still exhibits conditional heteroskedasticity or more general nonlinear dependence.}
 
\noindent Since the works of \cite{hamilton1988rational}, \cite{hamilton1989new}, \cite{mcculloch1994bayesian}, \cite{Cai1994}, \cite{Kim1994}, \cite{chib1996calculating} and \cite{Krolzig1997}, to site few,  the class of Markov regime-switching models  takes a large part in the literature of time series. 
Now we present some of the main works about ARHMC model estimation when the noise is strong. For the estimation of the class of the strong ARHMC processes, the commonly used estimation method is based on the maximum likelihood or the least squares estimators (see for instance : \cite{Cavi2016},  \cite{Cavi2014}, \cite{xie2008general}, \cite{douc2004asymptotic}, \cite{FG2004jtsa}, \cite{francq1998ergodicity} and \cite{francq1997white}).  The statistical properties of these models have also been studied, for instance, by \cite{Chiba2023}, \cite{Cavi2016},  \cite{Cavi2014}, \cite{billio1999bayesian}, \cite{douc2004asymptotic}, \cite{FG2004jtsa}, \cite{francq1998ergodicity}, \cite{francq1997white}, \cite{leroux1992maximum}, \cite{baum1966statistical} and \cite{petrie1969probabilistic} ). 
This body of work underscores the growing interest and ongoing advancements in Markov regime-switching models, as researchers continue to refine estimation techniques and broaden the applicability of these models. 

\noindent  As above-mentioned, the works on the statistical inference of ARHMC processes  are generally performed under the assumption that the errors are independents. This independence assumption is often considered too restrictive by practitioners. It precludes conditional heteroscedasticity and/or other forms of nonlinearity (see \cite{romano1996inference} and also \cite{fz05} for a review on ARMA models under the assumption that the errors are uncorrelated but not necessarily independent) which can not be generated by Markov regime switching models  with i.i.d. noises. Relaxing this independence assumption  allows to extend the range of application of the class of Markov regime switching models.
{For non-Gaussian innovations in the presence of a hidden Markov chain, standard least-squares and quasi-maximum likelihood approaches are neither applicable nor theoretically justified, since the latent chain is unobserved and no parametric form is imposed on the innovation distribution. To the best of our knowledge, no published method addresses this general setting. This is precisely the second major contribution of our work, which relies on a method-of-moments estimator that remains valid without a parametric specification of the innovation densities.}
Here, we focus on the estimation problem and statistical inference of a class of weak ARHMC$(p)$. To our knowledge, it is worth noting that very few studies have considered time series models with Markovian regime changes involving weak noise. Notable exceptions are \cite{mainassara2025estimating} who studied the estimation procedure and the asymptotic properties of a weak ARHMC$(1)$, 
\cite{boubacar2020estimation} who examined the asymptotic properties of the least squares estimator for a weak ARMA model with regime switching under the assumptions that the  realization of the Markov chain is observed. However, it is important to note also that in the model considered by \cite{boubacar2020estimation} the structure of the noise is not multiplicative. Instead, they assumed that the volatility was uniform within each segment of the time series. Finally \cite{francq2001stationarity} investigated the stationarity conditions of such models in a multivariate framework. The authors demonstrated that local stationarity of these processes is neither sufficient nor necessary to ensure a global stationarity. 

\noindent {In that respect, this paper is a solid extension of \cite{mainassara2025estimating}, incorporating a non centered autoregressive process of order $p\ge 1$, as opposed to a centered process of order $1$ in \cite{mainassara2025estimating}. The main contributions, contained in the forthcoming Theorems \ref{thm:2}, \ref{thm:3}, \ref{thm:6} and \ref{thm:7}, are a non trivial generalizations of Theorems 2 and 3 of  \cite{mainassara2025estimating}. Their proofs involve non trivial matrix analytic methods. This effort is rewarded in the applications, as the model captures more accurately the real data through the estimation.}

\noindent The paper is organized as follows. Section \ref{sec:2}
shows that the  moments estimator for the parameters of a centered weak ARHMC model is almost surely consistent and asymptotically normal under suitable strong-mixing conditions on the linear innovation process. The asymptotic variance of the proposed moments estimator may be very different in the weak and strong cases. Section \ref{sec:5} is devoted to the estimation of this covariance matrix. Section \ref{sec:6} extends the analysis to the case of non-centered process. The simulation studies and  illustrative applications on hourly meteorological data are presented and discussed, respectively, in  Sections \ref{sec:7} and \ref{sec:8}, which together corroborate the theoretical findings.
We provide a conclusion in Section \ref{conclusion} and finally the proofs of the main results are collected in Section \ref{proof}.

\section{Parametrization and Assumptions}
\label{sec:2}

\noindent  In order to study the process $(X_t)_{t\in\nbZ}$, it is convenient to write the model (\ref{eq:1}) in a vector autoregressive (hereafter VAR)  representation:
%
\begin{align*}
    \underline{X}_t = \mathbf{A}(\Delta_t)\underline{X}_{t-1} + \underline{f}(\Delta_t)\eta_t,\quad t\in\nbZ ,
\end{align*}
where
\small{\begin{multline}\label{def_theta_functions}
    \underline{X}_t : = \begin{pmatrix}
        X_t \\
        X_{t-1}\\
        \vdots \\
        X_{t-p+1}
    \end{pmatrix}\in\mathbb{R}^{p},
    \underline{f}(\Delta_t) : = \begin{pmatrix}
        f(\Delta_t) \\
        0 \\
        \vdots \\
        0
    \end{pmatrix}\in\mathbb{R}^p   
    \text{and} \  
    \mathbf{A}(\Delta_t) : = \begin{pmatrix}
        a_1(\Delta_t) &  & \cdots &  & a_p(\Delta_t) \\
        1 & 0 & \cdots &  & 0 \\
        0 & 1 & \cdots &  & 0 \\
        \vdots & \ddots & \ddots &  & \vdots \\
        0 &  & \cdots & 1 & 0 
    \end{pmatrix} \in\mathbb{R}^{p\times p}, 
\end{multline}}
\noindent for some function $\underline{f}(\cdot)$ (resp. $\mathbf{A}(\cdot)$) defined from $\calS$ to $\mathbb{R}^p$ (resp. to $\mathbb{R}^{p\times p}$). In the sequel, the parameter $\sigma^2_\eta$ is considered as a nuisance parameter.
The unknown parameter of interest $\theta_0 : = \left(a_i(s), p_{k,k'}, f(s), \ i = 1,\dots p, \ k,k',s = 1,\dots K \right)'$ is supposed to belong to the parameter space defined as
\begin{eqnarray*}
\Theta &:= &\biggl\{
 \theta = 
\left(\mathfrak{a}_i(s), \mathfrak{p}_{k,k'}, \mathfrak{f}(s), \ i = 1,\dots p, \ k,k',s = 1,\dots K \right)' 
\in \mathbb{R}^{Kp} \times {(0,1)}^{K(K-1)} \times \mathbb{R}^K; \\
&&\mathfrak{p}_{i,K} = 1 - \sum_{j=1}^{K-1} \mathfrak{p}_{i,j}\in(0,1), \ \text{and} \\ 
& &\quad \mathfrak{P}_{\theta} := (\mathfrak{p}_{ij})_{1 \leq i,j \leq K} \ \text{is an irreducible transition matrix}
\biggr\}\subset \nbR^{Kp+K^2}.
\end{eqnarray*}
\noindent Let $\mathcal{F}_{-\infty}^t$ and $\mathcal{F}_{t+h}^\infty$ be the $\sigma$-fields generated by $\{\eta_u : u\leq t\}$ and $\{\eta_u : u\geq t+h\}$ respectively.
The stationary process  \( (\eta_t)_{t\in\mathbb{Z}} \)  is not supposed to be an independent sequence. So
one needs to control its dependency by means of its strong mixing coefficients  \( (\alpha_\eta(h))_{h\in\mathbb{N}} \) (independent of $t\in\mathbb{Z}$) defined by
%
\begin{align}\label{eq:12}
\alpha_{\eta}(h) := \sup_{A\in \mathcal{F}_{-\infty}^t, B\in \mathcal{F}_{t+h}^{\infty}} |\mathbb{P}(A \cap B) - \mathbb{P}(A)\mathbb{P}(B)|.
\end{align}

\noindent Let \(\otimes\) denote the tensor product or Kronecker product and \(||\cdot||\) denote any operator norm on sets of \(p \times p\) or \(p \times 1\) matrices. To derive the conditions for the strict stationarity and the condition for the existence of second-order moments of model \eqref{eq:1} given in \cite{francq2001stationarity}, we also introduce the Lyapunov exponent \(\gamma\) defined as follows
\begin{align*}
    \gamma := \inf_{t \in \mathbb{N}\setminus\{0\}} \left\{ \mathbb{E} \frac{1}{t} \log ||\mathbf{A}(\Delta_t) \mathbf{A}(\Delta_{t-1}) \ldots \mathbf{A}(\Delta_1)|| \right\}.
\end{align*}
For all matrix and column vector functions $F:{\cal S}\longrightarrow \nbR^{m\times m}$ and $G:{\cal S}\longrightarrow \nbR^{m'}$ we will denote
\begin{eqnarray}
  Q_{F} &:= &\begin{pmatrix}
       p_{11}F(1) & p_{21}F(1) & \cdots & p_{K1}F(1) \\ 
       p_{12}F(2) & p_{22}F(2) & \cdots & p_{K2}F(2) \\ 
       \vdots & \vdots && \vdots \\
       p_{1K}F(K) & p_{2K}F(K) & \cdots &  p_{KK}F(K) \\ 
   \end{pmatrix}\in\nbR^{Km\times Km},\label{eq:1.3}\\
   \Pi_G&:=& \left(\pi(1)G(1)', \dots, \pi(K)G(K)'\right)'\in\nbR^{Km'}.\label{def_Pi_G}
\end{eqnarray}

\noindent Our main results are established under the following assumptions.
\begin{enumerate}
\item[$(\mathbf{A_1})$] The processes \((\eta_t)_{t\in\mathbb{Z}}\) and \((\Delta_t)_{t\in\mathbb{Z}}\) are {strictly} stationary and  $(\eta_t)_{t\in\mathbb{Z}}$ is an ergodic process.
%
\item[$(\mathbf{A_2})$] The processes $(\Delta_t)_{t\in\mathbb{Z}}$ and $(\eta_t)_{t\in\mathbb{Z}}$ are independent.
\item[$(\mathbf{A_3})$] We have $\mathbb{E}(|\eta_t|^{4+2\nu})<\infty$, $\displaystyle\sum_{h = 0}^{\infty}\alpha_{\eta}(h)^{\frac{\nu}{2+\nu}}<\infty$ and \ $\displaystyle\sum_{h = 0}^{\infty}\alpha_{\Delta}(h)^{\frac{\nu}{2+\nu}}<\infty$  for some $\nu>0$.
\item[$(\mathbf{A_4})$] We have $\theta_0\in\mathring{\Theta}$, where $\mathring{\Theta}$ denotes the interior of $\Theta$. 
\item[$(\mathbf{A_5})$] The Lyapunov exponent $\gamma$ is strictly negative.
\item[$(\mathbf{A_6})$] The spectral radius of $Q_{\bm{A}\otimes\bm{A}} \in \nbR^{Kp^2 \times Kp^2}$, defined in \eqref{eq:1.3} with $F(i)=\mathbf{A}(i) \otimes \mathbf{A}(i)$, $i\in {\cal S}$, is less than $1$.
\end{enumerate}
{Those assumptions are technical and cannot be verified in practice. They may be interpreted as follows. Assumption $(\mathbf{A_3})$ says that the mixing coefficients $\alpha_{\eta}(h)$ and $\alpha_{\eta}(h)$ decrease fast as $h\to\infty$, i.e. that the corresponding processes \((\eta_t)_{t\in\mathbb{Z}}\) and \((\Delta_t)_{t\in\mathbb{Z}}\) can be intuitively considered as independent asymptotically. Assumptions $(\mathbf{A_5})$ and $(\mathbf{A_6})$ are stability conditions that ensures the process $(\underline{X}_t)_{t\in\nbZ}$ in \eqref{def_theta_functions} to be finite almost surely and admits a stationary version, as proved thereafter in Proposition \ref{prop_stability}.}

\subsection{Parameter estimation of Model \eqref{eq:1}}
\label{sec:3}

\begin{pr1}\label{prop_stability}
Under the assumptions $(\mathbf{A_1})$, $(\mathbf{A_2})$ and $(\mathbf{A_5})$, from  \citet[Theorem 1.1, page 1715]{bougerol1992strict}  (see also \citet[Theorem 1, page 212]{brandt1986stochastic}), the series 
\begin{equation}
\label{eq:4}
\underline{X}_t = \sum_{k=0}^{\infty} \left( \prod_{j=0}^{k-1} \mathbf{A}(\Delta_{t-j}) \right) \underline{f}{(\Delta_{t-k}}){\eta}_{t-k}\in\nbR^p,\;\text{for all }t\in\mathbb{Z}
\end{equation}
converges almost surely and the process $(X_t)_{t\in\mathbb{Z}}$, defined as the first component of $\underline{X}_t$ is the unique strictly stationary solution of (\ref{eq:1}), with the usual convention that \(\prod_{j=\ell}^{\ell'} = 1\) for \(\ell, \ell' \in \mathbb{Z}\) and \(\ell' < \ell\).
\end{pr1}
\noindent In order to state our results, we will use the following  result which is a multivariate version  of the result obtained by \citet[Lemma 1]{francq2004estimation}.
\begin{l1}\label{lem:1}
Let $(\Delta_t)_{t\in\nbZ}$ be an irreducible, aperiodic and stationary Markov chain with state space ${\cal S}=\{1,\dots,K\}$, transition probabilities $(p_{ij})_{i,j = 1,\ldots,K}$, and stationary distribution $\pi = \{\pi(1),\dots,\pi(K)\}$. Then, for any $k \geq 1$, let  $F_1,\dots,F_k$ be the $d \times d$ matrices functions and $F_{k+1}$ be a $d \times 1$ vector function, defined on $\{1,\dots,K\}$. We have
\begin{equation}
\label{eq:3}
    \mathbb{E}(F_1(\Delta_{t-1})\dots F_k(\Delta_{t-k}) F_{k+1}(\Delta_{t-k-1})) = \left(\bm{1}'\otimes I_{{d}}\right)\left(\prod_{\ell=1}^k {Q}_{F_\ell}\right)\bm\Pi_{F_{k+1}}\in\nbR^d,
\end{equation}
where \(\bm{1}':= (1, \dots, 1)\) is a row vector of dimension \(K\). We recall that $Q_{F_\ell} $, $\ell=1,\ldots,k $ and $\bm\Pi_{F_{k+1}} $ are defined thanks to \eqref{eq:1.3} and \eqref{def_Pi_G}. 
\end{l1}
\noindent We denote by $\Vect(U)$ the vector obtained by stacking the columns of $U$.
\begin{t1}\label{thm:1}
Under Assumptions $(\mathbf{A_1})$, $(\mathbf{A_2})$, $(\mathbf{A_5})$ and $(\mathbf{A_6})$, the joint moments of the model (\ref{eq:1}) satisfy for all $k\in\mathbb{N}$,
\begin{equation}
\label{eq:2}
     C_{k,0} : = \mathbb{E}(\Vect\{\underline{X}_0\underline{X}_k'\}) = \sigma^2_\eta\left(\bm{1}'\otimes I_{p^2}\right)Q_{\mathbf{A}\otimes I_p}^k\left(I_{Kp^2}-Q_{\mathbf{A}\otimes \mathbf{A}}\right)^{-1}\bm\Pi_{\underline{f}\otimes\underline{f}} \in \nbR^{p^2},
\end{equation}
where \(I_{d}\) denotes the $d\times d$ identity matrix, \(\bm{1}':= (1, \dots, 1)\) is a row vector of dimension \(K\), \(Q_{\mathbf{A}\otimes I_p}\) is the \(Kp^2 \times Kp^2\) matrix defined by \eqref{eq:1.3} with $F(i)=\mathbf{A}(i) \otimes I_p$, $i\in {\cal S}$. Finally, \(\bm\Pi_{\underline{f}\otimes\underline{f}}\in \nbR^{Kp^2 }$ is defined as \eqref{def_Pi_G} with $ G(i)= \underline{f}(i)\otimes\underline{f}(i)\in \nbR^{p^2}\), $i\in {\cal S}$.
\end{t1}
\noindent The proof of this theorem is given in Section \ref{proofthm:1}.
\begin{r1}\label{sig=1}
When we assumes that $\sigma^2_\eta$ is known, the distributions of the process $(\eta_t)_{t\in\mathbb{Z}}$ can be standardized such that $\mathbb{E}(\eta_t^2)=\sigma^2_\eta=1$. Thus the parameter $\sigma^2_\eta$  vanishes in \eqref{eq:2} which reduced the number of estimated parameters.   In practice this parameter has
also to be estimated.

\end{r1}
\noindent 
In the following, for all $\theta = \left(\mathfrak{a}_i(s), \mathfrak{p}_{k,k'}, \mathfrak{f}(s), \ i = 1,\dots p, \ k,k',s = 1,\dots K \right)'\in \Theta$ with associated column vector and matrix functions $ \underline{\mathfrak{f}}(\cdot)$ and $\mathfrak{A}(\cdot)$ defined as in \eqref{def_theta_functions}, we define
\begin{equation}\label{def_C_k_theta}
C_{k,0}(\theta):=\sigma^2_\eta\left(\bm{1}'\otimes I_{p^2}\right)Q_{\mathfrak{A}\otimes I_p}^k\left(I_{Kp^2}-Q_{\mathfrak{A}\otimes \mathfrak{A}}\right)^{-1}\bm\Pi_{\underline{\mathfrak{f}}\otimes\underline{\mathfrak{f}}}
\end{equation}
so that $C_{k,0}=:C_{k,0}(\theta_0)$ when $\theta=\theta_0$ from \eqref{eq:2}. To estimate the parameter of interest $\theta_0$, we will adopt a similar estimation approach proposed by \cite{mainassara2025estimating} in the specific case where $p=1$. The method presented here for estimating $\theta_0$ is therefore a generalization of this method. 

\noindent Let \((X_1, \ldots, X_n)\) be a realization of length $n \in \mathbb{N}\setminus\{0\}$ of the nonanticipative stationary solution of Model \eqref{eq:1} and let \(N \in \mathbb{N}\setminus\{0\}\) be chosen such that \(p^2N \geq Kp + K^2\). We now introduce the following functions
\begin{equation*}
\bm{\Psi}^N :
    \left\lbrace
        \begin{aligned}
            & \mathring{\Theta} & \rightarrow & \ \mathbb{R}^{p^2N} \\
            & \theta & \mapsto & \ (C'_{1,0}(\theta), \dots, C'_{N,0}(\theta))'
        \end{aligned}
    \right., \quad\quad 
    \bm{F}^{N,n}: 
    \left\lbrace
        \begin{aligned}
            & \mathring{\Theta} & \rightarrow & \ \mathbb{R}^{p^2N} \\
            & \theta & \mapsto & \ (\{C_{1,0}(\theta)-\hat{C}_{1,0} \}', \dots, \{ C_{N,0}(\theta)-\hat{C}_{N,0}\}')',
        \end{aligned}
    \right.
\end{equation*}
where for all $1 \leq k \leq N$ we let
\begin{align*}
\hat{C}_{k,0} := (n-k)^{-1} \sum_{t=1}^{n-k} \text{Vec}(\underline{X}_{t}\underline{X}_{t+k}') = (n-k)^{-1} \sum_{t=1}^{n-k} \underline{X}_{t+k}\otimes\underline{X}_t\in\nbR^{p^2},
\end{align*} 
by using the elementary identity $\Vect\{ U_1V_1'\} =U_1\otimes V_1$ for $U_1$ and $V_1$ in $\nbR^{m}$
and $\hat{C}_{k,0}$ is an unbiased estimator of $C_{k,0}$.
We then introduce the estimation function $\bm{\mathcal{F}}^{N,n}(\cdot) \in \mathbb{R}^{Kp+K^2}$ defined as 
\begin{align}
\label{eq:1.8}
    \bm{\mathcal{F}}^{N,n}(\theta) := J_{\bm{\Psi}^N}(\theta)'\bm{F}^{N,n}(\theta), \ \forall \theta \in \Theta,
\end{align}
where $J_{\bm{\Psi}^N}(\theta)\in \nbR^{p^2N\times (Kp+K^2)}$ denotes the Jacobian matrix of the function $\bm{\Psi}^N$.

\noindent Thus, a moment estimator $\hat{\theta}_n$ of $\theta_0$ is given by a zero of the function $\bm{\mathcal{F}}^{N,n}(\cdot)$. To find this zero, we employ the Newton-Raphson method which is widely-used iterative numerical technique for finding roots of real-valued functions. The Newton-Raphson method refines an initial guess through successive iterations, converging to the actual solution under appropriate conditions. The algorithm is described as follows 

\begin{algorithm}[H]
\label{algo:algo_1}
  \SetAlgoLined
  \KwData{An initial parameter $\theta^0 \in \mathbb{R}^{K^2 + Kp}$, tolerance $\varepsilon > 0$}
  \KwResult{A value of $\theta\in\Theta$ such that $\|\bm{\mathcal{F}}^{N,n}(\theta)\| < \varepsilon$}

  \textbf{Initialization :} $\theta^0 \in \mathbb{R}^{K^2 + Kp}$\;
  \For{$k \geq 0$}{
    Compute $J_{\bm{\mathcal{F}}^{N,n}}(\theta^{(k)})$\;
    Update $\theta^{(k+1)}$ using the equation
    \[
    J_{\bm{\mathcal{F}}^{N,n}}(\theta^{(k)})(\theta^{(k+1)} - \theta^{(k)}) = -  \bm{\mathcal{F}}^{N,n}(\theta^{(k)})
    \]
    \If{$\|\bm{\mathcal{F}}^{N,n}(\theta^{(k+1)})\| < \varepsilon$}{
      \textbf{break} \; 
    }
  }
  \caption{Newton's method for finding root of $\bm{\mathcal{F}}^{N,n}(\cdot)$ with specified tolerance.}
\end{algorithm}

\subsection{Asymptotic properties}
\label{sec:4}
\noindent We now turn to the main results of this paper, namely the strong consistency and asymptotic normality of the estimator $\hat{\theta}_n$.


\begin{t1}
\label{thm:2}
Under the assumption that $J_{\bm{\Psi^{N,n}}}(\theta_0)$ is {of maximal rank $ K^2 + Kp$}, there exists a neighborhood $\mathcal{V}_{\theta_0}$ of $\theta_0$ in $\Theta$ and a unique sequence $(\hat{\theta}_n)_{n\in\mathbb{N}}$ taking values in $\mathcal{V}_{\theta_0}$ such that
\begin{align}
\label{eq:1.9}
    0 = J_{\bm{\Psi}^{N}}(\hat{\theta}_n)'\bm{F}^{N,n}(\hat{\theta}_n) = \bm{\mathcal{F}}^{N,n}(\hat{\theta}_n).
\end{align}
Furthermore, we have
\begin{align*}
    \hat{\theta}_n \xrightarrow[n\rightarrow\infty]{a.s} \theta_0.
\end{align*}
\end{t1}
\noindent The proof of this theorem is given in Section \ref{proofthm:2}.

\begin{t1}
\label{thm:3}
Assuming that the Assumptions \((\mathbf{A_1})\), \((\mathbf{A_2})\), \((\mathbf{A_3})\), \((\mathbf{A_4})\), $(\mathbf{A_5})$ and $(\mathbf{A_6})$ are satisfied, that the matrix $J_{\bm{\Psi}^{N}}(\theta_0)$ is of full rank, and defining \(\hat{\theta}_n\) as given in Equation (\ref{eq:1.9}), we have the following Central Limit Theorem 
\begin{align*}
\sqrt{n}(\hat{\theta}_n - \theta_0) \xrightarrow[n\rightarrow\infty]{\mathcal{D}} \mathcal{N}(0,\Xi^{-1}\bm{J}'\bm{I}\bm{J}\Xi^{-1}),
\end{align*}
where the matrices $\Xi$, $\bm I$ and $\bm J$ are defined as follows 
\begin{align}
\label{eq:000}
    \bm J & := J_{\bm\Psi^N}(\theta_0)\in\mathbb{R}^{p^2N\times (Kp+K^2)}, \notag\\
    \bm I & := \bm I^N(\theta_0) = \sum_{k=-\infty}^{\infty}\mathrm{Cov}\left(Z_t(\theta_0), Z_{t-k}(\theta_0)\right)\in\mathbb{R}^{p^2N\times p^2N},\notag\\
    \Xi & : = \Xi^N(\theta_0) = J_{\bm\Psi^N}(\theta_0)'J_{\bm\Psi^N}(\theta_0) \in\mathbb{R}^{p^2N\times p^2N}  
\end{align}
with
\begin{align}
\label{eq:001}
    Z_t:= Z_t(\theta_0) = \left(\{\underline{X}_{t+1}\otimes\underline{X}_t\}',\dots, \{\underline{X}_{t+N}\otimes\underline{X}_t\}'\right)'\in\mathbb{R}^{p^2N}.
\end{align}
\end{t1}
\noindent The proof of this theorem is given in Section \ref{proofthm:3}.

\subsection{Estimation of the asymptotic covariance matrix}
\label{sec:5}
\noindent An estimator $\hat{\bm{J}}_n$ of the matrix $\bm{J}$ in the context of our study is given by 
\begin{align*}
    \hat{\bm{J}}_n := J_{\bm{\Psi}^N}(\hat{\theta}_n),
\end{align*}
where $\hat{\theta}_n$ is a zero of the estimating function $\bm{\mathcal{F}}^{N,n}(\cdot)$  and obtained in practice via Algorithm \ref{algo:algo_1}.

\noindent {The estimation of the long-run variance (LRV) matrix $\bm{I}$ is more complicated. In the literature, two types of estimators are generally
employed: heteroskedasticity and autocorrelation consistent (HAC) estimators based on kernel methods (see \cite{newey},  \cite{A91}, \cite{PDR1995} and \cite{P2011} for general references; and see also \cite{VF2022}, \citet{CY2024} and \citet{LC2025} for recent references) and the spectral density estimators (see e.g. \citet{berk1974consistent} and \cite{den1996practitioner} for general references).}

The approach used {in this paper} to estimate the matrix $\bm{I}$ is based on the spectral density-based estimation as in \cite{mainassara2025estimating}. 

\noindent It is worth noting that such an approach has been widely adopted over the past decade in numerous articles to estimate the variance-covariance matrix when it can be expressed as a series. Among others, we can cite \cite{BMES2023}, \cite{BMES2021}, \cite{boubacar2020estimation}, \cite{BMCF2012}, \cite{BM2011}, \cite{BMF2011} and \cite{FRZ2005} for examples of this method.


\noindent Let us consider the stationary process $(\Upsilon_t)_{t \in \mathbb{Z}}$, defined as follows
\begin{align}
    \Upsilon_t := Z_t - \mathbb{E}\left(Z_t\right),\quad t\in\nbZ ,\label{eq:5.2}
\end{align}
where we recall that $Z_t:=Z_t(\theta_0) = \left(\{\underline{X}_{t+1}\otimes\underline{X}_t\}',\dots, \{\underline{X}_{t+N}\otimes\underline{X}_t\}'\right)'\in\mathbb{R}^{p^2N}$. The process $(\Upsilon_t)_{t\in\mathbb{Z}}$, as defined, is in $L^2$, is stationary and thus admits the following Wold decomposition
\begin{align}
\label{eq:5.3}
    \Upsilon_t = u_t + \sum_{i=1}^{\infty} \Phi_i u_{t-i},\quad t\in\nbZ ,
\end{align}
where $\Phi_i \in \mathbb{R}^{p^2N \times p^2N}$, $i\in\nbZ$, and  $(u_t)_{t \in \mathbb{Z}} \in \mathbb{R}^{p^2N}$ is a weak white noise with variance-covariance matrix $\Sigma_u := \Var(u_t)$.
We then make the following assumptions
\begin{enumerate}
    \item[$(\mathbf{A_7})$] The matrix $\Sigma_u$ is non-singular.
    \item[$(\mathbf{A_8})$] We have $\displaystyle\sum_{i=1}^{\infty} \|\Phi_i\| < \infty$ and $\det\left(I_{p^2N} + \displaystyle\sum_{i=1}^{\infty} {\Phi}_i z^i \right) \neq 0 \quad \text{if } |z| \leq 1$.
\end{enumerate}
{We recall from \citet[Theorem 5.7.1 p.187]{brockwell1991time} that the weak noise process $(u_t)_{t \in \mathbb{Z}}$ can be expressed in function of $(\Upsilon_t)_{t \in \mathbb{Z}}$ as $u_t= \Upsilon_t- {\cal P}_{t-1} (\Upsilon_t)$, where ${\cal P}_{t-1}$ is the projection on the space spanned by the random vectors $\Upsilon_s$, $s\le t-1$. Hence the matrix $\Sigma_u$ in $(\mathbf{A_7})$ is non singular if $\Upsilon_t$ is a linear combination of the past random vectors $\Upsilon_s$, $s\le t-1$, which is not possible in practice (except in rare cases) because of the Markov chain, as this would imply that $\Delta_t$ depends deterministically on $\Delta_s$, $s\le t-1$.}

\noindent Let us introduce the matrix valued function $\varphi:z\in\nbC \mapsto I_{p^2N} + \displaystyle\sum_{i=1}^{\infty} \varphi_i z^i \in \nbC^{p^2N\times p^2N}$ . Under assumptions $(\mathbf{A_7})$ and $(\mathbf{A_8})$, the process $(\Upsilon_t)_{t\in\mathbb{Z}}$ admits an $AR(\infty)$ representation of the form
\begin{align}
\label{eq:5.4}
  \bm{\varphi}(L)\Upsilon_t := \Upsilon_t - \sum_{i=1}^{\infty} \varphi_i \Upsilon_{t-i} = u_t, \quad u_t \perp \{\Upsilon_{t-1},\dots,\Upsilon_{t-i},\dots\}.
\end{align}
such that $\sum_{i=1}^{\infty}\|\varphi_i\| < \infty$ and $\det(\bm{\varphi}(z)) \neq 0$ if $|z| \leq 1$, where $L$ denotes the back-shift operator. By interpreting $(2\pi)^{-1}\bm{I}$ as the spectral density of $(\Upsilon_t)_{t\in\mathbb{Z}}$ evaluated at frequency $0$ (see \citet[page 459]{brockwell1991time}), we have
\begin{align*}
\label{eq:5.5}
  \bm{I} = \bm{\varphi}(1)^{-1}\Sigma_u\bm{\varphi}'(1)^{-1}.
\end{align*}



\noindent Consider the regression of $\Upsilon_t$ on $\Upsilon_{t-1},\dots,\Upsilon_{t-r}$ defined by
\begin{equation}\label{eq:22}
\Upsilon_t=\sum_{k=1}^{r}\varphi_{r,k}\Upsilon_{t-k}+u_{r,t},
\end{equation}
where $u_{r,t}$ is uncorrelated with $\Upsilon_{t-1},\dots,\Upsilon_{t-r}$. Since the process $(\Upsilon_t)_{t\in\mathbb{Z}}$ is not observable, we introduce $\hat{\Upsilon}_t$ obtained by replacing $\mathbb{E}\left(Z_t\right)$ by its empirical estimator $\hat{\bm{C}}_{1}^N := \left(\hat{C}'_{1,0}, \dots, \hat{C}'_{N,0}\right)'$ in Equation (\ref{eq:5.2}):
\begin{align*}
    \hat{\Upsilon}_t = Z_t - \hat{\bm{C}}_{1}^N,\quad t=1,\dots,n.
\end{align*}
\noindent Let $\hat{\varphi}_{1,r}, \dots, \hat{\varphi}_{r,r}$ be the coefficients of the multivariate regression of $\hat{\Upsilon}_t$ on $\hat{\Upsilon}_{t-1}, \dots, \hat{\Upsilon}_{t-r}$, for $t = 1, \dots, n - p^2N$. Let $\hat{u}_{r,t}$ be the residuals of this regression and  let $\hat{\Sigma}_{\hat{u}_r}$ be the empirical variance of $\hat{u}_{r,1},\dots,\hat{u}_{r,r}$. We define
\begin{equation}\label{def_phi_chapeau}
\hat{\varphi}:z\in \nbC\mapsto I_{p^2N} + \displaystyle\sum_{i=1}^{r} {\hat{\varphi}}_{i,r} z^i\in \nbC^{p^2N\times p^2N} .
\end{equation}
\noindent The asymptotic study of the estimator of $\bm{I}$ using the spectral density method is given in the following theorem.
\begin{t1}
\label{thm:4}
We assume that the conditions of Theorem \ref{thm:3} are satisfied. Furthermore, we assume that the process defined in (\ref{eq:5.2}) admits an $AR(\infty)$ representation as defined in (\ref{eq:5.4}), where $\|\varphi_i\| = \mathrm{o}(i^{-2})$ as $i \rightarrow \infty$,  $(\mathbf{A_7})$ and $(\mathbf{A_8})$ are satisfied. Additionally, we assume $\|\eta_t\|_{8+4\nu} < \infty$ for some $\nu > 0$. Under these conditions, the spectral estimator of the matrix $\bm{I}$ is defined as
\begin{align}\label{def_ISP}
  \hat{\bm{I}}^{SP} := \hat{\bm{\varphi}}(1)^{-1}\hat{\Sigma}_{\hat{u}_r}\hat{\bm{\varphi}}'(1)^{-1} 
\end{align}
Then the following convergence in probability holds
$$
\hat{\bm{I}}^{SP} \xrightarrow[n\rightarrow\infty]{\nbP}\bm{\varphi}(1)^{-1}\Sigma_u\bm{\varphi}'(1)^{-1}= \bm{I}
$$
where $r$ is such that $r = r(n) \rightarrow \infty$ and $r^3/n \rightarrow 0$ as $n \rightarrow \infty$.
\end{t1}
\noindent   The proof of this theorem is given in Section \ref{proofthm:4}. {Going back to the required assumptions, and in view of the expression of the estimator \eqref{def_ISP} of the matrix $I$, one observes that $(\mathbf{A_7})$ and $(\mathbf{A_8})$ may in practice be verified by checking the estimators computed with the data, i.e. that $\hat{\Sigma}_{\hat{u}_r}$ is invertible, that $\sum_{i=1}^r \| \hat{\varphi}_{i,r}\|$ is bounded in $r\ge 1$, and that the equation $\det \hat{\varphi}(z)=0$ does not admit any solution $z\in \nbC$ with $|z|\le 1$, where $\hat{\varphi}(z)$ is defined in \eqref{def_phi_chapeau}. }
\begin{r1}
{
For the calculation of \(\hat{I}^{SP}\), we used the \texttt{statsmodels} function from the \texttt{Python} package \textbf{VAR}. The order \(r\) of the AR model can be automatically selected by AIC (Akaike Information Criterion).
Other information criteria can also be used: BIC, FPE, HQ. 
}
\end{r1}
\section{Estimating non-centered ARHMC$(p)$}
\label{sec:6}
\noindent
In this section, we extend the analysis of Model~\eqref{eq:1} to the non-centered case.

\subsection{Parameter estimation}
\noindent
Let \((Y_t)_{t \in \mathbb{Z}}\) be a real-valued stochastic process exhibiting regime shifts governed by an unobserved Markov chain \((\Delta_t)_{t \in \mathbb{Z}}\) defined in Section \ref{sec:2}.
For a fixed integer \(p \geq 1\), the dynamics of the process \((Y_t)\) within each regime follow the equation
\begin{align}
\label{eq:43}
    Y_t - m(\Delta_t) = \sum_{i=1}^{p} a_i(\Delta_t) \left\{ Y_{t-i} - m(\Delta_{t-i}) \right\} + f(\Delta_t)\eta_t,
\end{align}
where the weak white noise process \((\eta_t)_{t \in \mathbb{Z}}\), the functions $a_i(\cdot)$ for all $1 \leq i \leq p$ and $f(\cdot)$
are given in Section \ref{sec:2}.
%
\noindent
In this formulation, the  parameter vector of interest is defined by \(\vartheta_0 := (\theta_0, \underline{\mathfrak{m}}_0)\), with \(\underline{\mathfrak{m}}_0 := (m(1), \dots, m(K))' \in \mathbb{R}^K\) and we recall that $\theta_0$ is defined in Section \ref{sec:2}. {Such a model is relevant, as the process oscillates around a tendency $m(\Delta_t)$ that thus depends on the state of the Markov chain, which  captures a more complex and accurate behaviour of the data since a set of data issued from some natural phenomenon needs not be centered and even has a tendency that depends on external (climatic or economic) factors. This is precisely this approach that was taken in the application to wind speed real data in the forthcoming Section \ref{sec:8}, where the estimation of those tendecies is paramount for the differentiated series.}

The parameter \(\vartheta_0\) belongs to the extended parameter space
\[
\mathbf{\Theta}: = \Theta \times \mathbb{R}^K,
\]
where \(\Theta\) is the parameter space introduced in Section \ref{sec:2}.

\noindent
As for Model~\eqref{eq:1}, we can also write the model \eqref{eq:43} in vector form as follows 
\begin{align}
\label{eq:ARHMC-vector}
    \underline{Y}_t - \underline{m}(\Delta_t) = \mathbf{A}(\Delta_t)\left\{ \underline{Y}_{t-1} - \underline{m}(\Delta_{t-1}) \right\} + \underline{f}(\Delta_t)\eta_t, \quad t \in \mathbb{Z},
\end{align}
where
\[
\underline{Y}_t := (Y_t, \dots, Y_{t-p+1})' \in \mathbb{R}^p, \quad 
\underline{m}(\Delta_t) := (m(\Delta_t), \dots, m(\Delta_{t-p+1}))' \in \mathbb{R}^p,
\]
and where the functions \(\mathbf{A} : \mathcal{S} \to \mathbb{R}^{p \times p}\) and \(\underline{f} : \mathcal{S} \to \mathbb{R}^p\) are defined as in Section \ref{sec:2}.



\begin{pr1}
Under Assumptions $(\mathbf{A_1})$, $(\mathbf{A_2})$ and $(\mathbf{A_5} )$, from \cite{brandt1986stochastic} the series 
\begin{equation}
\label{eq:44}
\underline{Y}_t-\underline{m}(\Delta_t) = \sum_{k=0}^{\infty} \left( \prod_{j=0}^{k-1} \mathbf{A}(\Delta_{t-j}) \right) \underline{f}{(\Delta_{t-k}}){\eta}_{t-k} = \underline{X}_t\in\nbR^p,\, \text{ for all } t\in\mathbb{Z}
\end{equation}
converges almost surely and the process $(Y_t-m(\Delta_t))_{t\in\mathbb{Z}}$, defined as the first component of $\underline{Y}_t-\underline{m}(\Delta_t)$ is the unique strictly stationary solution of (\ref{eq:43}), with the usual convention that \(\prod_{j=\ell}^{\ell'} = 1\) for \(\ell, \ell' \in \mathbb{Z}\) and \(\ell' < \ell\).
\end{pr1}
\begin{t1}
\label{thm:5}
Under Assumptions $(\mathbf{A_1})$, $(\mathbf{A_2})$, $(\mathbf{A_5})$ and $(\mathbf{A_6})$, the joint moments of the model (\ref{eq:43}) satisfy for all $k\in\mathbb{N}$,
\begin{equation}
\label{eq:2bis}
     \mathbb{E}(\Vect\{\underline{Y}_0\underline{Y}_k'\}) = \mathbb{E}(\Vect\{\underline{X}_0\underline{X}_k'\}) + (\bm{1}'\otimes I_{p^2})Q_{\underline{m}\otimes I_p}\Pi_{\underline{m}},
\end{equation}
where 
\(Q_{\underline{m}\otimes I_p}\) is the \(Kp^2 \times Kp\) matrix defined by \eqref{eq:1.3} with $F(i)=\underline{m}(i) \otimes I_p$, $i\in {\cal S}$ and \(\Pi_{\underline{m}}\in \nbR^{Kp}$ is defined as \eqref{def_Pi_G} with $ G(i)= \underline{m}(i)\in \nbR^{p}\), $i\in {\cal S}$.
\end{t1}
\noindent   The proof of this theorem is given in Section \ref{proofthm:5}.

 \noindent
In the following, for \(\vartheta \in \boldsymbol{\Theta}\), we denote by \(\mathfrak{C}_{k,0}(\vartheta)\), the joint moment \(\mathbb{E}(\underline{Y}_k \otimes \underline{Y}_0)\) evaluated at a given \(\vartheta\). For convenience, we use the simplified notation \(\mathfrak{C}_{k,0} := \mathfrak{C}_{k,0}(\vartheta_0)\) when the true parameter \(\vartheta_0\) is considered.

\noindent
We assume to have at our disposal the observations $(Y_1,\dots,Y_n)$. Let $M\in\nbN$ such that $p^2M\geq Kp+K^2+K$. We now define the following functions 
\begin{align}
\label{eq:47}
\bm{\Phi}^M :
\begin{cases}
    \mathring{\boldsymbol{\Theta}} \rightarrow \mathbb{R}^{p^2 M}, \\
    \vartheta \mapsto \big(\mathfrak{C}_{1,0}(\vartheta)', \ldots, \mathfrak{C}_{M,0}(\vartheta)'\big)',
\end{cases}
\qquad
\bm{G}^{M,n} :
\begin{cases}
    \mathring{\boldsymbol{\Theta}} \rightarrow \mathbb{R}^{p^2 M}, \\
    \vartheta \mapsto \bm{\Phi}^M(\vartheta)-\hat{\bm{\Phi}}^M,
\end{cases}
\end{align}
where the empirical counterpart \(\hat{\bm{\Phi}}^M(\vartheta_0)\) is defined by
\[
\hat{\bm{\Phi}}^M := \big(\hat{\mathfrak{C}}_{1,0}', \ldots, \hat{\mathfrak{C}}_{M,0}'\big)' \quad \text{with} \quad \hat{\mathfrak{C}}_k := \frac{1}{n-k} \sum_{t=1}^{n-k} \underline{Y}_{t+k} \otimes \underline{Y}_t, \ 1\leq k\leq M.
\]
\noindent We then introduce the estimation function 
    \begin{align*}
\label{eq:}
    \bm{\mathcal{G}}^{M,n}(\vartheta) := J_{\bm{\Phi}^M}(\vartheta)'\bm{G}^{M,n}(\vartheta), \ \vartheta \in \Theta.
\end{align*}
\begin{r1}
\label{rem:1}
\noindent
Observing that any element \(\vartheta \in \boldsymbol{\Theta}\) can be decomposed into a pair \((\theta, \underline{\mathfrak{m}}) = (\theta,0)+(0,\underline{\mathfrak{m}})\), where \(\theta \in \Theta\) and \(\underline{\mathfrak{m}} \in \mathbb{R}^K\), it follows that the functions defined in Equation~\eqref{eq:47} can be rewritten accordingly as 
\begin{align*}
\bm{\Phi}^{M}(\vartheta) = \bm{\Psi}^{M}(\theta)+\zeta^M(\underline{\mathfrak{m}}),
\end{align*}
where the vector \(\zeta^M(\underline{\mathfrak{m}}) := ((\bm{1}' \otimes I_{p^2}) Q_{\underline{\mathfrak{m}} \otimes I_p} \Pi_{\underline{\mathfrak{m}}}, \ldots, (\bm{1}' \otimes I_{p^2}) Q_{\underline{\mathfrak{m}} \otimes I_p} \Pi_{\underline{\mathfrak{m}}})' \in \mathbb{R}^{p^2M}\)  
and where \(\theta\) (resp. \(\underline{\mathfrak{m}}\)) is identified with the vector \((\theta, 0)\) (resp. \((0, \underline{\mathfrak{m}})\)).
\end{r1}
\noindent
As stated in Section \ref{sec:3}, a moment estimator of the true parameter \(\vartheta_0\) is defined as a zero of the function \(\bm{\mathcal{G}}^{M,n}(\cdot)\). To compute this solution, we rely on the Newton-Raphson method, following the same strategy used in the centered case (see Section \ref{sec:3}). The algorithm remains identical to Algorithm~\ref{algo:algo_1}, with the only modification is by replacing \(\bm{\mathcal{F}}^{M,n}(\cdot)\) by \(\bm{\mathcal{G}}^{M,n}(\cdot)\). The initial point \(\vartheta^0\) should be chosen in \(\mathbb{R}^{K^2 + Kp + K}\).
\subsection{Asymptotic properties} 
 \noindent We denote by \(\hat{\vartheta}_n\) the moment estimator of \(\vartheta_0\) obtained through Algorithm~\ref{algo:algo_1}.

 \noindent
The main results namely, the consistency and the asymptotic normality of the estimator \(\hat{\vartheta}_n\) can be viewed as a natural extension of the results established in Section \ref{sec:4} for the centered case.

\noindent We now be able to state the following consistency result.
\begin{t1}\label{thm:6}
Assume that the Jacobian matrix \(J_{\bm{\Phi}^M}(\vartheta_0)\) is of full rank. Then, there exists a neighborhood \(\mathfrak{V}_{\vartheta_0} \subset \boldsymbol{\Theta}\) of \(\vartheta_0\) and a sequence \((\hat{\vartheta}_n)_{n \in \mathbb{N}}\) taking values in \(\mathfrak{V}_{\vartheta_0}\) such that
\begin{equation}
\label{eq:48}
    \bm{\mathcal{G}}^{M,n}(\hat{\vartheta}_n) 
    = J_{\bm{\Phi}^M}(\hat{\vartheta}_n)' \bm{G}^{M,n}(\hat{\vartheta}_n) 
    = 0.
\end{equation}
Furthermore, we have
\begin{equation}
\label{eq:49}
    \hat{\vartheta}_n\xrightarrow[n\rightarrow\infty]{a.s}\vartheta_0
\end{equation}
\end{t1}
\subsubsection*{Proof.} 
\noindent The proof is the same as  Theorem~\ref{thm:2}, so it is omitted. It is sufficient to apply the same arguments by using the function $\bm{\mathcal{G}}^{M,n}(\cdot)$ in place of $\bm{\mathcal{F}}^{N,n}(\cdot)$. \qed

\begin{t1}
\label{thm:7}
Assume that Assumptions \((\mathbf{A_1})\) through \((\mathbf{A_6})\) hold, that the matrix $J_{\bm{\Phi}^M}(\vartheta_0)$ is of full rank. Let \(\hat{\vartheta}_n\) be the moment estimator of $\vartheta_0$ as defined in Equation (\ref{eq:48}), we have the following Central Limit Theorem 
\begin{align*}
\sqrt{n}(\hat{\vartheta}_n - \vartheta_0) \xrightarrow[n\rightarrow\infty]{\mathcal{D}} \mathcal{N}(0,\mathfrak{S}^{-1}\mathfrak{J}'\mathfrak{T}\mathfrak{J}'\mathfrak{S}^{-1}),
\end{align*}
where matrices $\mathfrak{S}, \mathfrak{J}$ and $\mathfrak{T}$ are defined as follows
\begin{align}
    \mathfrak{J}: & =  J_{\Phi^M}(\vartheta_0) = (J_{\Psi^M}(\theta_0),J_{\zeta^M}(\underline{\mathfrak{m}}_0)) \notag\\
    \mathfrak{S}: & = (J_{\Phi^M}(\vartheta_0))'J_{\Phi^M}(\vartheta_0) = \mathfrak{I'\mathfrak{I}}\notag\\
    \mathfrak{T}: & = \sum_{k=-\infty}^{\infty} \Cov(U_{t,M},U_{t-k,M}),\,\text{ with }\, U_{t,M} =  \bigl(\{\underline{Y}_{t+1} \otimes \underline{Y}_t\}',\, \ldots,\, \{\underline{Y}_{t+M} \otimes \underline{Y}_t\}' \bigr)'.
\end{align}
\end{t1}
\subsubsection*{Proof.} 
\noindent The proof of this theorem is the same as  Theorem~\ref{thm:3} and it is omitted.  The only notable difference lies in the higher dimensionality involved, due to the presence of the additional parameter \(\underline{\mathfrak{m}}_0\). \qed

\begin{r1}
The estimation of the matrix \(\mathfrak{I}\) can be carried out using a standard plug-in approach. More precisely, we construct \(\hat{\mathfrak{I}}_n\) by substituting the true parameter \(\vartheta_0\) with its consistent estimator \(\hat{\vartheta}_n\) in the theoretical expression of \(\mathfrak{I}\). This allows for a fully data-driven version of the matrix that retains the desired asymptotic properties under regularity conditions. Regarding the estimation of the long-run variance matrix \(\mathfrak{T}\), the methodology follows identically the procedure outlined in Section~\ref{sec:5}. The steps described therein remain applicable in the non-centered case as well. The only modification required is to work with the process \(U_{t,M}\) in place of the vector \(Z_t\), as the structure of the underlying process changes accordingly. This substitution ensures that the empirical covariance structure captures the correct second-order dependence of the transformed data under the general model formulation.
\end{r1}
\section{Numerical illustrations}
\label{sec:7}

\noindent In this section, we present a numerical illustration of the models introduced in Equations~(\ref{eq:1}) and \eqref{eq:43}. These numerical
illustrations  are made with the \texttt{Python} software. In Table~\ref{tab:noise_cases}, we summarize the different noise cases examined in this study. The Weak 2 noise is an extension of a noise process in \cite{romano1996inference}. Contrary to the Weak 1 and Weak 3 processes, the Weak 2 noise is not a martingale difference sequence for which the limit theory is more classical. We also note that in the following simulations the distributions of $(\eta_t)_{t\in\mathbb{Z}}$ for the Weak 2 and Weak 3 noises are standardized such that $\sigma^2_\eta=\mathbb{E}(\eta_t^2)=1$.
 The linear innovation $(\eta_t)_{t\in\nbZ}$ is driven by a process $(w_t)_{t\in\nbZ}$ generated from the standard normal distribution, $w_t \stackrel{\mathcal{D}}{=} \mathcal{N}(0,1)$. 

\begin{table}[H]
\centering
{
\begin{tabular}{p{2.0cm} p{4.4cm} p{4.5cm} p{4.1cm}}
\toprule
\textbf{Noise type} & \textbf{Expression for $\eta_t$} & \textbf{Description} & \textbf{Plausible application} \\
\midrule
Strong 
& $\eta_t = w_t$
& Benchmark specification with independent and identically distributed innovations.
& Standard reference case; suitable when the noise is reasonably modeled as i.i.d. \\ \\

Weak 1
& $\eta_t = w_t w_{t-1}$
& Uncorrelated but dependent noise, obtained through a multiplicative interaction with the previous innovation; it is a martingale difference sequence.
& Stylized dependence structure illustrating nonlinear effects; may mimic residual dependence in transformed financial or environmental series. \\ \\

Weak 2
& $\eta_t = \frac{1}{\sqrt{3}} w_t^2 w_{t-1}$
& Uncorrelated but dependent noise generated by a nonlinear interaction between the current innovation and the previous one; it is not a martingale difference sequence.
& Stylized dependence structure illustrating higher-order nonlinear effects; may mimic residual dependence in transformed financial, environmental or energy series. \\ \\

Weak 3
& $\vphantom{\dfrac{1}{1}}\begin{aligned}
\eta_t
&= \sqrt{\frac{1-a_0-\beta_0}{\omega_0}}\, h_t^{1/2} w_t, \\
h_t
&:= \omega_0 + a_0 \eta_{t-1}^2 + \beta_0 h_{t-1}
\end{aligned}$
& Normalized GARCH-type noise with conditional heteroskedasticity; it is also a martingale difference sequence.
& Financial log-returns, wind speed or wind-power series, and other data exhibiting volatility clustering or time-varying variability.  \\
\bottomrule
\end{tabular}
}
\caption{Different innovation processes considered in the simulation study, together with their interpretation and possible fields of application.}
\label{tab:noise_cases}
\end{table}

\subsection{The centered case}

\noindent Recall that for $p=1$, numerical investigations of consistency and asymptotic normality under various noise structures (strong white noise and weak white noise) have already been conducted in \cite{mainassara2025estimating}. Accordingly, this section replicates that approach for the case $p=2$. As noted at the beginning of this paper, imposing constraints on the transition probabilities significantly reduces the number of parameters to be estimated. Specifically, when we consider a two-regime setting ($K=2$), the parameter of interest, $\theta_0$, becomes an 8-dimensional vector ($\theta_0 \in \mathbb{R}^8$). Thus, only eight parameters need to be estimated in total. 
Fig~\ref{fig: simulation_X} below illustrates the dynamics of the process $(X_t)_{t\in\mathbb{Z}}$ under the influence of the various noise types listed in Table~\ref{tab:noise_cases}.

\begin{figure}[H]
    \centering
    \includegraphics[width=1\textwidth]{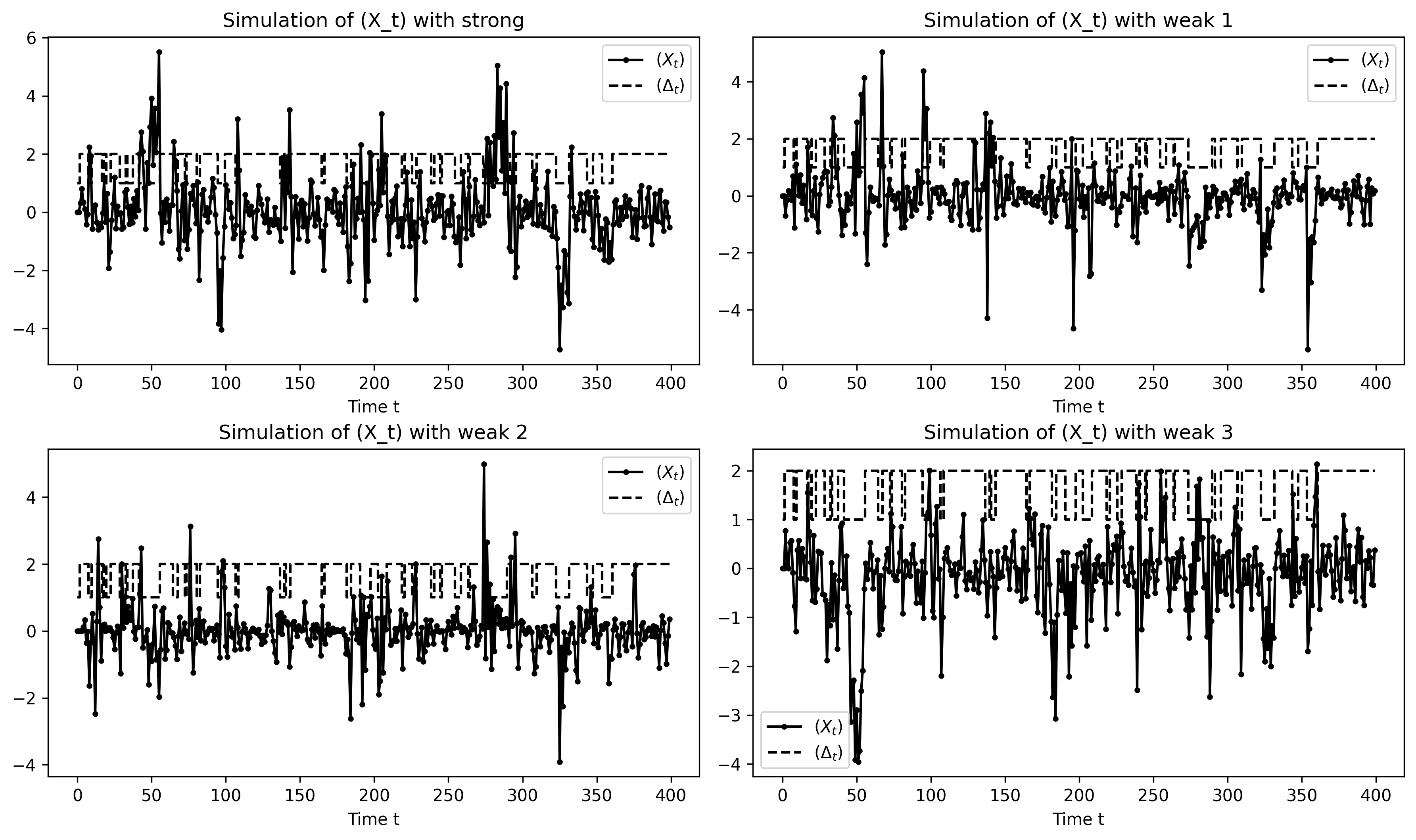}
    \caption{Simulation of length 400 for model (\ref{eq:1}).}
    \label{fig: simulation_X}
\end{figure}
\medskip

\begin{figure}[H]
    \centering

    \begin{subfigure}[b]{0.48\textwidth}
        \centering
        \includegraphics[width=\textwidth]{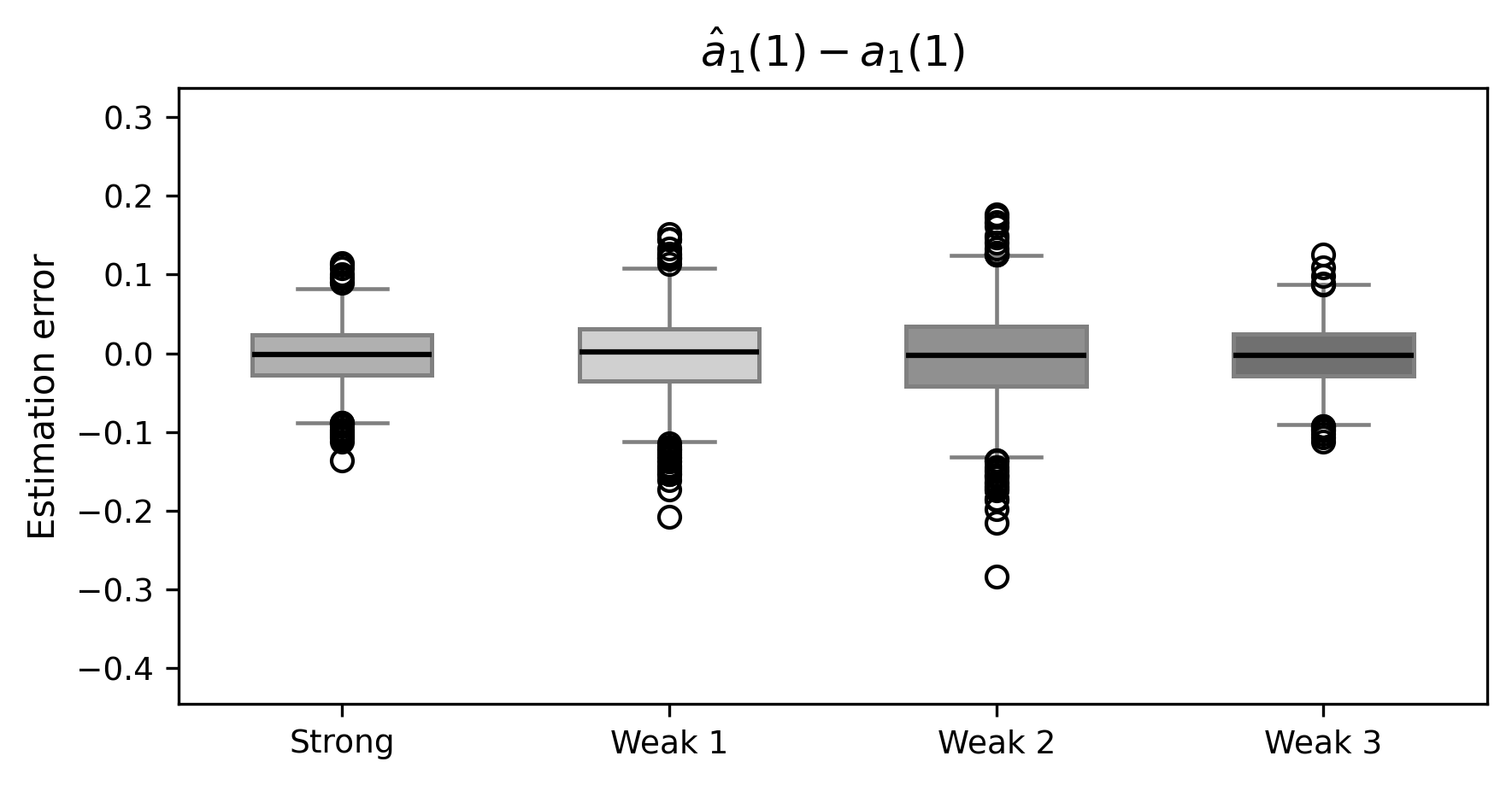}
        \label{fig:boxplot_a}
    \end{subfigure}
    \hfill
    \begin{subfigure}[b]{0.48\textwidth}
        \centering
        \includegraphics[width=\textwidth]{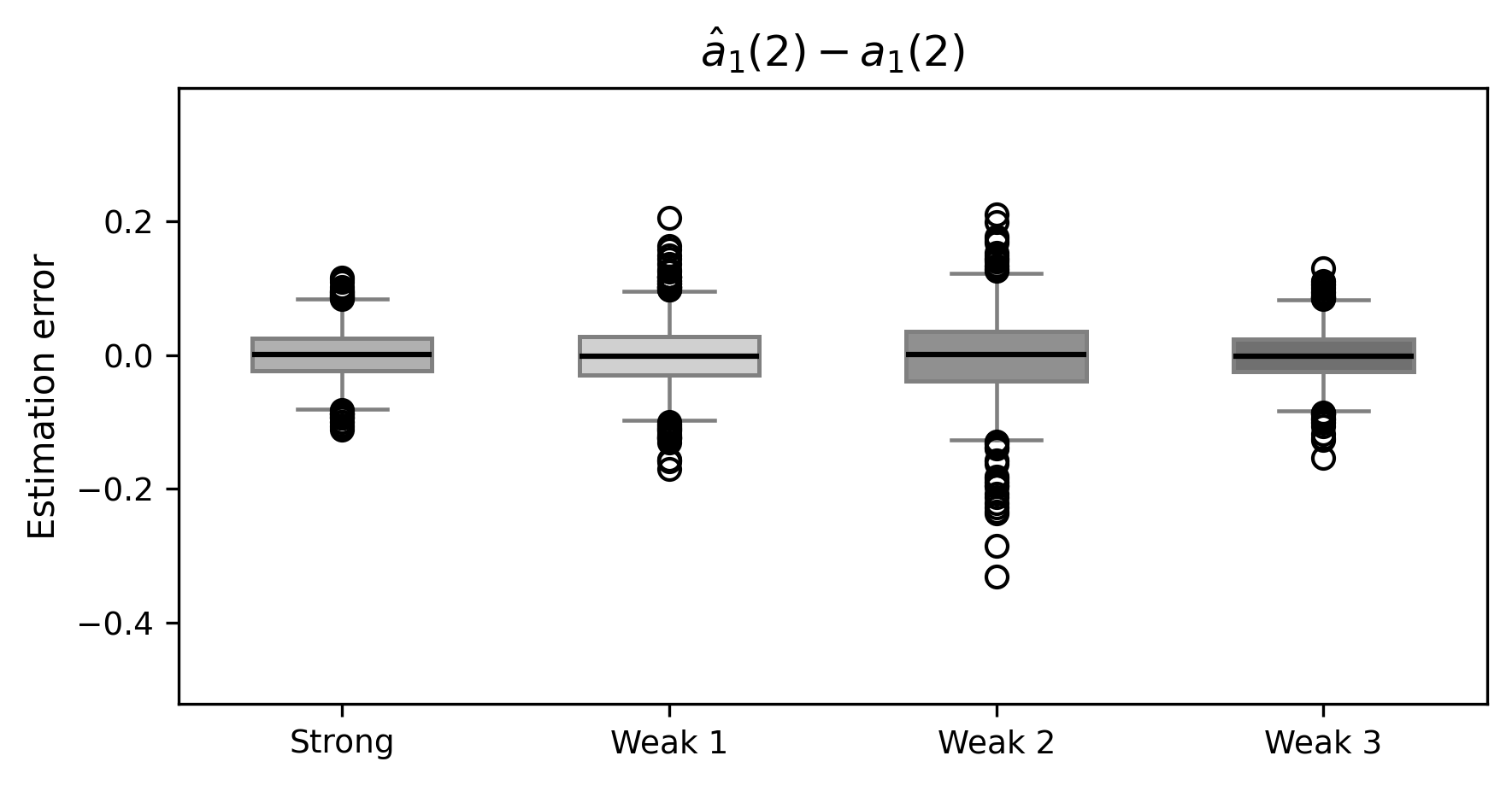}
        \label{fig:boxplot_b}
    \end{subfigure}

    \vspace{0.25cm}

    \begin{subfigure}[b]{0.48\textwidth}
        \centering
        \includegraphics[width=\textwidth]{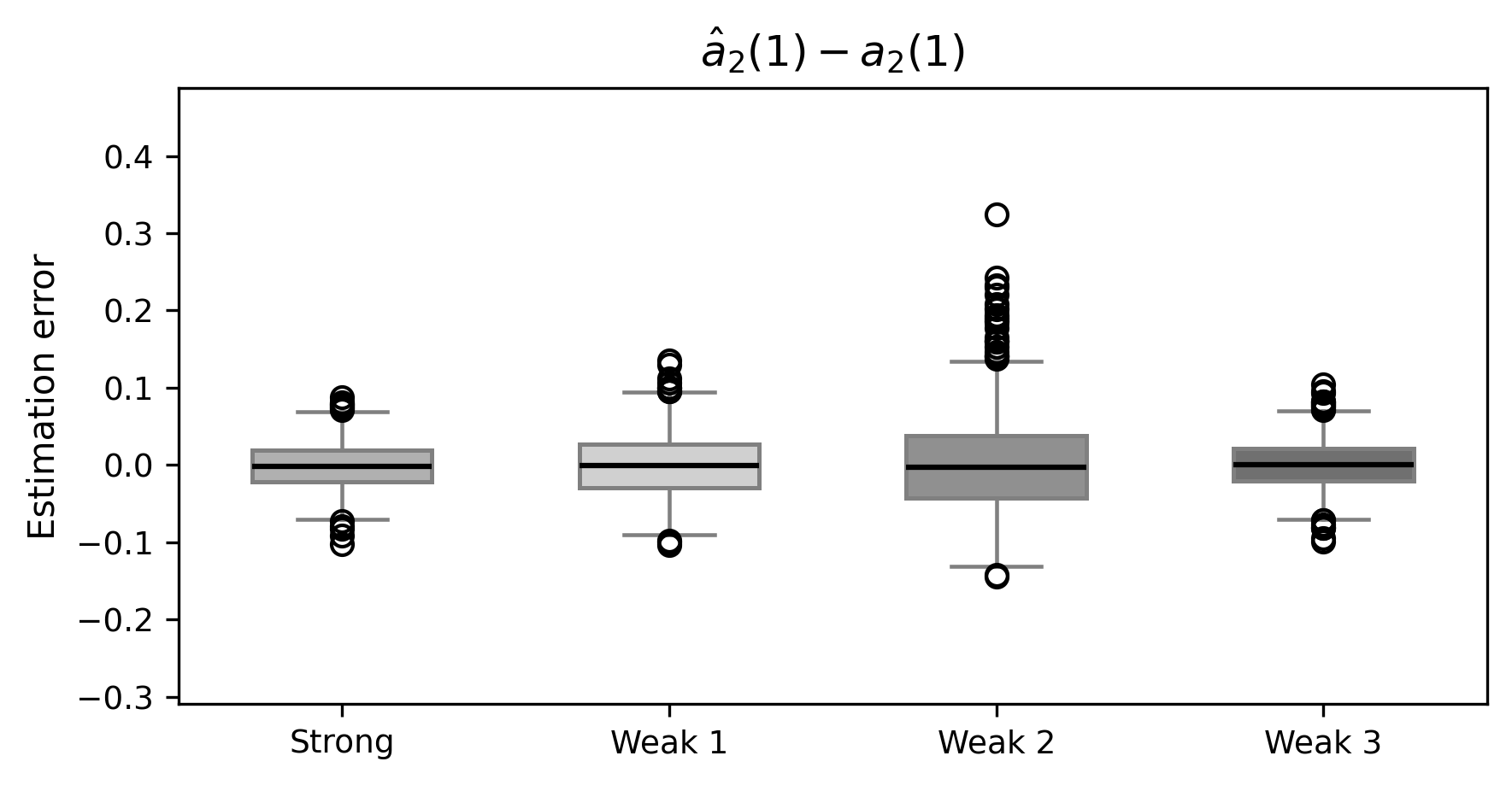}
        \label{fig:boxplot_c}
    \end{subfigure}
    \hfill
    \begin{subfigure}[b]{0.48\textwidth}
        \centering
        \includegraphics[width=\textwidth]{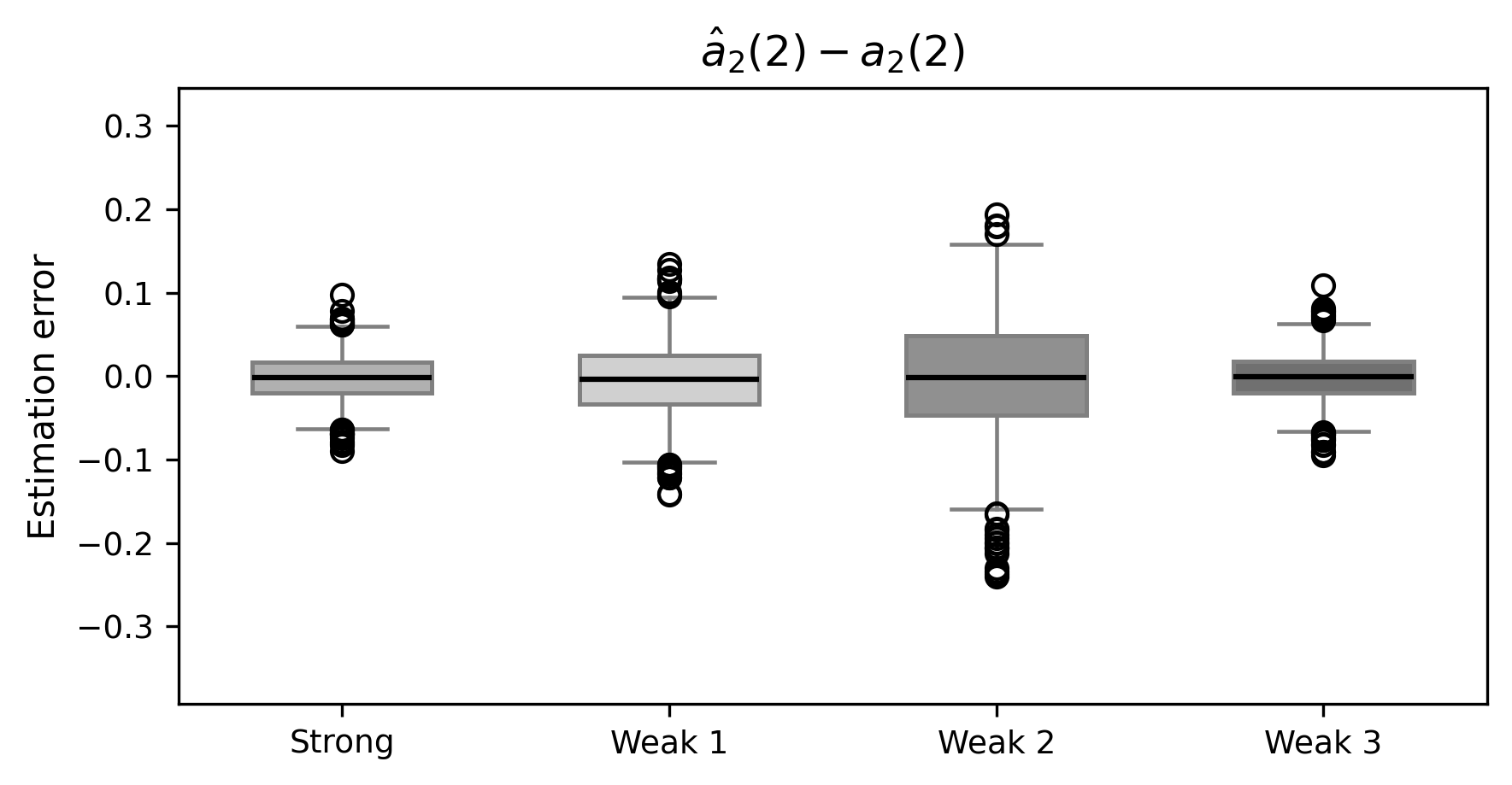}
        \label{fig:boxplot_d}
    \end{subfigure}

    \vspace{0.25cm}

    \begin{subfigure}[b]{0.48\textwidth}
        \centering
        \includegraphics[width=\textwidth]{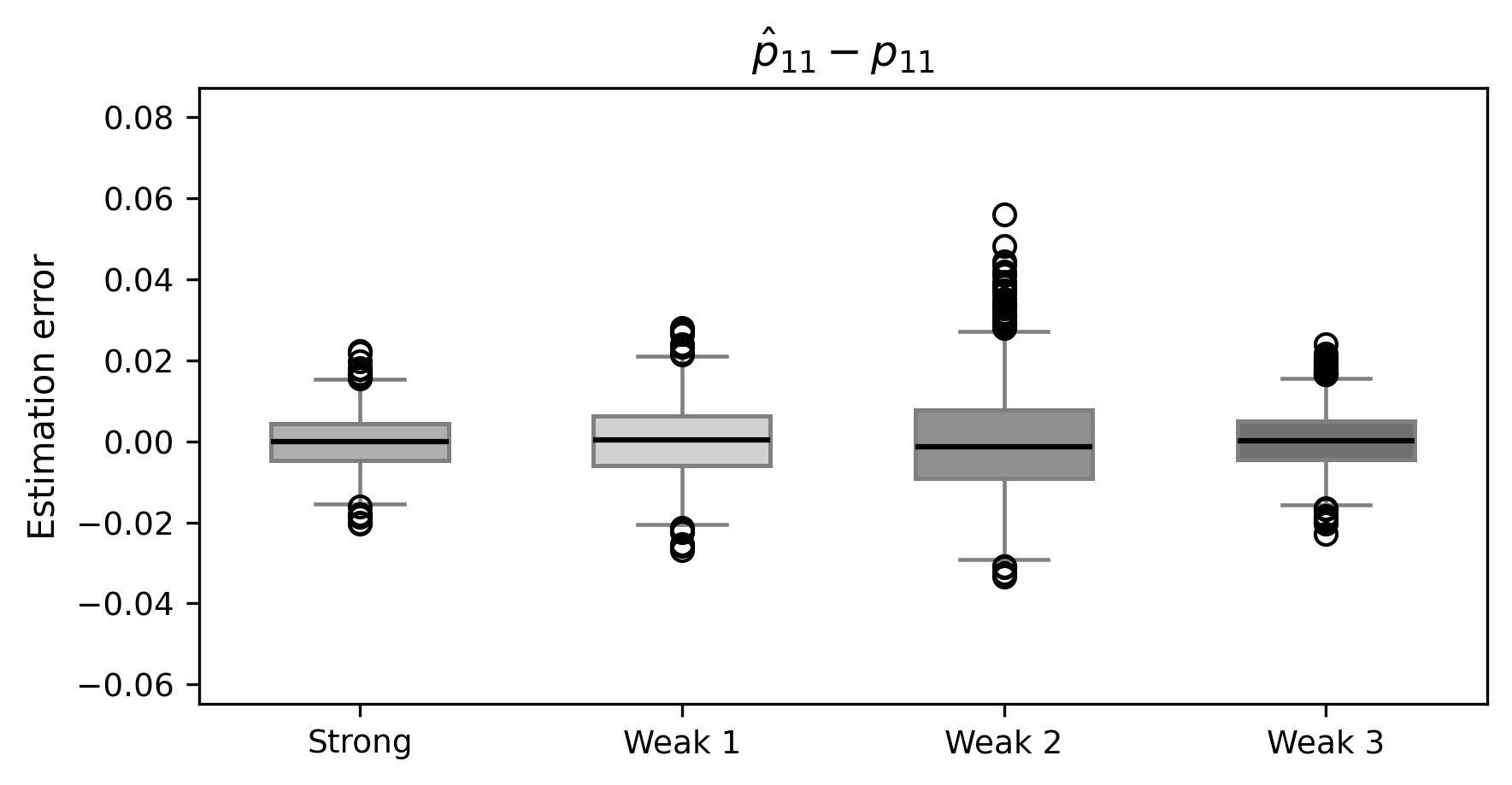}
        \label{fig:boxplot_e}
    \end{subfigure}
    \hfill
    \begin{subfigure}[b]{0.48\textwidth}
        \centering
        \includegraphics[width=\textwidth]{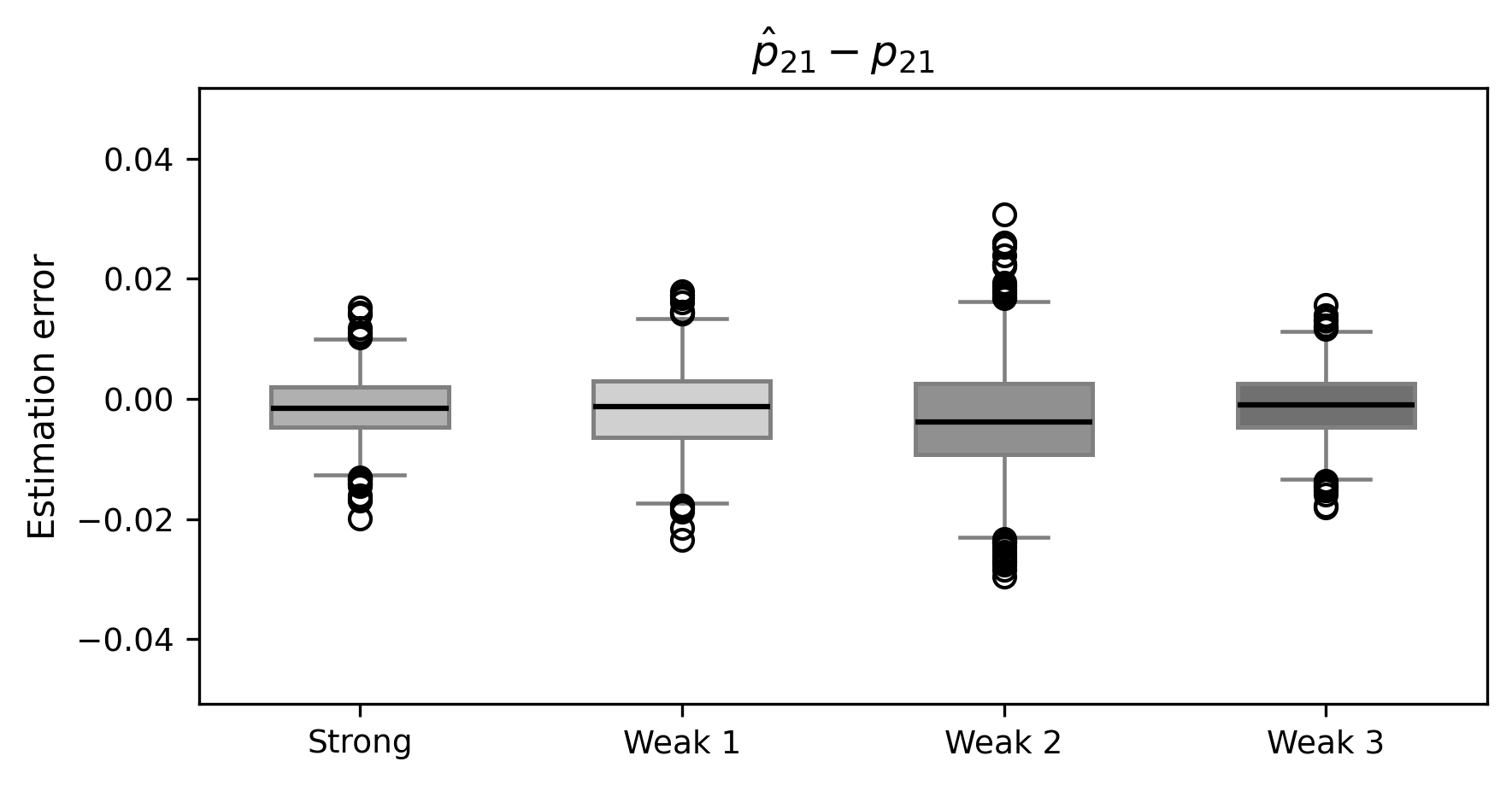}
        \label{fig:boxplot_f}
    \end{subfigure}

    \vspace{0.25cm}

    \begin{subfigure}[b]{0.48\textwidth}
        \centering
        \includegraphics[width=\textwidth]{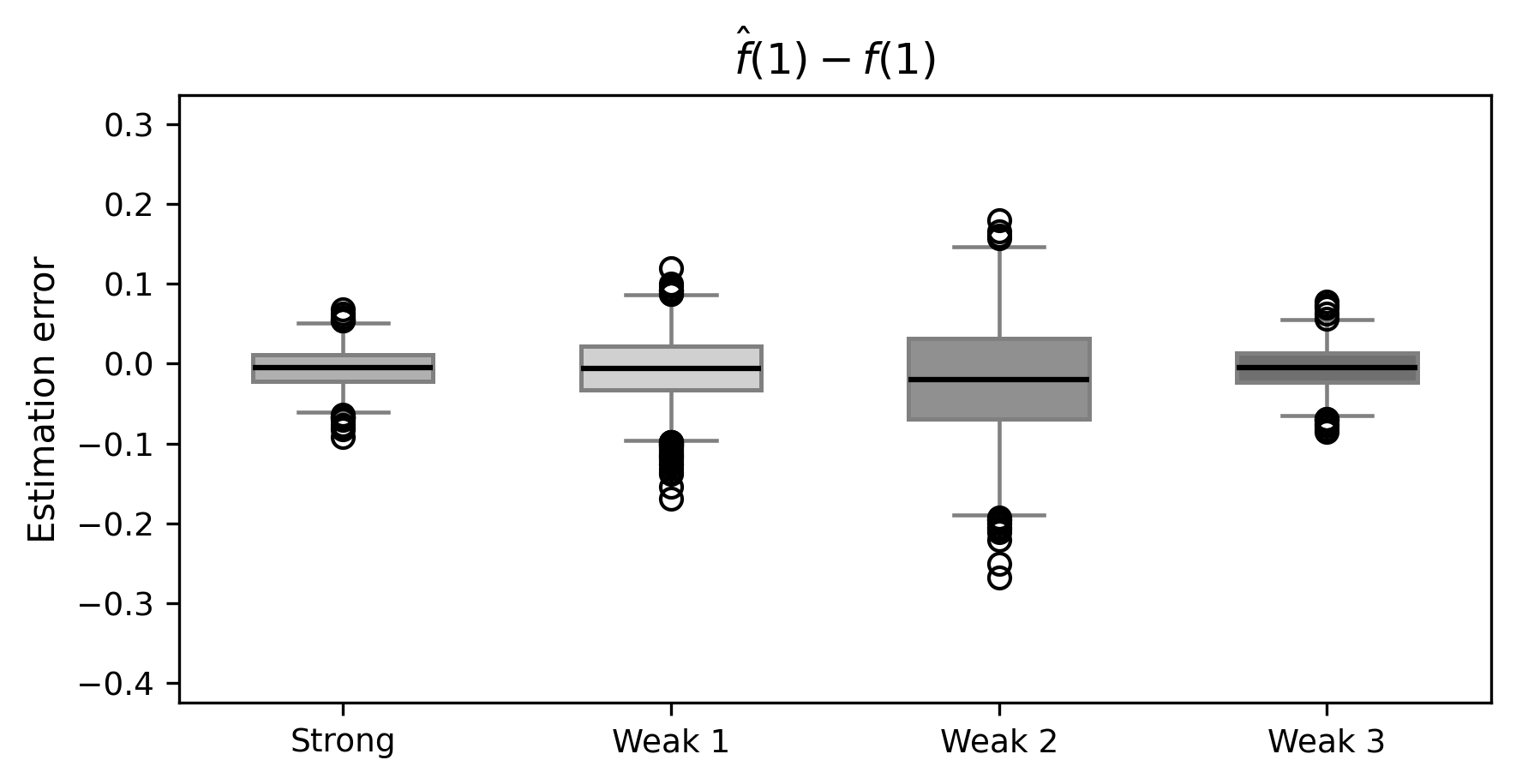}
        \label{fig:boxplot_g}
    \end{subfigure}
    \hfill
    \begin{subfigure}[b]{0.48\textwidth}
        \centering
        \includegraphics[width=\textwidth]{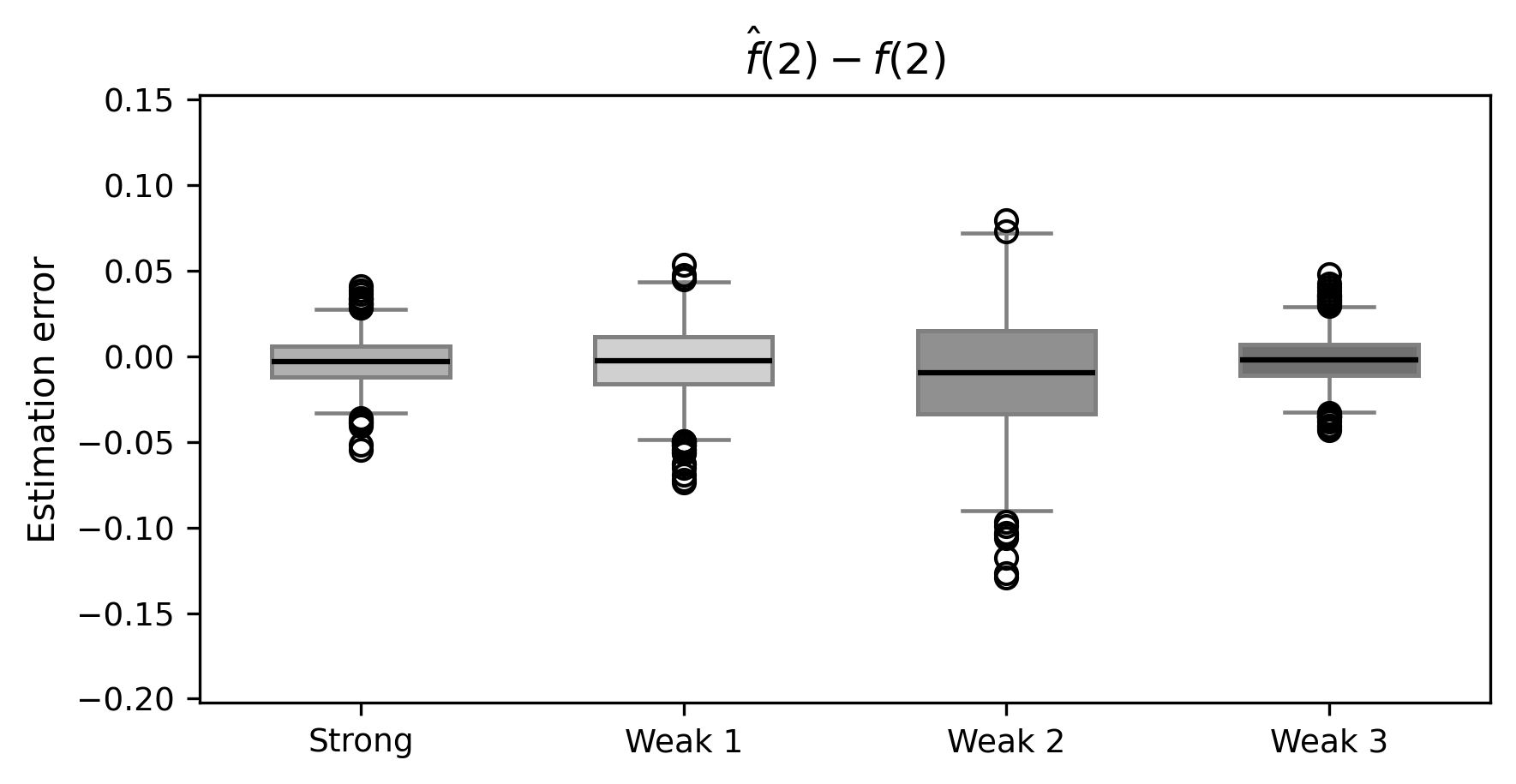}
        \label{fig:boxplot_h}
    \end{subfigure}

    \caption{{Moment estimators obtained from $N=1,000$ independent simulations of the model \eqref{eq:1} of size $n=2,000$ and with unknown parameter $\theta_0=(0.2, 0.3, 0.5, -0.4, 0.6, 0.3, 1.5, 0.5)$,  when the noise \((\eta_t)\) is defined respectively as \textit{strong}, \textit{weak 1}, \textit{weak 2} and \textit{weak 3}. The distribution of the estimation error \(\hat{\theta}_n(i) - \theta_0(i)\) for \(i = 1, \ldots, 8\) in box-and-whisker plots as a function of these four types of noise.
    }}
    \label{fig: errors}
\end{figure}

\begin{figure}[H]
    \centering
    \includegraphics[width=0.9\textwidth]{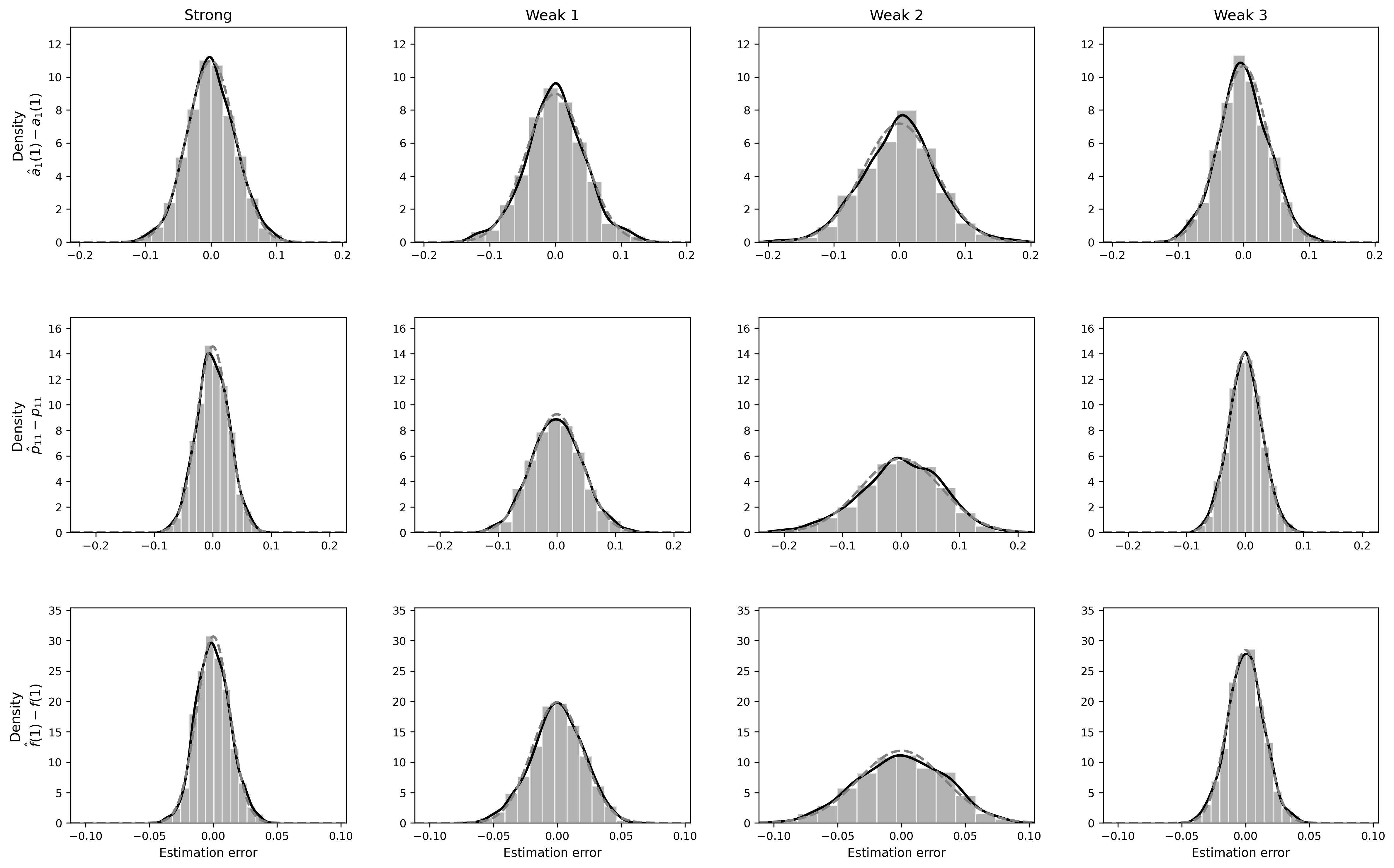}
    \caption{{Empirical distributions of the estimation errors for three selected parameters, displayed across the four noise specifications (strong, weak 1, weak 2, and weak 3). The first row corresponds to $\hat{a}_1(1)-a_1(1)$, the second to $\hat{p}_{11}-p_{11}$, and the third to $\hat{f}(1)-f(1)$. In each panel, the solid curve represents the kernel density estimate, while the dashed curve denotes the Gaussian density with the same empirical mean and variance.}}
    \label{fig: density}
\end{figure}

\begin{figure}[H]
    \centering
    \includegraphics[width=0.95\textwidth]{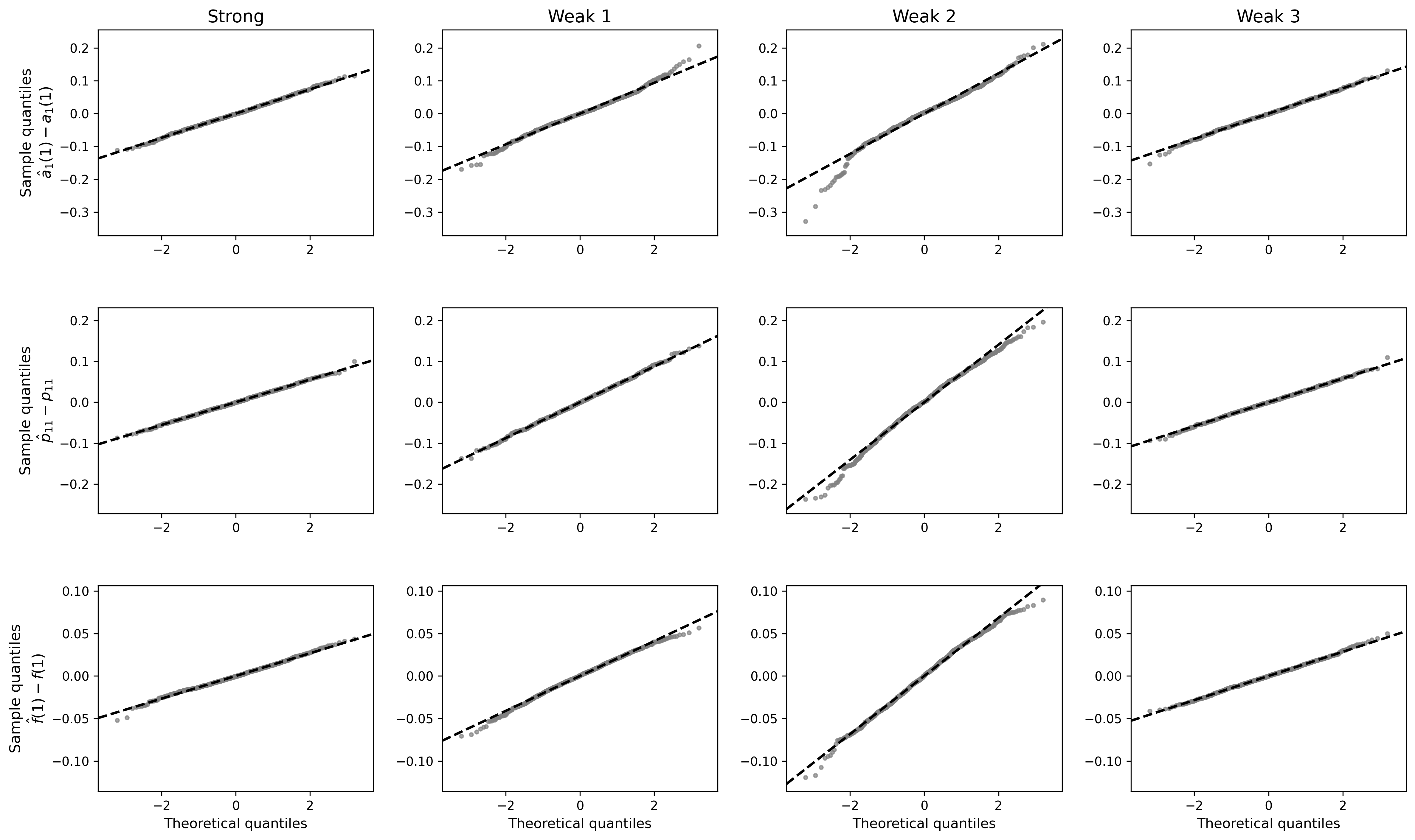}
    \caption{{Q--Q plots of the estimation errors for three selected parameters under the four noise specifications (strong, weak 1, weak 2, and weak 3). The first row corresponds to $\hat{a}_1(1)-a_1(1)$, the second to $\hat{p}_{11}-p_{11}$ and the third to $\hat{f}(1)-f(1)$. The dashed line is the Gaussian reference line.}}
    \label{fig: qqplot}
\end{figure}

{
\noindent We simulated $T = 1,000$ independent trajectories of model~(\ref{eq:1}), each with lengths $n = 2,000$ and $n = 300$, respectively, based on the types of noises defined in Table~\ref{tab:noise_cases}. For each trajectory, the parameter vector $\theta_0$ was estimated using the moments method described in Section~\ref{sec:2} across the different noise types listed in Table~\ref{tab:noise_cases} to illustrate various scenarios. The procedure is as follows. We begin with the vector 
$\theta_0 = (0.2,0.3,0.5,-0.4,0.6,0.3,1.5,0.5)'$
which we randomly initialize while ensuring that it satisfies the stationarity conditions $(\mathbf{A_5})$ and $(\mathbf{A_6})$.
Next, we simulate $T$ trajectories using a specific noise type from Table~\ref{tab:noise_cases} via the estimation function $\mathcal{F}^{N,n}(\cdot)$ defined in Equation~(\ref{eq:1.8}). Finally, for each trajectory, we estimate the parameter vector $\theta_0$. This approach allows us to assess the performance of our estimation method under different noise conditions. By varying the noise structure and trajectory length, we can evaluate the robustness and accuracy of the parameter estimates obtained through the method of moments.


\noindent Tab~\ref{tab:strong} through \ref{tab:weak_3} provide an overview of the statistical properties observed in our simulations of model~(\ref{eq:1}) under the various noise types introduced in Tab~\ref{tab:noise_cases}. Specifically, each table presents the mean value over $T$ replications of each parameter in \(\theta_0\), as well as its standard deviation (\emph{Std}) and bias  values recorded. These metrics capture both the central tendency and range of the parameter estimates across the simulations. 

As expected, Tables~\ref{tab:strong} through \ref{tab:weak_3} show that the bias and the \emph{Std} decrease when the size of the sample increases.

\noindent Fig.~\ref{fig: errors} through~\ref{fig: qqplot} respectively display the box plots, the empirical distributions and Q--Q plots of the estimation errors associated with the components of the parameter vector $\theta_0$. More precisely, these graphics illustrate the discrepancies
\[
\hat{{a}}_i(s) - {a}_i(s), \quad 
\hat{{p}}_{i,1} - {p}_{i,1}, \quad 
\hat{{f}}(s) - {f}(s), \quad 
\text{for } i,s \in \{1,2\}.
\]
Here, $\hat{{a}}_i(s)$, $\hat{{p}}_{i,1}$ and $\hat{{f}}(s)$ represent the moment estimators of the true parameters ${a}_i(s)$, ${p}_{i,1}$ and ${f}(s)$, respectively. These analyses are performed under the different noises structures listed in Tab~\ref{tab:noise_cases}. 
Each noise category is incorporated into the model from Equation~(\ref{eq:1}) with \(p=2\). 
%
\noindent Fig.~\ref{fig: errors} through~\ref{fig: qqplot} compare the distribution of the moment estimators of size $n=2,000$ in the strong and weak noise cases. 
The distributions of $\hat{{a}}_i(s)$, $ 
\hat{{p}}_{i,1}$ and $
\hat{{f}}(s)$ for $i,s \in \{1,2\}$ are similar in the all cases, whereas the moment estimators of $\hat{\theta}_n$ are more accurate in the strong and  the weak 3 cases than in the weak 1 and the weak 2 ones. Remark also that the empirical distributions of $\hat{\theta}_n$  in the weak 1 case are more accurate than the weak 2 ones. We draw the conclusion that the estimations of $\hat{{a}}_i(s)$, $ 
\hat{{p}}_{i,1}$ and $
\hat{{f}}(s)$ for $i,s \in \{1,2\}$ are more difficult under some complexities dependencies (see weak 2 which is not a martingale difference).
}

\begin{table}[H]
\centering
\setlength{\tabcolsep}{6pt} 
\renewcommand{\arraystretch}{1.0} 
{
\begin{tabular}{llccccccccc}
\toprule
&  & ${a}_1(1)$ & ${a}_1(2)$ & ${a}_2(1)$ & ${a}_2(2)$ & ${p}_{11}$ & ${p}_{21}$ & ${f}(1)$ & ${f}(2)$ \\
\cmidrule(lr){3-10}
& $\theta_0$ & 0.2 & 0.3 & 0.5 & -0.4 & 0.6 & 0.3 & 1.5 & 0.5 \\
\cmidrule(lr){3-10}
\multicolumn{10}{c}{$n = 300$} \\ 
\cmidrule(lr){1-10}
& Mean  & 0.19190  & 0.30030  & 0.49285  & -0.40999  &  0.60076 & 0.29453  &  1.47025 & 0.48980 \\
& Bias  & -0.00809  & 0.00030  & -0.00714  &  -0.00999 & 0.00076  & -0.00546  & -0.02975  & -0.01019 \\
& Std   & 0.10160  & 0.11687  &  0.10045 & 0.08699  & 0.03516  & 0.03048  & 0.09165  & 0.09483 \\
\midrule
\multicolumn{10}{c}{$n = 2,000$} \\ 
\cmidrule(lr){1-10}
& Mean  & 0.19743  & 0.29996  &  0.49748 & -0.39840  & 0.59779  &  0.29956 & 1.49050  & 0.49356 \\
& Bias  & -0.00256  & -0.00003  & -0.00251  & 0.00159  & -0.00220  & -0.00043  & -0.00949  &  -0.00643\\
& Std   & 0.06098  & 0.07952  & 0.08308  &  0.06283 & 0.03129  & 0.02863  & 0.07252  & 0.10020 \\
\bottomrule
\end{tabular}
}
\caption{Summary statistics of parameter estimates for model~(\ref{eq:1}) under \textit{strong} noise, based on 1000 replications for sequence lengths of 300 and 2000, respectively.}
\label{tab:strong}
\end{table}

\begin{table}[H]
\centering
\setlength{\tabcolsep}{6pt} 
\renewcommand{\arraystretch}{1.0} 
{
\begin{tabular}{llccccccccc}
\toprule
&  & ${a}_1(1)$ & ${a}_1(2)$ & ${a}_2(1)$ & ${a}_2(2)$ & ${p}_{11}$ & ${p}_{21}$ & ${f}(1)$ & ${f}(2)$ \\
\cmidrule(lr){3-10}
& $\theta_0$ & 0.2 & 0.3 & 0.5 & -0.4 & 0.6 & 0.3 & 1.5 & 0.5 \\
\cmidrule(lr){3-10}
\multicolumn{10}{c}{$n = 300$} \\ 
\cmidrule(lr){1-10}
& Mean  &  0.19231 &  0.29569 & 0.48963  & -0.41799  & 0.59750  &  0.29087 & 1.45162  & 0.47075 \\
& Bias  & -0.00768  & -0.00430  & -0.01036  & -0.01799  & -0.00249  & -0.00912  &  -0.04837 & -0.02924 \\
& Std   & 0.12510  & 0.14345  & 0.11426  & 0.11735  & 0.03758  & 0.03070  & 0.11880  &  0.10252 \\
\midrule
\multicolumn{10}{c}{$n = 2,000$} \\ 
\cmidrule(lr){1-10}
& Mean  & 0.19986  &  0.29892 & 0.49733  &  -0.40085 & 0.59758  & 0.29906  & 1.48271  & 0.49215 \\
& Bias  &  -0.00013  & -0.00107  & -0.00266  & -0.00085  & -0.00241  & -0.00093  & -0.01728  & -0.00784 \\
& Std   & 0.06982  & 0.08071  & 0.08466  & 0.07061  & 0.03196  & 0.02896  & 0.08656  & 0.09513 \\
\bottomrule
\end{tabular}
}
\caption{Summary statistics of parameter estimates for model~(\ref{eq:1}) under \textit{weak 1} noise, based on 1000 replications for sequence lengths of 300 and 2000, respectively.}
\label{tab:weak_1}
\end{table}
\begin{table}[H]
\centering
\setlength{\tabcolsep}{6pt} 
\renewcommand{\arraystretch}{1.0} 
{
\begin{tabular}{llccccccccc}
\toprule
&  & ${a}_1(1)$ & ${a}_1(2)$ & ${a}_2(1)$ & ${a}_2(2)$ & ${p}_{11}$ & ${p}_{21}$ & ${f}(1)$ & ${f}(2)$ \\
\cmidrule(lr){3-10}
& $\theta_0$ & 0.2 & 0.3 & 0.5 & -0.4 & 0.6 & 0.3 & 1.5 & 0.5 \\
\cmidrule(lr){3-10}
\multicolumn{10}{c}{$n = 300$} \\ 
\cmidrule(lr){1-10}
& Mean  & 0.18643 & 0.29236 & 0.49498 & -0.40938 & 0.59612 & 0.28459 & 1.40663 & 0.45605 \\
& Bias  & -0.01357 & -0.00764 & -0.00502 & -0.00938 & -0.00388 & -0.01541 & -0.09337 & -0.04395 \\
& Std   & 0.12593 & 0.13976 & 0.12284 & 0.14257 & 0.02736 & 0.01836 & 0.14365 & 0.06698 \\
\midrule
\multicolumn{10}{c}{$n = 2,000$} \\ 
\cmidrule(lr){1-10}
& Mean  & 0.19810 & 0.29489 & 0.50304 & -0.40338 & 0.60005 & 0.29684 & 1.48162 & 0.49093 \\
& Bias  & -0.00190 & -0.00511 & 0.00304 & -0.00338 & 0.00005 & -0.00316 & -0.01838 & -0.00907 \\
& Std   & 0.06506 & 0.05981 & 0.06133 & 0.07048 & 0.01315 & 0.00953 & 0.07180 & 0.03295 \\
\bottomrule
\end{tabular}
}
\caption{Summary statistics of parameter estimates for model~(\ref{eq:1}) under \textit{weak 2} noise, based on 1000 replications for sequence lengths of 300 and 2000, respectively.}
\label{tab:weak_2}
\end{table}

\begin{table}[H]
\centering
\setlength{\tabcolsep}{6pt} 
\renewcommand{\arraystretch}{1.0} 
{
\begin{tabular}{llccccccccc}
\toprule
&  & ${a}_1(1)$ & ${a}_1(2)$ & ${a}_2(1)$ & ${a}_2(2)$ & ${p}_{11}$ & ${p}_{21}$ & ${f}(1)$ & ${f}(2)$ \\
\cmidrule(lr){3-10}
& $\theta_0$ & 0.2 & 0.3 & 0.5 & -0.4 & 0.6 & 0.3 & 1.5 & 0.5 \\
\cmidrule(lr){3-10}
\multicolumn{10}{c}{$n = 300$} \\ 
\cmidrule(lr){1-10}
& Mean  & 0.19388  & 0.29890  & 0.49810  &  -0.40704 &  0.59849 &  0.29399 &  1.46606 & 0.48382 \\
& Bias & -0.00612 & -0.00110 & -0.00189 & -0.00705 & -0.00151 & -0.00601 & -0.03393 & -0.01618 \\
& Std & 0.10792 & 0.11803 & 0.10554 & 0.09093 & 0.03492 & 0.02935 & 0.10019 & 0.09659  \\
\midrule
\multicolumn{10}{c}{$n = 2,000$} \\ 
\cmidrule(lr){1-10}
& Mean  & 0.20010 & 0.29755 & 0.49553 & -0.39804 & 0.59915 & 0.29837 & 1.48813 & 0.49680 \\
& Bias  & 0.00010 & -0.00245 & -0.00447 & 0.00196 & -0.00085 & -0.00163 & -0.01187 & -0.00320 \\
& Std   & 0.05859 & 0.07784 & 0.08022 & 0.06295 & 0.03084 & 0.02975 & 0.07486 & 0.09333 \\
\bottomrule
\end{tabular}
}
\caption{Summary statistics of parameter estimates for model~(\ref{eq:1}) under \textit{weak 3} noise, where $(\omega_0, \alpha_0, \beta_0) = (0.2,\ 0.1,\ 0.5)$. The statistics are based on 1000 replications for sequence lengths of 300 and 2000, respectively.}
\label{tab:weak_3}
\end{table}

\subsection{The non-centered case}
\begin{figure}[H]
    \centering

    \begin{subfigure}[b]{0.48\textwidth}
        \centering
        \includegraphics[width=\textwidth]{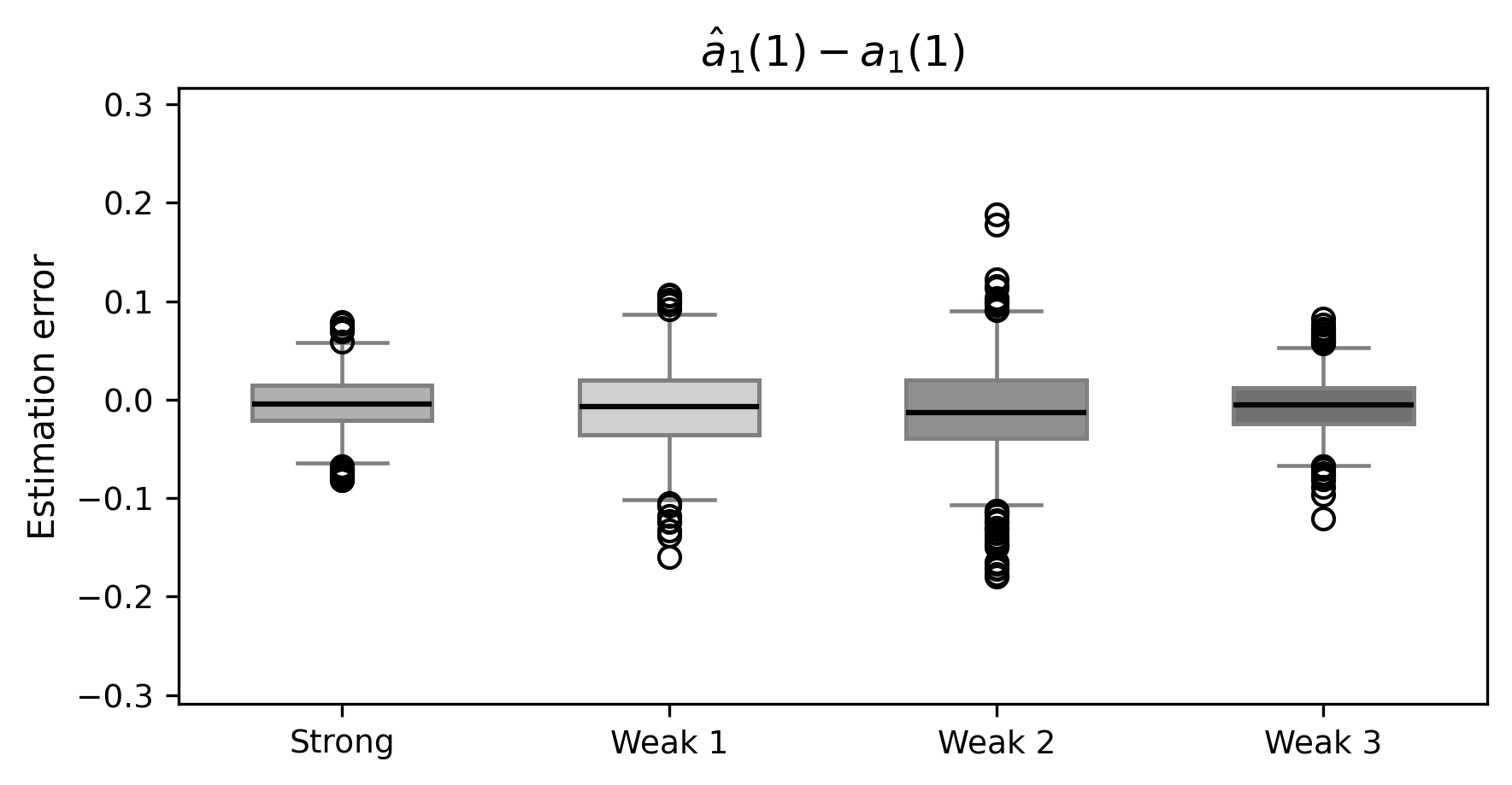}
    \end{subfigure}
    \hfill
    \begin{subfigure}[b]{0.48\textwidth}
        \centering
        \includegraphics[width=\textwidth]{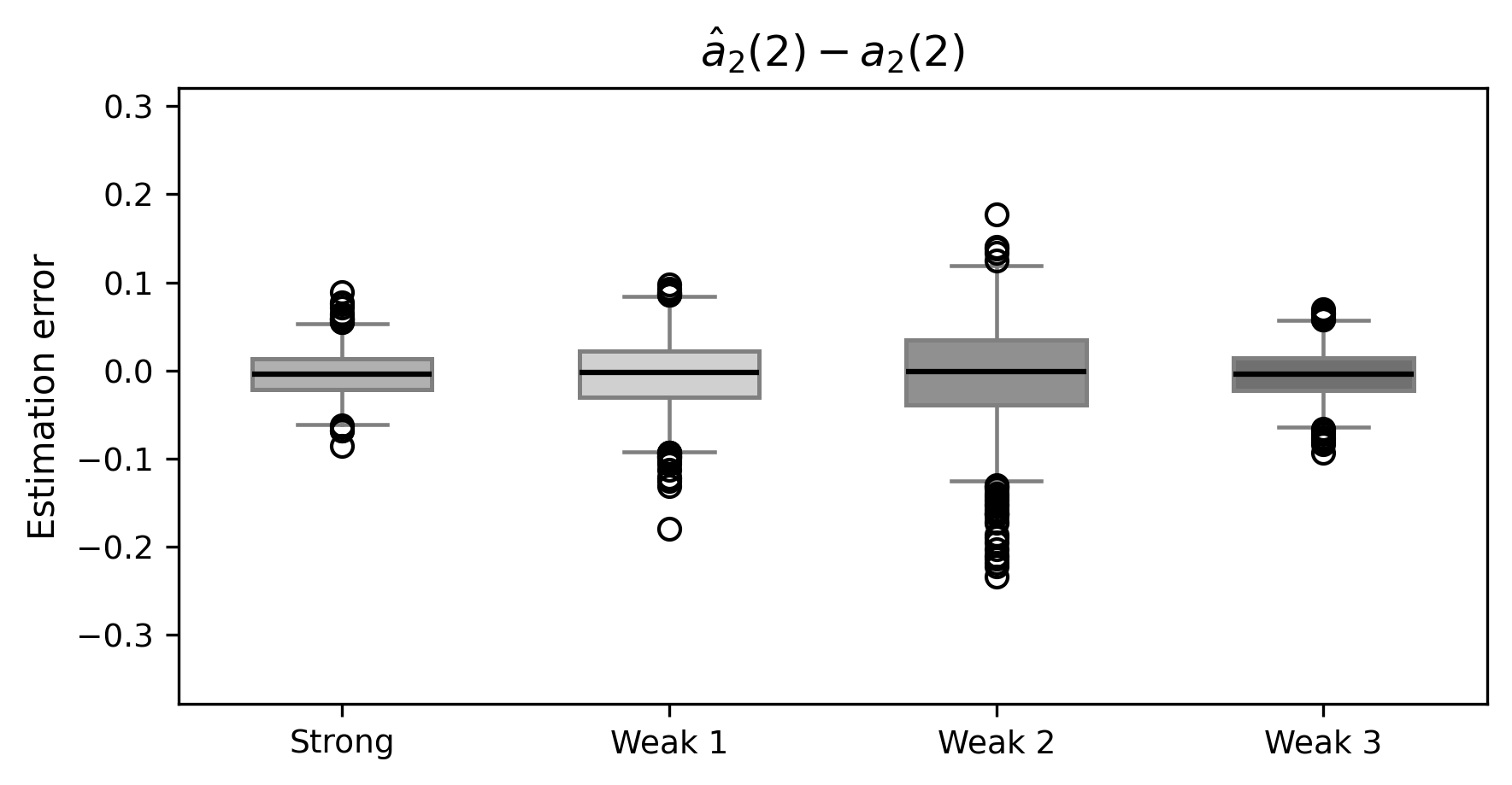}
    \end{subfigure}

    \vspace{0.25cm}

    \begin{subfigure}[b]{0.48\textwidth}
        \centering
        \includegraphics[width=\textwidth]{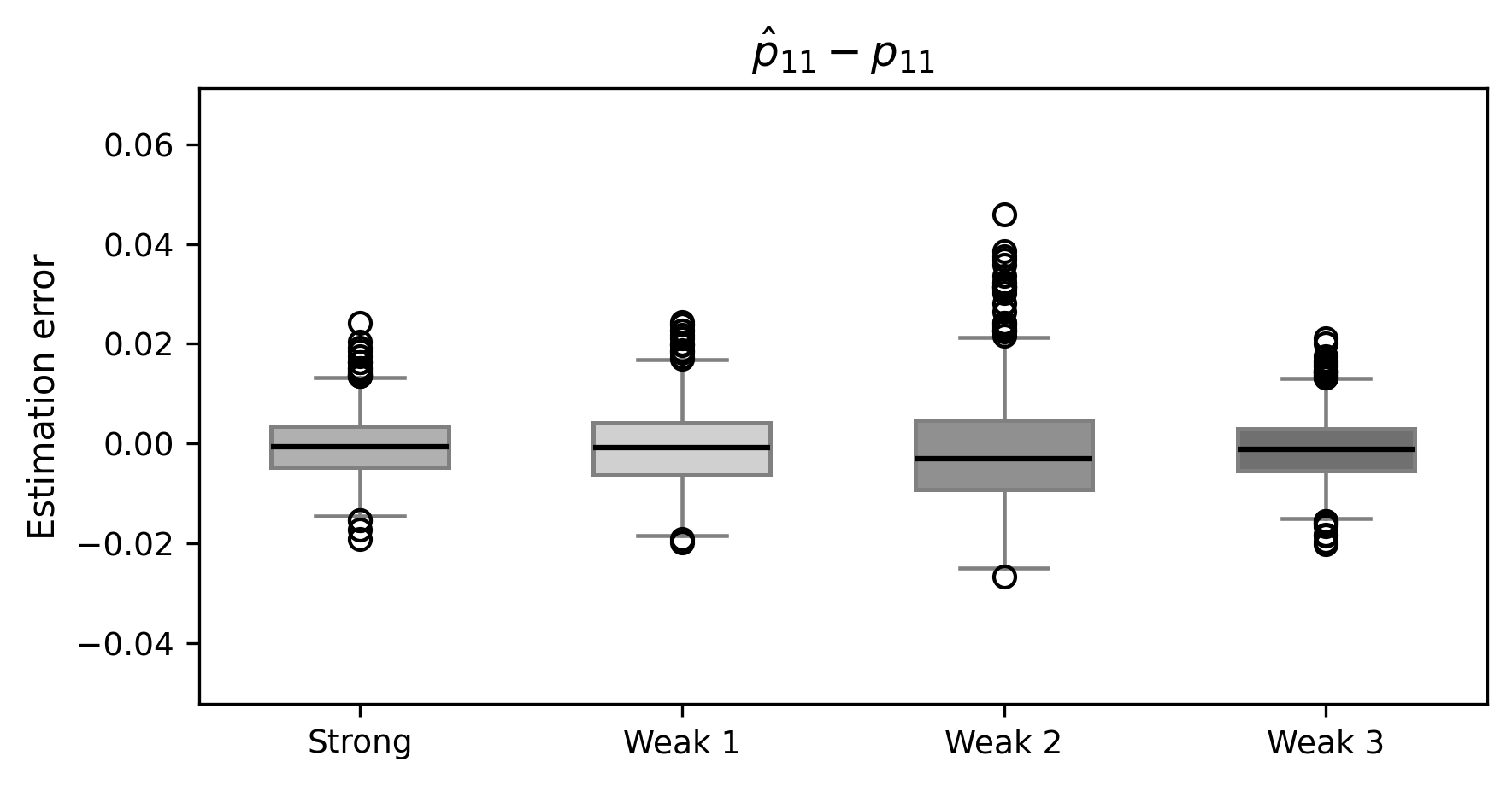}
    \end{subfigure}
    \hfill
    \begin{subfigure}[b]{0.48\textwidth}
        \centering
        \includegraphics[width=\textwidth]{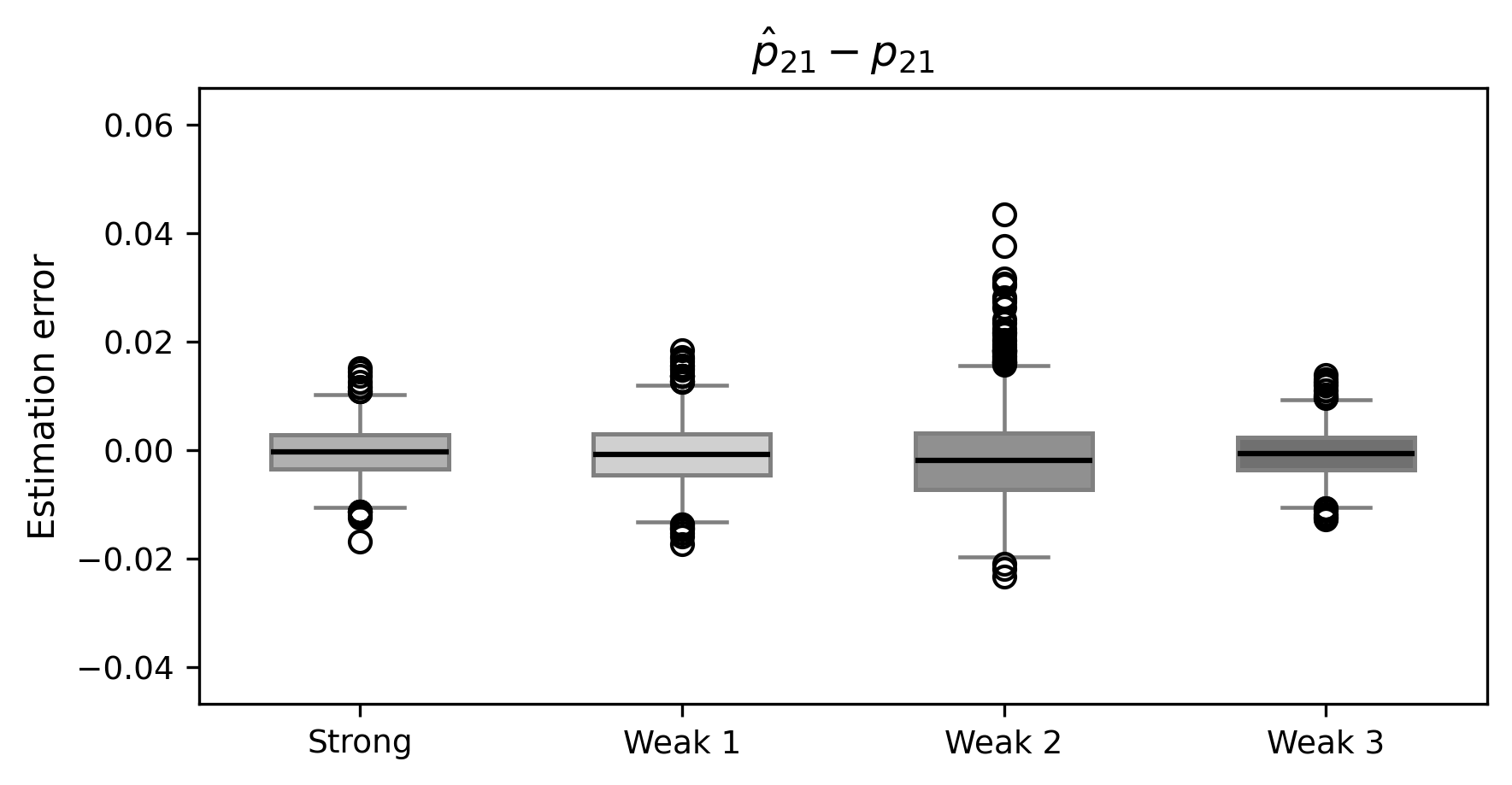}
    \end{subfigure}

    \vspace{0.25cm}

    \begin{subfigure}[b]{0.48\textwidth}
        \centering
        \includegraphics[width=\textwidth]{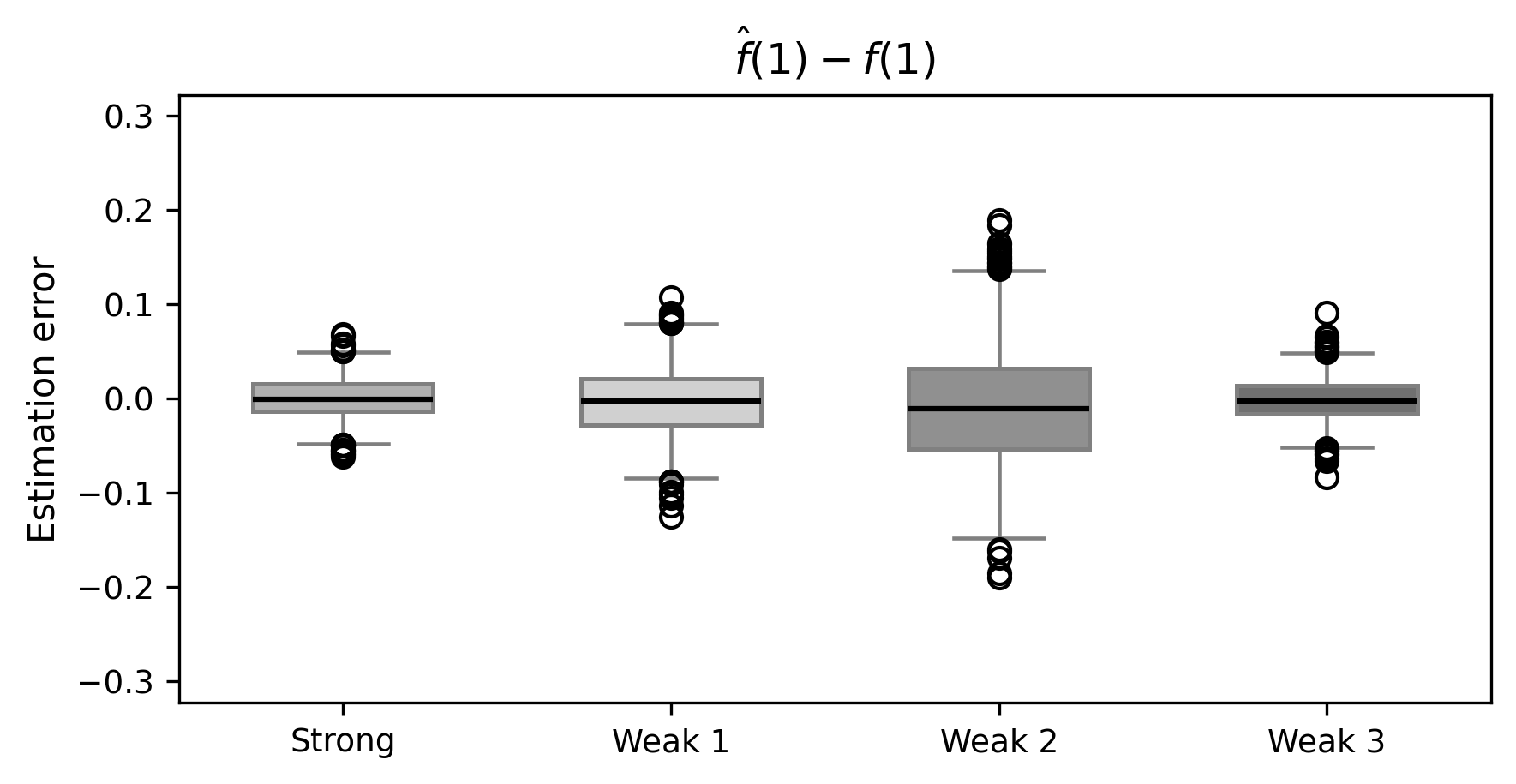}
    \end{subfigure}
    \hfill
    \begin{subfigure}[b]{0.48\textwidth}
        \centering
        \includegraphics[width=\textwidth]{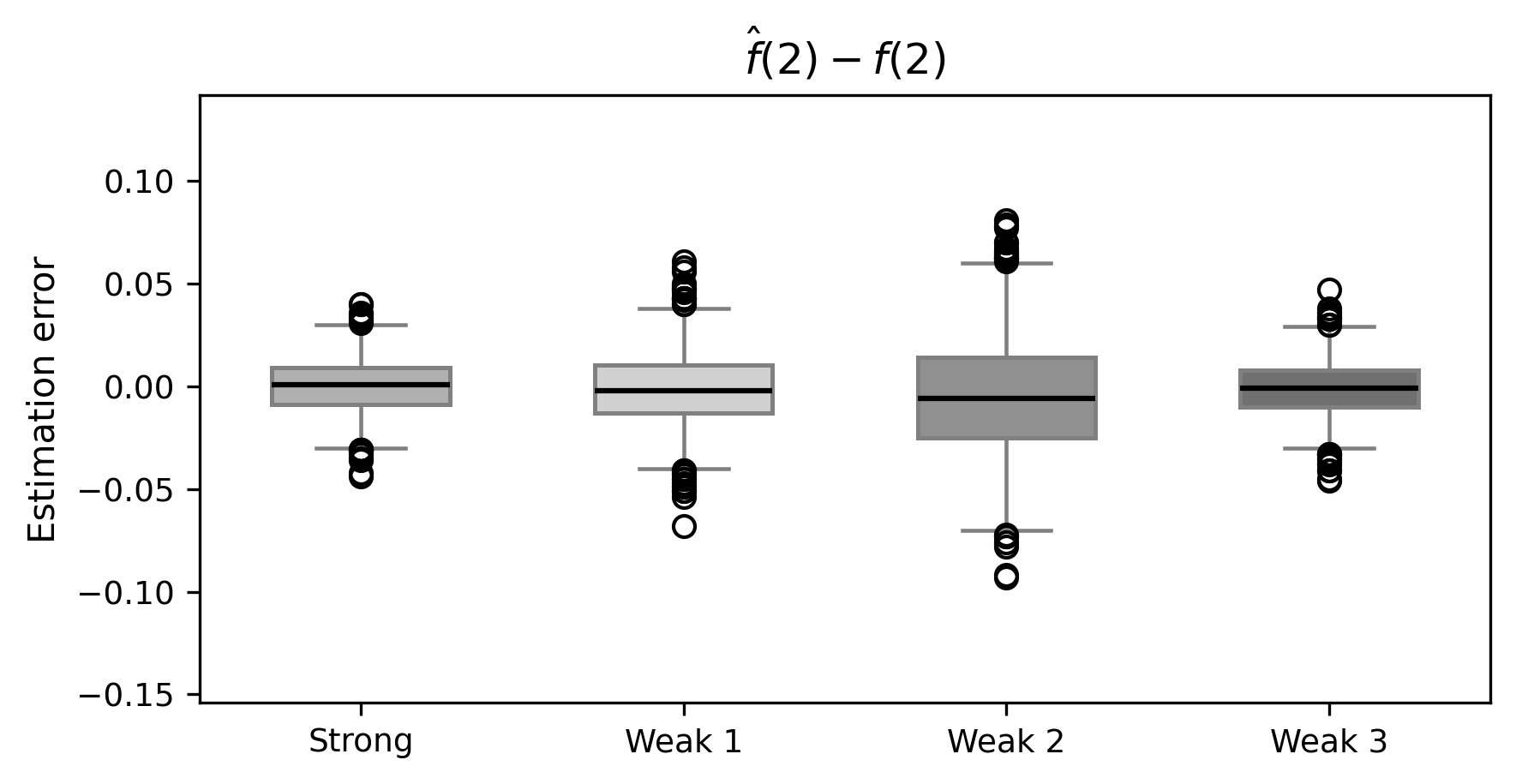}
    \end{subfigure}

    \vspace{0.25cm}

    \begin{subfigure}[b]{0.48\textwidth}
        \centering
        \includegraphics[width=\textwidth]{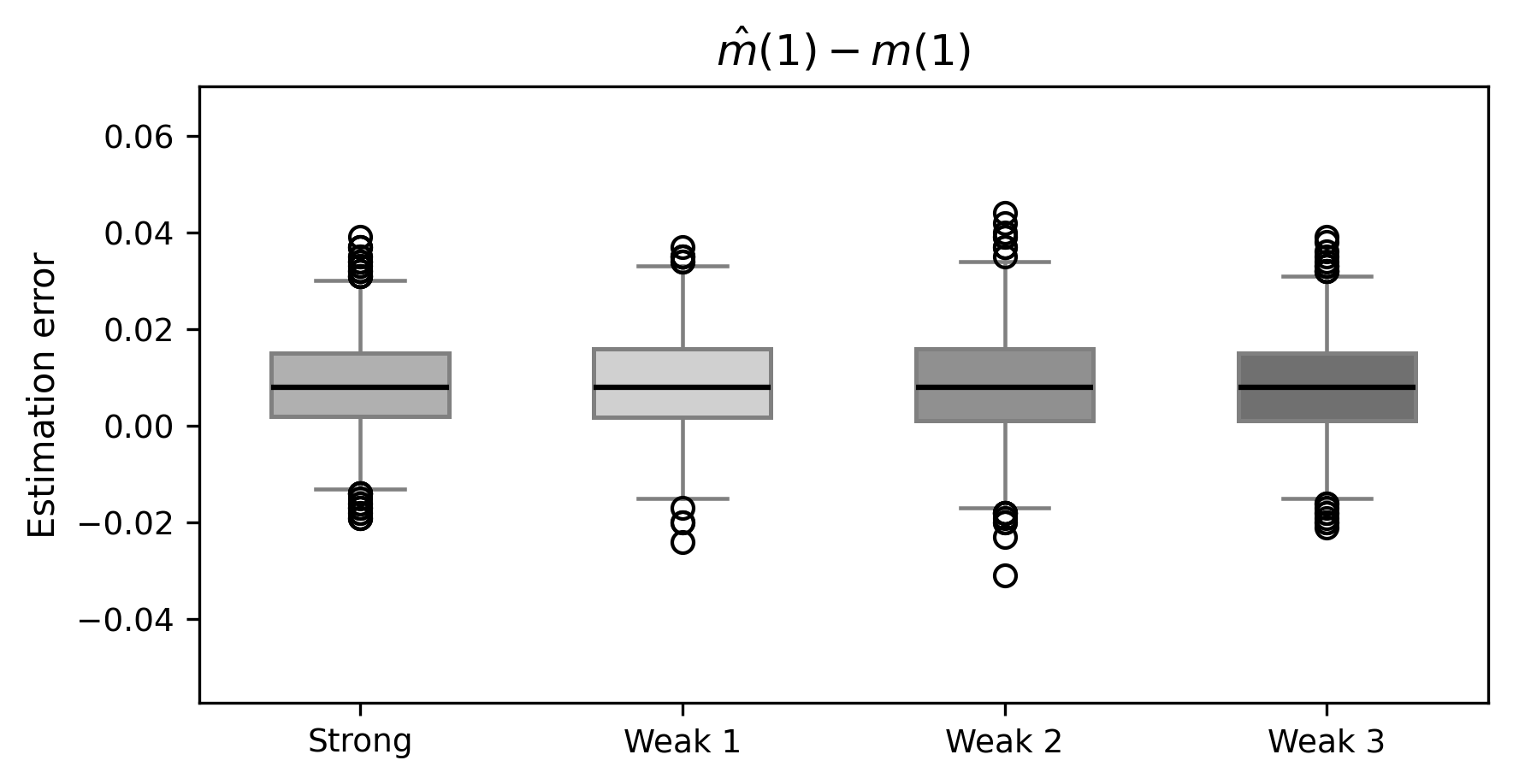}
    \end{subfigure}
    \hfill
    \begin{subfigure}[b]{0.48\textwidth}
        \centering
        \includegraphics[width=\textwidth]{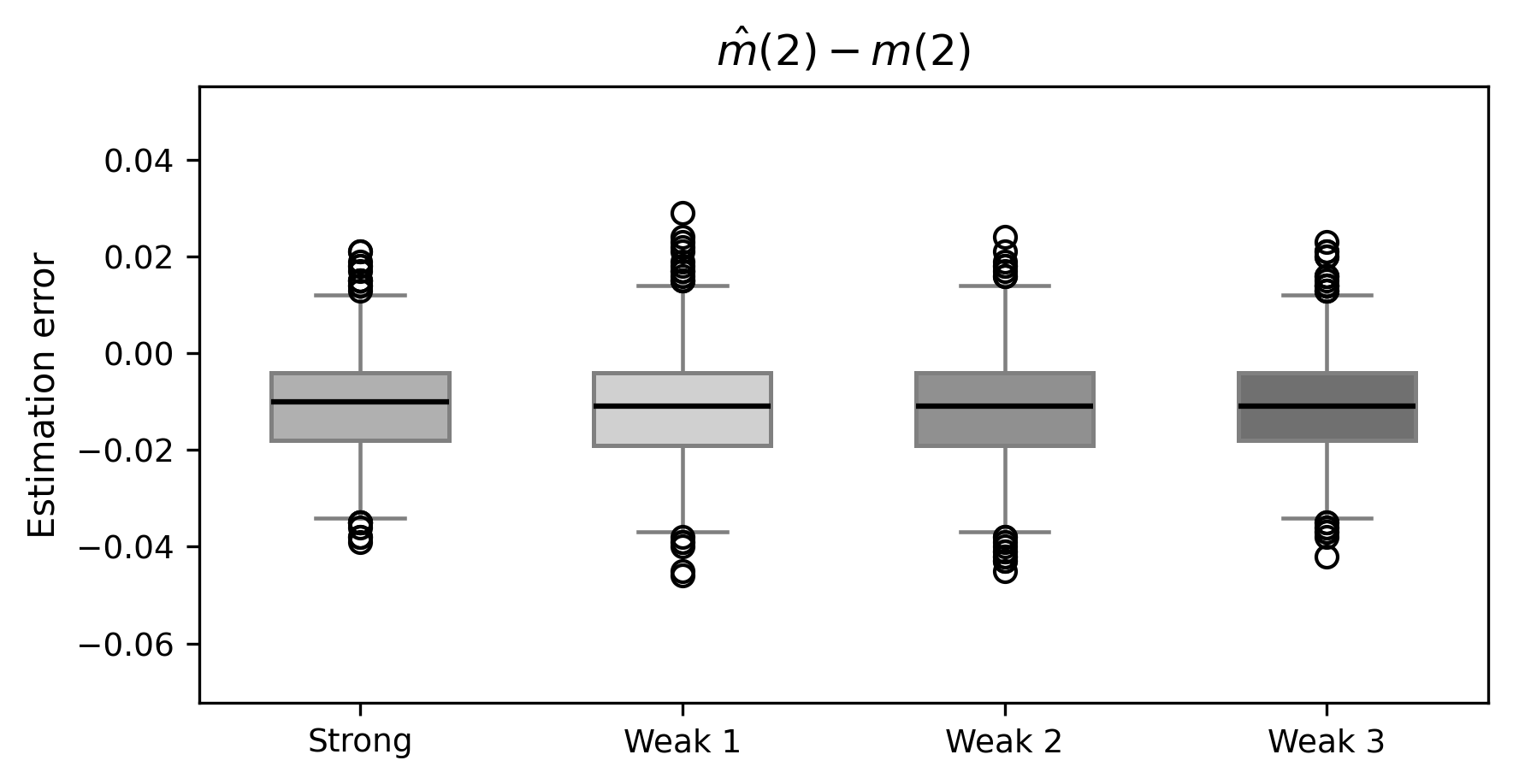}
    \end{subfigure}

    \caption{{Moment estimators obtained from $N=1,000$ independent simulations of the model \eqref{eq:43} of size $n=2,000$ and with unknown parameter $\theta_0=(0.2, 0.3, 0.5, -0.4, 0.6, 0.3, 1.5, 0.5, -0.16,0.16)$,  when the noise \((\eta_t)\) is defined respectively as \textit{strong}, \textit{weak 1}, \textit{weak 2} and \textit{weak 3}. The distribution of the estimation error \(\hat{\theta}_n(i) - \theta_0(i)\) for \(i = 1, \ldots, 8\) in box-and-whisker plots as a function of these four types of noise.
    }
    }
    \label{fig: errorsnc}
\end{figure}

\begin{figure}[H]
    \centering
    \includegraphics[width=0.9\textwidth]{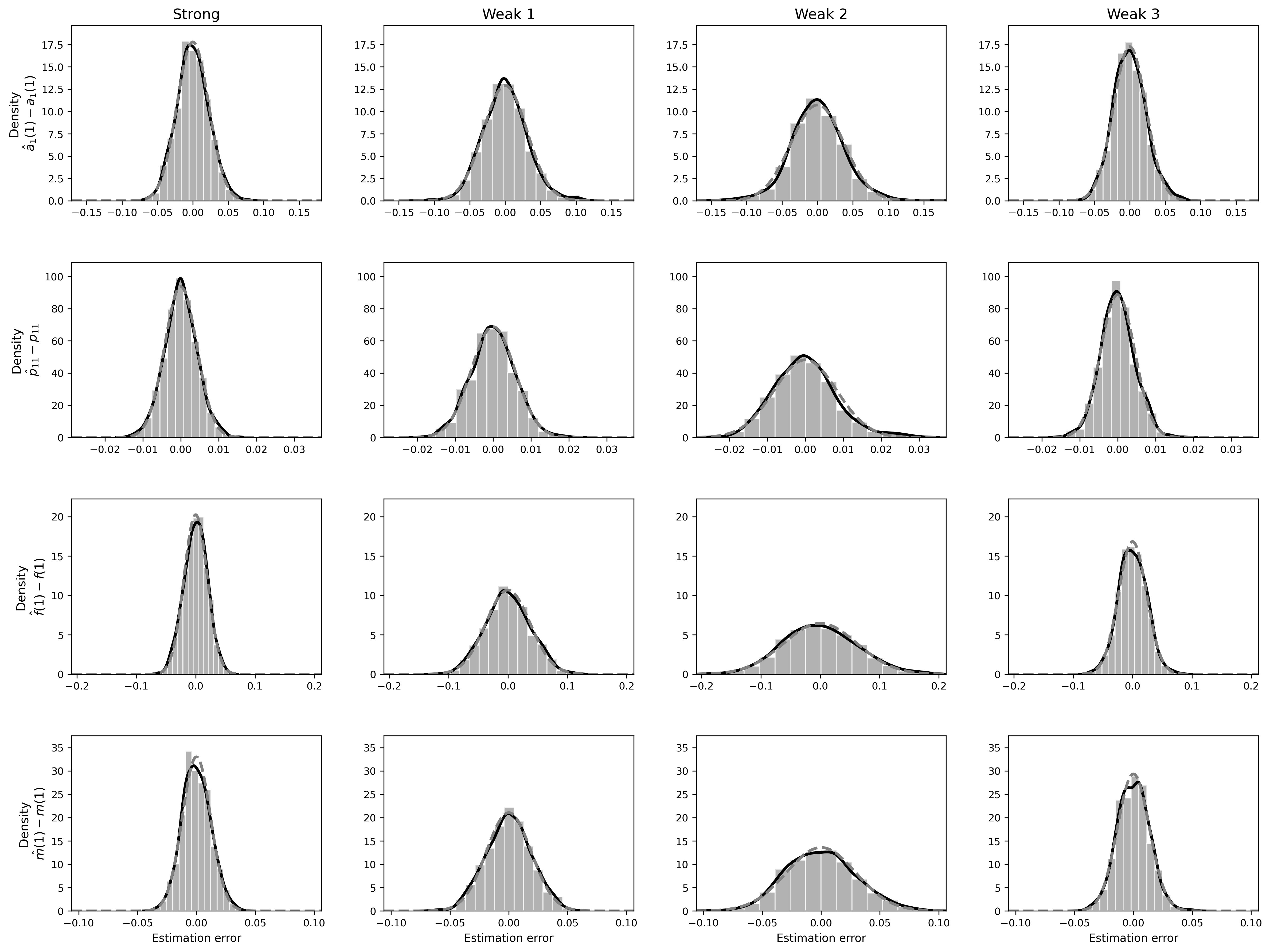}
    \caption{{Empirical distributions of the estimation errors for four selected parameters in the non-centered ARHMC(2) model, under the four noise specifications (strong, weak 1, weak 2, and weak 3). The first row corresponds to $\hat{a}_1(1)-a_1(1)$, the second to $\hat{p}_{11}-p_{11}$, the third to $\hat{f}(1)-f(1)$, and the fourth to $\hat{m}(1)-m(1)$. In each panel, the solid curve represents the kernel density estimate, while the dashed curve denotes the Gaussian density with the same empirical mean and variance.}}
    \label{fig: densitync}
\end{figure}

\begin{figure}[H]
    \centering
    \includegraphics[width=0.9\textwidth]{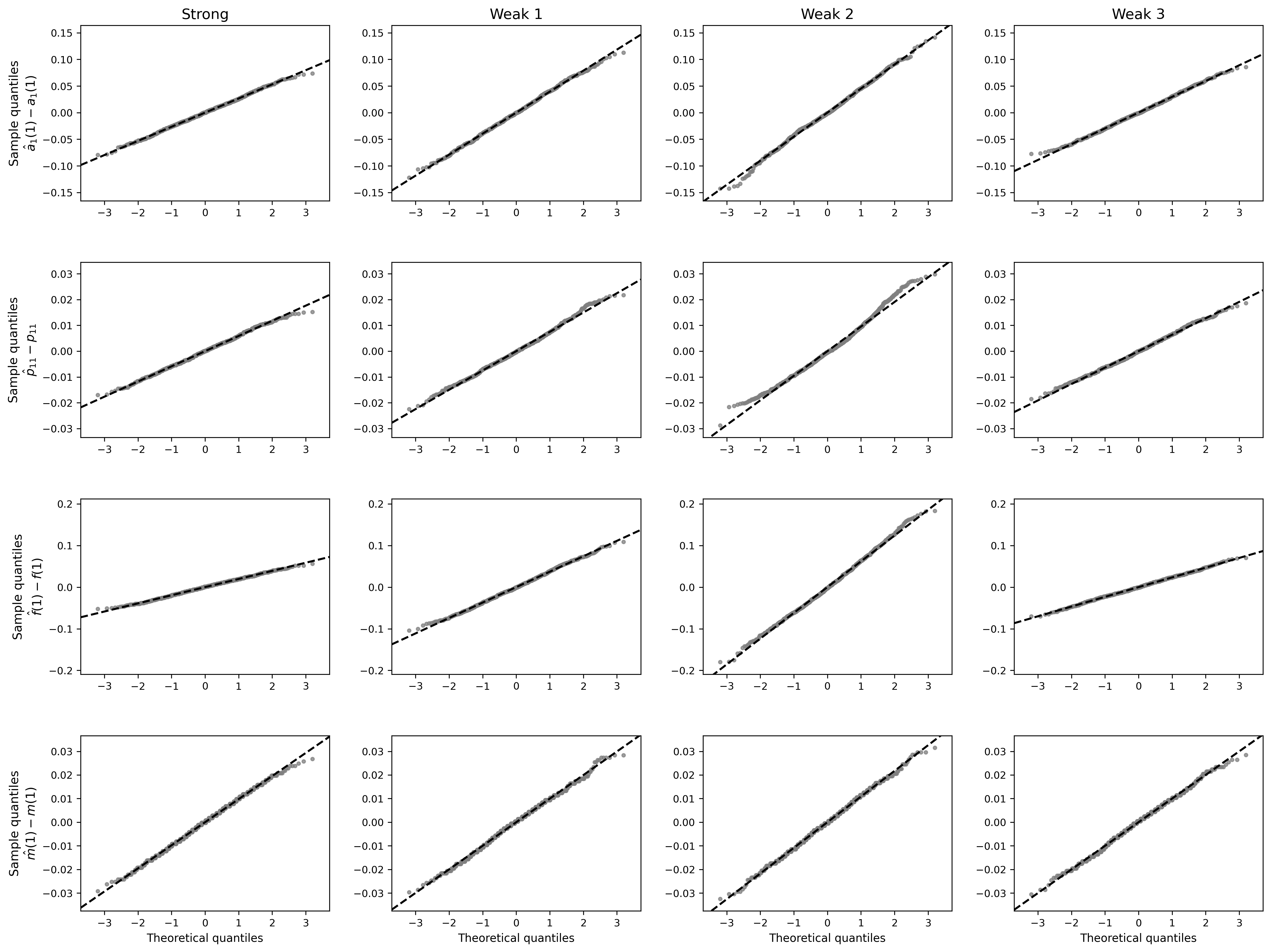}
    \caption{{Q--Q plots of the estimation errors for four selected parameters in the non-centered ARHMC(2) model under the four noise specifications (strong, weak 1, weak 2, and weak 3). The first row corresponds to $\hat{a}_1(1)-a_1(1)$, the second to $\hat{p}_{11}-p_{11}$, the third to $\hat{f}(1)-f(1)$, and the fourth to $\hat{m}(1)-m(1)$. The dashed line is the Gaussian reference line.}}
    \label{fig: qqplotnc}
\end{figure}

\noindent Now we are going to repeat the same experiments on Model \eqref{eq:43}. Tab~\ref{tab:strongnc} through \ref{tab:weak3nc} present the results of moments method estimator of the parameter \(\vartheta_0\) now includes both the structural parameters and the regime-specific means. Fig.~\ref{fig: errorsnc} through~\ref{fig: qqplotnc} compare the distribution of the moment estimators in the strong and weak noise cases. The conclusions are similar to the preview Section where the model is centered.


\begin{figure}[H]
    \centering
    \includegraphics[width=1\textwidth]{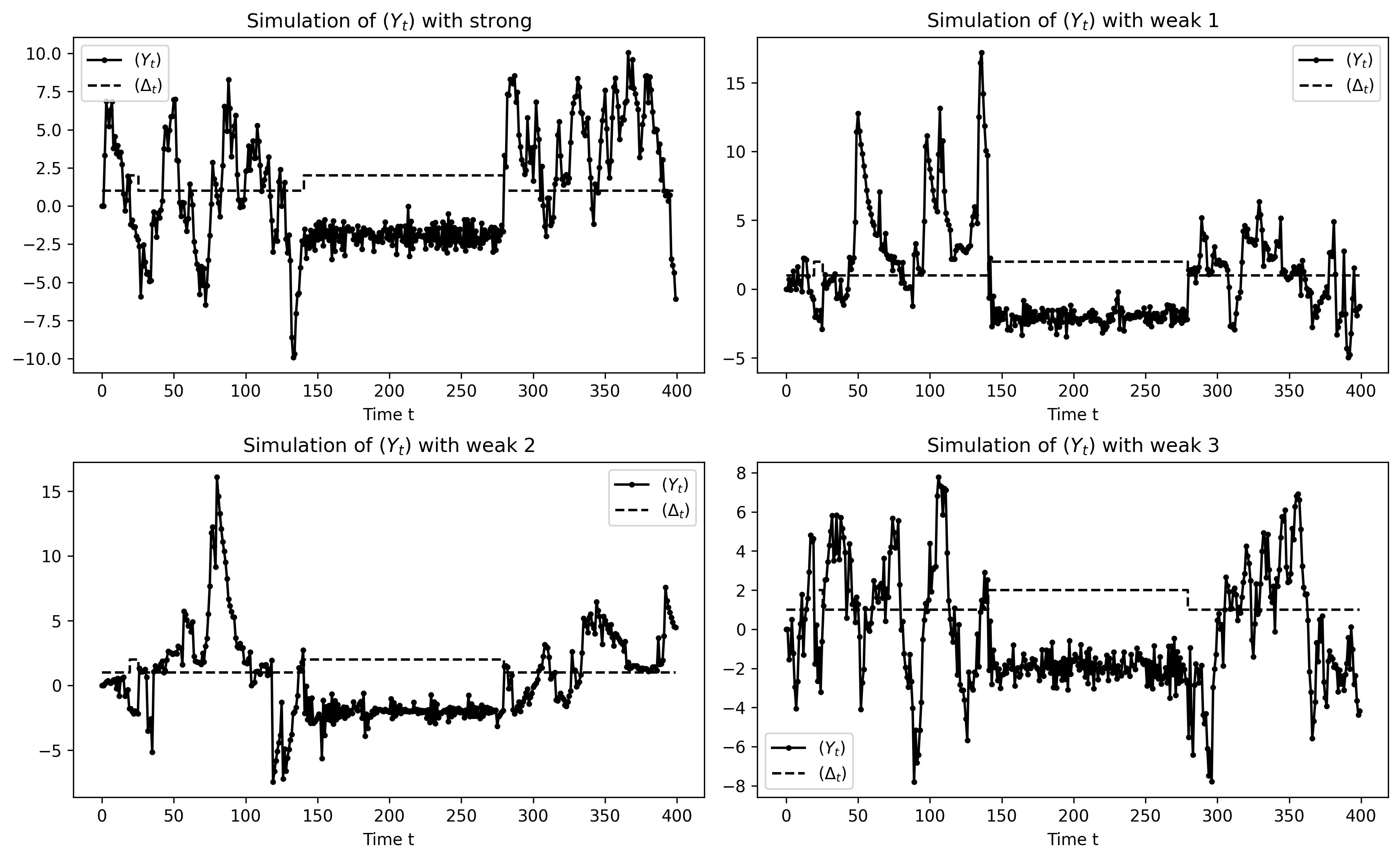}
    \caption{Simulation of length 400 for model  (\ref{eq:43}) using the different noises in Tab \ref{tab:noise_cases} with $\vartheta = ({a}_1(1) = 0.9, {a}_1(2) = -0.3, {a}_2(1) = 0.0, {a}_2(2) = 0.4; {p}_{1,1} = 0.98, {p}_{2,2} = 0.98; {f}(1) = 1.5, {f}(2) = 0.5; {m}(1) = 1.5, {m}(2) = -2.0)$.}
    \label{fig: simulation_Y}
\end{figure}

\begin{table}[H]
\centering
\setlength{\tabcolsep}{3pt} 
\renewcommand{\arraystretch}{1.0} 
{
\begin{tabular}{llccccccccccc}
\toprule
&  & ${a}_1(1)$ & ${a}_1(2)$ & ${a}_2(1)$ & ${a}_2(2)$ & ${p}_{11}$ & ${p}_{21}$ & ${f}(1)$ & ${f}(2)$ & ${m}(1)$ & ${m}(2)$ \\
\cmidrule(lr){3-12}
& $\vartheta_0$ & 0.2 & 0.3 & 0.5 & -0.4 & 0.6 & 0.3 & 1.5 & 0.5 & -0.16 & 0.16 \\
\cmidrule(lr){3-12}
\multicolumn{12}{c}{$n = 300$} \\ 
\cmidrule(lr){1-12}
& Mean  & 0.18562  & 0.30432  &  0.49470 & -0.40967  & 0.59905  & 0.29830  & 1.49269  & 0.49629 & -0.14950 & 0.14668  \\
& Bias  & -0.01437  & 0.00432  & -0.00529 & -0.00967 & -0.00094 & -0.00169 & -0.00730 & -0.00370 & 0.01049 & -0.01331  \\
& Std   & 0.06426 & 0.06049 & 0.06435 & 0.05988 & 0.01386 & 0.00921 & 0.05207 & 0.02449 & 0.01389 & 0.01649\\
\midrule
\multicolumn{12}{c}{$n = 2000$} \\ 
\cmidrule(lr){1-12}
& Mean  & 0.19531  & 0.30183  & 0.49669  & -0.40364  & 0.59931 &  0.29960 & 1.49901  & 0.49958 & -0.15137 & 0.14874 \\
& Bias  & -0.00468  &  0.00183  & -0.00330 & -0.00364 & -0.00068 & -0.00039 & -0.00098 & -0.00041 & 0.00862 & -0.01125\\
& Std   & 0.02639 & 0.02270 & 0.02694 & 0.02336 & 0.00620 & 0.00447 & 0.02103 & 0.01293 & 0.01018 & 0.01045\\
\bottomrule
\end{tabular}
}
\caption{Summary statistics of parameter estimates for model~(\ref{eq:43}) under \textit{strong} noise, based on 1000 replications for sequence lengths of 300 and 2000, respectively.}
\label{tab:strongnc}
\end{table}

\begin{table}[H]
\centering
\setlength{\tabcolsep}{3pt} 
\renewcommand{\arraystretch}{1.0} 
{
\begin{tabular}{llccccccccccc}
\toprule
&  & ${a}_1(1)$ & ${a}_1(2)$ & ${a}_2(1)$ & ${a}_2(2)$ & ${p}_{11}$ & ${p}_{21}$ & ${f}(1)$ & ${f}(2)$ & ${m}(1)$ & ${m}(2)$ \\
\cmidrule(lr){3-12}
& $\vartheta_0$ & 0.2 & 0.3 & 0.5 & -0.4 & 0.6 & 0.3 & 1.5 & 0.5 & -0.16 & 0.16 \\
\cmidrule(lr){3-12}
\multicolumn{12}{c}{$n = 300$} \\ 
\cmidrule(lr){1-12}
& Mean  & 0.17942  &  0.30454 & 0.49373  & -0.41421 & 0.59734 &  0.29590 & 1.47820 & 0.48911 & -0.14981 & 0.14669 \\
& Bias  &  -0.02057 & 0.00454  & -0.00626 & -0.01421 & -0.00265 &-0.00409  & -0.02179 &  -0.01088 &0.01018 & -0.01330 \\
& Std   & 0.08923 & 0.09386 & 0.08461 & 0.09362 & 0.01799 & 0.01330 & 0.08809 & 0.04030 & 0.01467 & 0.01659 \\
\midrule
\multicolumn{12}{c}{$n = 2000$} \\ 
\cmidrule(lr){1-12}
& Mean  & 0.19495  &  0.30006 &  0.49589 & -0.40468  & 0.59904 & 0.29924  & 1.49638  & 0.49791 & -0.15101 & 0.14855 \\
& Bias  &  -0.00504 & 0.00006  & -0.00410 & -0.00468 &  -0.00095 & -0.00075 & -0.00361 & -0.00208 & 0.00898 & -0.01144 \\
& Std   & 0.03884 & 0.03049 & 0.03436 & 0.03634 & 0.00767 & 0.00592 & 0.03740 & 0.01896 & 0.01053 & 0.01062\\
\bottomrule
\end{tabular}
}
\caption{Summary statistics of parameter estimates for model~(\ref{eq:43}) under \textit{weak 1} noise, based on 1000 replications for sequence lengths of 300 and 2000, respectively.}
\label{tab:weak1nc}
\end{table}

\begin{table}[H]
\centering
\setlength{\tabcolsep}{3pt} 
\renewcommand{\arraystretch}{1.0} 
{
\begin{tabular}{llccccccccccc}
\toprule
&  & ${a}_1(1)$ & ${a}_1(2)$ & ${a}_2(1)$ & ${a}_2(2)$ & ${p}_{11}$ & ${p}_{21}$ & ${f}(1)$ & ${f}(2)$ & ${m}(1)$ & ${m}(2)$ \\
\cmidrule(lr){3-12}
& $\vartheta_0$ & 0.2 & 0.3 & 0.5 & -0.4 & 0.6 & 0.3 & 1.5 & 0.5 & -0.16 & 0.16 \\
\cmidrule(lr){3-12}
\multicolumn{12}{c}{$n = 300$} \\ 
\cmidrule(lr){1-12}
& Mean  & 0.16738 & 0.30234 & 0.49253 & -0.41373 & 0.59402 & 0.29299 & 1.44968 & 0.47517 & -0.14904 & 0.14531 \\
& Bias  & -0.03262 & 0.00234 & -0.00747 & -0.01373 & -0.00598 & -0.00701 & -0.05032 & -0.02483 & 0.01096 & -0.01469 \\
& Std   & 0.09246 & 0.12393 & 0.10506 & 0.12818 & 0.02133 & 0.01744 & 0.12029 & 0.05580 & 0.01479 & 0.01736 \\
\midrule
\multicolumn{12}{c}{$n = 2000$} \\ 
\cmidrule(lr){1-12}
& Mean  & 0.19316 & 0.30149 & 0.49803 & -0.41100 & 0.59870 & 0.29856 & 1.49392 & 0.49675 & -0.15116 & 0.14850 \\
& Bias  & -0.00684 & 0.00149 & -0.00197 & -0.01100 & -0.00130 & -0.00144 & -0.00608 & -0.00325 & 0.00884 & -0.01150 \\
& Std   & 0.04492 & 0.04080 & 0.05170 & 0.06652 & 0.01095 & 0.00871 & 0.06387 & 0.03011 & 0.01091 & 0.01084 \\
\bottomrule
\end{tabular}
}
\caption{Summary statistics of parameter estimates for model~(\ref{eq:43}) under \textit{weak 2} noise, based on 1000 replications for sequence lengths of 300 and 2000, respectively.}
\label{tab:weak2nc}
\end{table}

\begin{table}[H]
\centering
\setlength{\tabcolsep}{3pt} 
\renewcommand{\arraystretch}{1.0} 
{
\begin{tabular}{llccccccccccc}
\toprule
&  & ${a}_1(1)$ & ${a}_1(2)$ & ${a}_2(1)$ & ${a}_2(2)$ & ${p}_{11}$ & ${p}_{21}$ & ${f}(1)$ & ${f}(2)$ & ${m}(1)$ & ${m}(2)$ \\
\cmidrule(lr){3-12}
& $\vartheta_0$ & 0.2 & 0.3 & 0.5 & -0.4 & 0.6 & 0.3 & 1.5 & 0.5 & -0.16 & 0.16 \\
\cmidrule(lr){3-12}
\multicolumn{12}{c}{$n = 300$} \\ 
\cmidrule(lr){1-12}
& Mean  & 0.19125 & 0.30689 & 0.49567 & -0.41114 & 0.59914 & 0.29730 & 1.49062 & 0.49529 & -0.14955 & 0.14668 \\
& Bias  & -0.00876 & 0.00689 & -0.00433 & -0.01114 & -0.00086 & -0.00270 & -0.00938 & -0.00471 & 0.01045 & -0.01332 \\
& Std   & 0.07021 & 0.06582 & 0.06981 & 0.06453 & 0.01494 & 0.00947 & 0.05405 & 0.02471 & 0.01434 & 0.01712 \\
\midrule
\multicolumn{12}{c}{$n = 2000$} \\ 
\cmidrule(lr){1-12}
& Mean  & 0.19380 & 0.30092 & 0.49728 & -0.40338 & 0.59934 & 0.29957 & 1.49929 & 0.49955 & -0.15139 & 0.14865 \\
& Bias  & -0.00620 & 0.00092 & -0.00272 & -0.00338 & -0.00066 & -0.00043 & -0.00071 & -0.00045 & 0.00861 & -0.01135 \\
& Std   & 0.02678 & 0.02472 & 0.02959 & 0.02618 & 0.00642 & 0.00467 & 0.02399 & 0.01373 & 0.01024 & 0.01016 \\
\bottomrule
\end{tabular}
}
\caption{Summary statistics of parameter estimates for model~(\ref{eq:43}) under \textit{weak 3} noise, based on 1000 replications for sequence lengths of 300 and 2000, respectively.}
\label{tab:weak3nc}
\end{table}

\section{Application to real data}
\label{sec:8}
\noindent The empirical analysis in this study is based on hourly wind speed observations collected in the Berlin region for the entire month of January 2024. Let \( H_t \), for \( t = 1, \dots, 744 \), denote the wind speed at 100 meters above ground level, measured in kilometers per hour (km/h). These data are obtained from the \textit{Open-Meteo.com} database. The time series \( \{H_t\} \) exhibits considerable intra-month variability, with distinct temporal patterns. A regime of low to moderate wind activity dominates the first half of the month, whereas the second half is characterized by higher volatility and stronger wind events. The maximum wind speed is recorded at 63.8 km/h on January 24, 2024 at 15:00 and the minimum reaches 1.1 km/h on January 17, 2024 at 22:00. These observations highlight the presence of sharp transitions in wind intensity, which may be indicative of underlying regime-switching behavior. To better characterize these local fluctuations and remove the level component of the series, we consider the first-order differenced process defined as \( \Delta H_t = H_{t+1} - H_t \) for \( t = 1, \dots, 743 \). This transformation centers the data and enhances the visibility of short-term dynamic structures, such as volatility clustering or abrupt directional changes, which are often crucial in the identification and estimation of models incorporating latent regimes or state-dependent dynamics, such as Hidden Markov Models.

\begin{figure}[H]
    \centering
    \includegraphics[width=1.0\textwidth]{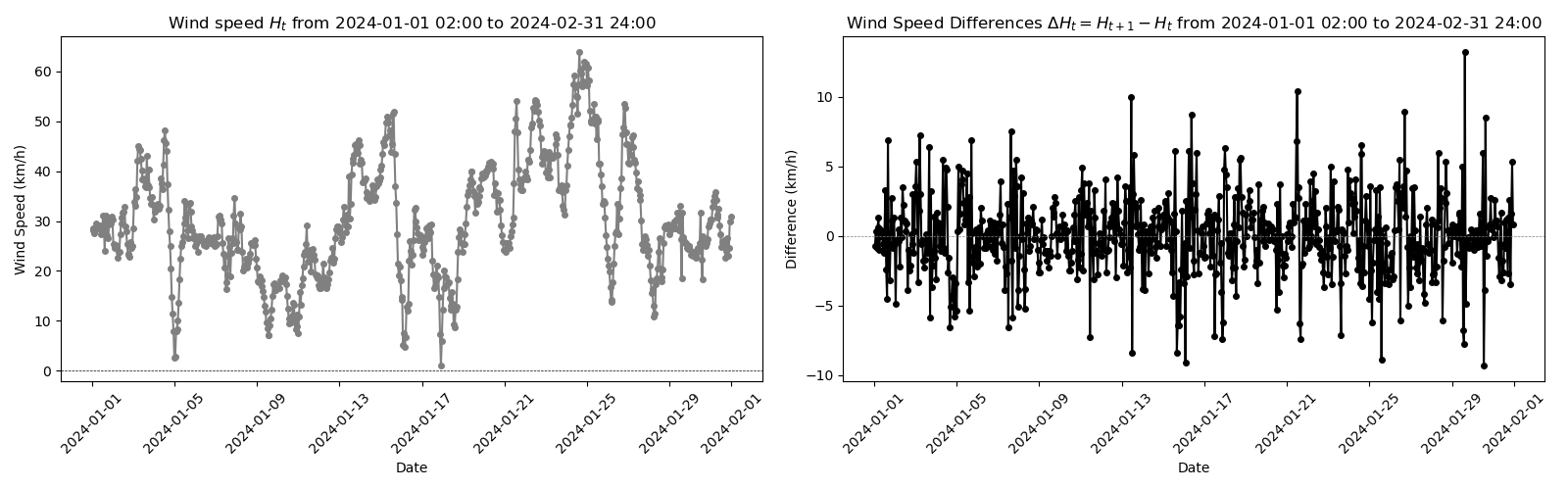}
    \caption{Wind speed series at 100 meters for January 2024 in Berlin. The left panel shows the original hourly wind speed measurements, while the right panel displays the first-order differences, defined as $\Delta H_t = H_{t+1} - H_t$. 
This transformation highlights short-term fluctuations and removes the level component of the original series.}
    \label{fig:wind_diff}
\end{figure}
\noindent In the absence of formal model selection tools, the choice of the number of regimes \( K \) in the weak ARHMC(\( p \)) model is guided by the interpretability and statistical coherence of the estimated parameters. To explore the appropriate number of regimes $K$ for model~(\ref{eq:43}), we conducted parameter estimation under two different specifications: $K = 2$ and $K = 3$, with $p=2$. In this section we estimated the parameter  $\sigma^{2}_{\eta}$ . The corresponding estimation outcomes are reported in Tables~\ref{tab:K2_resultss} and \ref{tab:K3_resultss}. This comparative approach is aimed at identifying the regime count that best captures the underlying structure of the data.

\begin{table}[H]
\centering
\setlength{\tabcolsep}{4pt}
\renewcommand{\arraystretch}{1.0}
\begin{tabular}{lccccccccccc}
\toprule
\toprule
\textbf{Parameter} & $\sigma^{2}_{\eta}$ & $\mathfrak{a}_1(1)$ & $\mathfrak{a}_1(2)$ & $\mathfrak{a}_2(1)$ & $\mathfrak{a}_2(2)$ & $\mathfrak{p}_{11}$ & $\mathfrak{p}_{21}$ & $\mathfrak{f}(1)$ & $\mathfrak{f}(2)$ & $\mathfrak{m}(1)$ & $\mathfrak{m}(2)$\\
\cmidrule(lr){2-12}
\textbf{Estimate}  & 0.1363 & 1.5000 & 0.0320 & -0.8250 & 0.5152 & 0.5476 & 0.9996 & 0.1668 & 4.4549 & 0.8200 & -1.0892 \\
\bottomrule
\bottomrule
\end{tabular}
\caption{Estimation of the parameters of model (\ref{eq:43}) based on the differenced wind speed data $\{ H_{t+1} - H_t\}_{t=1,\dots, 743}$ with $K=2$ and $p=2$.}
\label{tab:K2_resultss}
\end{table}

\begin{table}[H]
\centering
\setlength{\tabcolsep}{2pt}
\renewcommand{\arraystretch}{1.0}
\begin{tabular}{lccccccccccccc}
\toprule
\toprule
& \textbf{Parameter:} 
& $\sigma^{2}_{\eta}$ 
& $\mathfrak{a}_1(1)$ & $\mathfrak{a}_1(2)$ & $\mathfrak{a}_1(3)$ 
& $\mathfrak{a}_2(1)$ & $\mathfrak{a}_2(2)$ & $\mathfrak{a}_2(3)$ 
& $\mathfrak{p}_{11}$ & $\mathfrak{p}_{12}$ & $\mathfrak{p}_{13}$ 
& $\mathfrak{p}_{21}$ \\
\cmidrule(lr){3-13}
& \textbf{Estimate:} 
& 0.1452
& 1.5000 & -0.1005 & 0.8228 
& -0.7802 & 0.5728 & -0.1733 
& 0.4112 & 0.3390 & 0.2498 
& 0.0003 \\
\midrule
& \textbf{Parameter:} 
& $\mathfrak{p}_{22}$ & $\mathfrak{p}_{23}$ & $\mathfrak{p}_{31}$ & $\mathfrak{p}_{32}$ & $\mathfrak{p}_{33}$ 
& $\mathfrak{f}(1)$ & $\mathfrak{f}(2)$ & $\mathfrak{f}(3)$ 
& $\mathfrak{m}(1)$ & $\mathfrak{m}(2)$ & $\mathfrak{m}(3)$ \\
\cmidrule(lr){3-13}
& \textbf{Estimate:} 
& 0.9994 & 0.0003 & 0.9761 & 0.0081 & 0.0158 
& 0.1067 & 5.6259 & 0.1701 
& 0.2730 & -0.3820 & -1.4009 \\
\bottomrule
\bottomrule
\end{tabular}
\caption{Estimation of the parameters of model (\ref{eq:43}) based on the differenced wind speed data $\{ H_{t+1} - H_t\}_{t=1,\dots, 743}$ with $K=3$ and $p=2$.}
\label{tab:K3_resultss}
\end{table}

\noindent
From Tables~\ref{tab:K2_resultss} and~\ref{tab:K3_resultss} we see that the two-regime adjustment ($K=2$) already yields an interpretable description of the differenced wind--speed series. The estimated transition probabilities exhibit strong persistence: a trajectory currently in a \emph{quiet} (low-volatility) regime 1 remains there with probability $\mathfrak{p}_{11}\simeq 0.55$, whereas one in a \emph{volatile} (turbulent or high-volatility) regime 2 almost surely jumps back to regime 1 ($\mathfrak{p}_{21}\simeq 0.9996$). The volatilities parameters differ substantially across regimes ($\mathfrak{f}(1)\approx 0.17$ versus $\mathfrak{f}(2)\approx 4.45$), and the regime-specific means are well separated ($\mathfrak{m}(1)\approx 0.82$ and $\mathfrak{m}(2)\approx -1.09$). The estimated white noise variance is 0.1363.
Finally, although the estimated coefficient $\mathfrak{a}_1(1)=1.5$ exceeds unity and is locally explosive, this does \emph{not} violate the global stationarity of $\{\Delta H_t\}$, which is governed by the full switching mechanism. {In fact, using the estimated parameters for $K=2$, we obtained $\rho(Q_{\mathbf A\otimes \mathbf A})=0.7939<1$ and a numerical approximation of the Lyapunov exponent $\gamma\approx -0.1338<0$, so that assumptions (A5) and (A6) are satisfied.} Similar \enquote{border-of-unit-disk} coefficients have been reported in \cite{mainassara2025estimating} and the overall stationarity condition stated by \citet[page~343]{francq2001stationarity} remains satisfied.
Adding a third regime ($K=3$) only increases complexity without providing additional insight. One of the new states is virtually transient: its probability of leaving regime 2 at the next step is $1-\mathfrak{p}_{22}\approx 0.9997$, so it never constitutes a persistent phase of the process. Moreover, only one volatilities parameter ($\mathfrak{f}(2)\approx 5.63$) truly differs from the two quiet ones, which suggests over-parametrization and redundancy. {For completeness, the same verification for $K=3$ gives $\rho(Q_{\mathbf A\otimes \mathbf A})=0.6539<1$ and $\gamma\approx -0.2131<0$, so that (A5) and (A6) also hold in that case.} We therefore retain the $K=2$ adjustment, as it strikes the best balance between parsimony and explanatory power and captures the alternation of low-volatility and high-volatility phases in the increment series $\{\Delta H_t\}$.

\section{Conclusion}\label{conclusion}
\noindent In this paper we extended the ARHMC model studied in \cite{mainassara2025estimating}, which was limited to first-order dynamics (\(p=1\)), to the general case of any autoregressive order \(p\in\mathbb{N}\setminus\{0\}\). We first generalized Lemma 1 in \cite{francq2004estimation} to a multivariate setting, which enabled us to estimate the parameters of Models~\eqref{eq:1} (centered case)  and \eqref{eq:43} (non centered case) by using the moment method. In additional we established the consistency and the asymptotic normality of the proposed estimators. We also propose a consistent estimator of
the asymptotic variance matrix of the moment estimators in the two cases. In Monte Carlo experiments we demonstrated that the moment estimators of the centered and non centered models provide some satisfactory results, at least for the models considered in
our study. We also applied our findings to real-world data following an approach similar to that of \cite{francq1997white} and \cite{mainassara2025estimating} to determine the optimal number of regimes, represented by the parameter \(K\), in order to efficiently fit the data. 
In future works, we intent to study how the existing identification (see \cite{yac_jtsa},  \cite{BMK2016}) and diagnostic checking (see \cite{BM2011}, \cite{FRZ2005}) procedures should be
adapted to the present model. 
{It would also be of interest to extend the present framework toward more flexible regime-switching structures. In particular, one possible direction would be to investigate whether weak ARHMC($p$) models can be adapted to settings with a more flexible, possibly infinite, regime structure as in \cite{BauwensCarpantierDufays2017}, or to models with observation-driven transition mechanisms as in \cite{KamgaingOmbaoDavis2009}. Such extensions would considerably enlarge the scope of the present methodology and raise new questions on identifiability, estimation, and asymptotic inference in weak-noise settings.}

\section{Proofs}\label{proof}

\subsection{Proof of Theorem \ref{thm:1}}\label{proofthm:1}
\noindent Using the independence between the processes $(\eta_t)_{t\in\nbZ}$ and $(\Delta_t)_{t\in\nbZ}$ (see Assumption $(\mathbf{A_2}))$ and the fact that $(\eta_t)_{t\in\nbZ}$ is a centered process, we have for all $k \geq 0$
{\small
\begin{align*}
&    \mathbb{E}\left(\Vect{\{\underline{X}_0\underline{X}_k'\}}\right) = \ \mathbb{E}\left\{\left(\sum_{i_2=0}^{\infty}\left(\prod_{j_2=0}^{i_2-1}\mathbf{A}(\Delta_{k-j_2})\right)\underline{f}(\Delta_{k-i_2})\eta_{k-i_2}\right)\otimes \left(\sum_{i_1=0}^{\infty}\left(\prod_{j_1=0}^{i_1-1}\mathbf{A}(\Delta_{-j_1})\right) \underline{f}(\Delta_{-i_1}){\eta}_{-i_1}\right)\right\} \\
   &  = \ \sum_{i_1,i_2 = 0}^{\infty} \mathbb{E}\left\{\left(\prod_{j_2 = 0}^{i_2-1}\mathbf{A}(\Delta_{k-j_2})\right)\otimes\left(\prod_{j_1=0}^{i_1-1}\mathbf{A}(\Delta_{-j_1})\right)\left(\underline{f}(\Delta_{k-i_2})\otimes \underline{f}(\Delta_{-i_1})\right)\right\}\mathbb{E}\left(\eta_{k-i_2}{\eta}_{-i_1}\right)\mathds{1}_{\{k-i_2=-i_1\}} \\
&      =  \ \sum_{i_2 = k}^{\infty} \mathbb{E}\left\{\left(\prod_{j_2 = 0}^{i_2-1}\mathbf{A}(\Delta_{k-j_2})\right)\otimes\left(\prod_{j_1=0}^{i_2-k-1}\mathbf{A}(\Delta_{-j_1})\right)\left(\underline{f}(\Delta_{k-i_2})\otimes \underline{f}(\Delta_{k-i_2})\right)\right\}\mathbb{E}\left({\eta}_{k-i_2}{\eta}_{k-i_2}\right) \\ 
&     = \ \sigma^2_\eta\sum_{i_2=k}^{\infty}\mathbb{E}\left\{\left(\left[\left(\prod_{j_2=0}^{k-1}\mathbf{A}(\Delta_{k-j_2})\right)\left(\prod_{j_2=k}^{i_2-1}\mathbf{A}(\Delta_{k-j_2})\right)\right]\otimes \left(\prod_{j_1=k}^{i_2-1}\mathbf{A}(\Delta_{k-j_1})\right)\right)\left(\underline{f}(\Delta_{k-i_2})\otimes \underline{f}(\Delta_{k-i_2})\right)\right\} \\ 
&=  \ \sigma^2_\eta\sum_{i_2=k}^{\infty}\mathbb{E}\left\{\left(\prod_{j=0}^{k-1}(\mathbf{A}(\Delta_{k-j})\otimes I_p)\right)\left(\prod_{j=k}^{i_2-1}\left(\mathbf{A}(\Delta_{k-j})\otimes\mathbf{A}(\Delta_{k-j})\right)\right)\left(\underline{f}(\Delta_{k-i_2})\otimes \underline{f}(\Delta_{k-i_2})\right)\right\},
\end{align*}
\noindent where the equalities are obtained from the elementary identities $\Vect\{ U_1V_1'\} =U_1\otimes V_1$ for $U_1$ and $V_1$ in $\nbR^{m}$ and $(U \otimes V)(W \otimes S) = (UW) \otimes (VS)$ for $U$, $V$, $W$, $S$ in $\mathbb{R}^{m \times m}$. 
Finally, in view of Lemma \ref{lem:1} and under  Assumption ($\mathbf{A_6}$) which entails the convergence of the series $\sum_{j=0}^\infty Q_{\mathbf{A}\otimes \mathbf{A}}^j$, we obtain for all $k\geq 0$ that
\begin{align*}
     C_{k,0}(\theta_0) &: = \mathbb{E}\left(\Vect{\{\underline{X}_0\underline{X}_k'\}}\right) = \sigma^2_\eta\sum_{i_2\geq k}\left(\bm{1}'\otimes I_{p^2}\right)Q_{\mathbf{A}\otimes I_p}^k Q_{\mathbf{A}\otimes \mathbf{A}}^{i_2-k}\bm\Pi_{\underline{f}\otimes\underline{f}} \\
     &= \sigma^2_\eta(\bm{1}'\otimes I_{p^2})Q_{\mathbf{A}\otimes I_p}^k(I_{Kp^2}-Q_{\mathbf{A}\otimes \mathbf{A}})^{-1}\bm\Pi_{\underline{f}\otimes\underline{f}}.
\end{align*}
This completes the proof.
\qed 

\subsection{Skech of the proof of Theorem \ref{thm:2}}\label{proofthm:2}
\noindent The proof of this theorem is similar to the proof of Theorem 1 in \cite{mainassara2025estimating} and it
is omitted. The main ingredient is the application of the implicit function theorem to the function $\Lambda$ defined by
\begin{equation*}
\Lambda :
    \left\lbrace
        \begin{aligned}
            & \mathring\Theta \times \mathbb{R}^{K^2+Kp} & \rightarrow & \ \quad \mathbb{R}^{K^2+Kp} \\
            & \quad (\theta, \zeta) & \mapsto & \quad J_{\bm\Psi^{N}}(\theta)'\left(\zeta-\bm\Psi^N(\theta)\right).
        \end{aligned}
    \right.
\end{equation*}
in order to exhibit the estimator $\hat{\theta}_n$.
\qed
\subsection{Sketch of proof of Theorem \ref{thm:3}}\label{proofthm:3}
\noindent Here, we will outline the different steps to prove Theorem \ref{thm:3}. A more detailed approach can be found in \citet[Proof of Theorem 3]{mainassara2025estimating}.

\begin{l1}[\text{\cite{davydov1968convergence}}]\label{lem:2}
Let \(X\) and \(Y\) be two random variables, and let \(\sigma(X)\) and \(\sigma(Y)\) be the \(\sigma\)-fields generated by \(X\) and \(Y\) respectively. Consider three strictly positive numbers \(p\), \(q\), and \(r\) such that \(p^{-1} + q^{-1} + r^{-1} = 1\). Then,
\[
|\Cov(X,Y)| \leq C_0 \|X\|_p \|Y\|_q \left\{\alpha\left(\sigma(X), \sigma(Y)\right)\right\}^\frac{1}{r},
\]
where \(\|\cdot\|_p\) denotes the \(L^p\)-norm, \(C_0\) is a universal constant and 
$\alpha\left(\sigma(X), \sigma(Y)\right)$ denotes the strong mixing coefficient between the $\sigma$-fields $\sigma(X)$ and $\sigma(Y)$ generated by $X$ and $Y$.
\end{l1}

\noindent Let \( Q_{\underline{\mathbf{A}}^{8+4\nu}} \) be the \( K \times K \) matrix defined by  \eqref{eq:1.3} with $ F(i)=\|\mathbf{A}(i)\|^{8+4\nu}$, $i=1,\ldots,K$. 

\begin{l1}\label{lem:3}
We assume that the spectral radius of the matrix \( Q_{\underline{\mathbf{A}}^{8+4\nu}} \) is less than 1. Under the assumptions \((\mathbf{A_1})\), \((\mathbf{A_3})\) and \((\mathbf{A_6})\), for \(\nu > 0\), we have
\begin{align}
\label{eq:3.12}
    \mathbb{E}\left(\left\|\left( \prod_{j=0}^{k-1} \mathbf{A}(\Delta_{t-j}) \right) \underline{f}(\Delta_{t-k})\right\|^{8+4\nu}\right) \leq C\rho^k,
\end{align}
where \( C \) is a positive universal constant and \( \rho \) is a constant in \( (0,1) \).
\end{l1}
\subsubsection*{Proof.}
\noindent Using Lemma \ref{lem:1}, we have for all $t \in \mathbb{Z}$,

\begin{align*}
   \mathbb{E}\left(\left\|\left(\prod_{j=0}^{k-1}\mathbf{A}(\Delta_{t-j})\right)\underline{f}(\Delta_{t-k})\right\|^{8+4\nu}\right) \leq & \ \mathbb{E}\left(\prod_{j=0}^{k-1}\left\|\mathbf{A}(\Delta_{t-j})\right\|^{8+4\nu}\left\|\underline{f}(\Delta_{t-k})\right\|^{8+4\nu}\right)\\ = & \ \bm{1}'Q_{\underline{\mathbf{A}}^{8+4\nu}}^{k}\bm{\pi}_{\underline{f}^{8+4\nu}}
\end{align*}
where $\bm{\pi}_{\underline{f}^{8+4\nu}}$ is a vector of size $K\times 1$ and is defined in \eqref{def_Pi_G} with $G(i)=\|\underline{f}(i)\|^{8+4\nu}$, $i=1,\ldots,K$.

\noindent The conclusion follows by using the same reasoning as in \citet[proof of Lemma 3]{mainassara2025estimating}.

\qed

\noindent In the sequel, to simplify the notation, we define for all $(t,k)\in\mathbb{Z}\times\mathbb{N}$
\begin{equation}
\label{eq:c14}
    \bm{c}_t^k = \bm{c}(\Delta_t, \dots, \Delta_{t-k}) := \left( \prod_{j=0}^{k-1} \mathbf{A}(\Delta_{t-j}) \right) \underline{f}(\Delta_{t-k}),
\end{equation}
with the convention $\bm{c}_t^k=1$ if $k>t$.

\begin{l1}\label{lem:4}
Let Assumptions of Theorem~\ref{thm:3} be satisfied. Then the following convergence holds
$$\bm{I}^N(\theta_0) : = \displaystyle\lim_{n\rightarrow \infty}\Var\Big(\sqrt{n}\bm{F}^{N,n}(\theta_0)\Big)\in\mathbb{R}^{p^2N\times p^2N}.$$
\end{l1}
\subsubsection*{Proof.}
\noindent Using the expression given for \(\hat{C}_{k,0}\), the definition of \(\bm{F}^{N,n}\) and the stationarity of the process \((X_t)_{t \in \mathbb{Z}}\), we have
\begin{align*}
    \bm{I}^{N,n} : = \Var(\sqrt{n}\bm{F}^{N,n}(\theta_0)) = \frac{1}{n}\sum_{t=1}^{n}\sum_{s=1}^{n}\Cov(Z_t,Z_s) = \sum_{k=-n+1}^{n-1}\frac{\left(n-|k|\right)}{n}\Cov(Z_t,Z_{t-k}),
\end{align*}
\noindent where we recall that $Z_t$ is defined by (\ref{eq:001}). Denoting by \(Z_t[m]\) the \(m\)-th block of \(Z_t(\theta_0)\), we now introduce 
\begin{align*}
   \bm{d}_{k}(l,r)=\bm{d}_{k}(l,r) : =  \Cov(Z_t[l],Z_{t-k}[r]) = \Cov(\underline{X}_{t+l}\otimes \underline{X}_t, \underline{X}_{t-k+r}\otimes \underline{X}_{t-k}).
\end{align*}
\noindent Substituting $\underline{X}_t$ by its expression given in Equation (\ref{eq:4}) and using the elementary property $(U \otimes V)(W \otimes S) = (UW) \otimes (VS)$, we have
\begin{align*}
    \bm{d}_{k}(l,r) = \sum_{i_1,i_2,i_3,i_4 = 0}^{\infty}\xi_{i_1,i_2,i_3,i_4,k}(l,r)(\theta_0)
\end{align*}
where
\begin{align*}
 \xi_{i_1,i_2,i_3,i_4,k}(l,r)(\theta_0) = & \ \Cov(\bm{c}_{t+l}^{i_1}\eta_{t+l-i_1}\otimes \bm{c}_t^{i_2}\eta_{t-i_2}, \bm{c}^{i_3}_{t-k+r}\eta_{t-k+r-i_3}\otimes \bm{c}^{i_4}_{t-k}\eta_{t-k-i_4}) \\ = & \ \mathbb{E}\left[\left(\bm{c}^{i_1}_{t+l}\eta_{t+l-i_1}\otimes \bm{c}^{i_2}_t\eta_{t-i_2}\right)(\bm{c}^{i_3}_{t-k+r}\eta_{t-k+r-i_3}\otimes \bm{c}^{i_4}_{t-k}\eta_{t-k-i_4'})\right] \\ & -  \ \mathbb{E}\left(\bm{c}^{i_1}_{t+l}\eta_{t+l-i_1}\otimes \bm{c}_{t}^{i_2}\eta_{t-i_2}\right) \times \left[\mathbb{E}\left(\bm{c}^{i_3}_{t-k+r}\eta_{t-k+r-i_3}\otimes \bm{c}^{i_4}_{t-k}\eta_{t-k-i_4}\right)\right]' \\ 
 = & \ \mathbb{E}\left[\left(\bm{c}^{i_1}_{t+l}\otimes \bm{c}_{t}^{i_2}\right)\left(\bm{c}_{t-k+r}^{i_3'}\otimes \bm{c}_{t-k}^{i_4'}\right)\right] \times \mathbb{E}\left(\eta_{t+l-i_1}\eta_{t-i_2}\eta_{t-k+r-i_3}\eta_{t-k-i_4}\right) \\& -  \ \mathbb{E}\left(\bm{c}_{t+l}^{i_1}\otimes \bm{c}_{t}^{i_2}\right)\times \left[\mathbb{E}\left(\bm{c}_{t-k+r}^{i_3}\otimes \bm{c}_{t-k}^{i_4}\right)\right]' \times \mathbb{E}\left(\eta_{t+l-i_1}\eta_{t-i_2}\right) \mathbb{E}\left(\eta_{t-k+r-i_3}\eta_{t-k-i_4}\right)\\ = & \ \mathbb{E}\left((\bm{c}_{t+l}^{i_1}\otimes\bm{c}_{t}^{i_2})(\bm{c}^{i_3'}_{t-k+r}\otimes \bm{c}_{t-k}^{i_4'})\right) \times \Cov(\eta_{t+l-i_1}\eta_{t-i_2},\eta_{t-k+r-i_3}\eta_{t-k-i_4}) \\& +  \ \Cov(\bm{c}_{t+l}^{i_1}\otimes \bm{c}_{t}^{i_2},\bm{c}_{t-k+r}^{i_3}\otimes\bm{c}_{t-k}^{i_4}) \times \mathbb{E}\left(\eta_{t+l-i_1}\eta_{t-i_2}\right)\times \mathbb{E}\left(\eta_{t-k+r-i_3}\eta_{t-k-i_4}\right).
\end{align*}
and $c_t^k$ is given by (\ref{eq:c14}). 

\noindent Using the inequality \(\|A \otimes B\| \le \|A\| \,\|B\|\) for any matrices \(A\) and \(B\), in view of Lemma~\ref{lem:3} and applying the Cauchy-Schwarz inequality, we obtain
\begin{align}
\label{eq:3.13}
    \Bigl\|\mathbb{E}\Bigl[(\bm{c}_{t+l}^{i_1} \otimes \bm{c}_{t}^{i_2})
        (\bm{c}_{t-k+r}^{i_3}{}' \otimes \bm{c}_{t-k}^{i_4}{}')\Bigr]\Bigr\|
    &\;\le\;
    \mathbb{E}\Bigl[\bigl\|
        (\bm{c}_{t+l}^{i_1}\,\bm{c}_{t-k+r}^{i_3}{}')
        \otimes
        (\bm{c}_{t}^{i_2}\,\bm{c}_{t-k}^{i_4}{})'\bigr\|\Bigr]
    \notag\\
    &\;\le\;
    \mathbb{E}\bigl[\|\bm{c}_{t}^{i_2}\|^4\bigr]^{\tfrac14}\,
    \mathbb{E}\bigl[\|\bm{c}_{t+l}^{i_1}\|^4\bigr]^{\tfrac14}\,
    \mathbb{E}\bigl[\|\bm{c}_{t-k}^{i_4}\|^4\bigr]^{\tfrac14}\,
    \mathbb{E}\bigl[\|\bm{c}_{t-k+r}^{i_3}\|^4\bigr]^{\tfrac14}
    \notag\\
    &\;\le\;
    C\,\rho^{\,i_1 + i_2 + i_3 + i_4},
\end{align}
where \(C\) is a positive constant and \(0 < \rho < 1\).

\noindent First, suppose that $k\geq 0$, for all $l,r \in\{1,\dots, K^2+Kp\}$ and in view of (\ref{eq:3.13}), we have
\begin{align*}
   \|\bm{d}_k(l,r)\| \leq  \sum_{i_1,i_2,i_3,i_4 = 0}^{\infty} \|\xi_{i_1,i_2,i_3,i_4,k}(l,r)(\theta_0)\| \leq  
   u_1 + u_2 + u_3 + u_4 + u_5 + v_1 + v_2 + v_3
\end{align*}
where 
\begin{align*}
  u_1 & := \sum_{i_1 > \lfloor k/2 \rfloor} \sum_{i_2 = 0}^{\infty} \sum_{i_3 = 0}^{\infty} \sum_{i_4 = 0}^{\infty} C\rho^{i_1+i_2+i_3+i_4} \left|\text{Cov}(\eta_{t+l-i_1}\eta_{t-i_2}, \eta_{t-k+r-i_3}\eta_{t-k-i_4})\right|, \\
  u_2 & := \sum_{i_1 = 0} \sum_{i_2 > \lfloor k/2 \rfloor}\sum_{i_3 = 0}^{\infty} \sum_{i_4 = 0}^{\infty} C\rho^{i_1+i_2+i_3+i_4} \left|\text{Cov}(\eta_{t+l-i_1}\eta_{t-i_2}, \eta_{t-k+r-i_3}\eta_{t-k-i_4})\right|, \\
  u_3 & := \sum_{i_1 = 0}^{\infty} \sum_{i_2 = 0}^{\infty} \sum_{i_3 > \lfloor k/2 \rfloor} \sum_{i_4 = 0}^{\infty} C\rho^{i_1+i_2+i_3+i_4} \left|\text{Cov}(\eta_{t+l-i_1}\eta_{t-i_2}, \eta_{t-k+r-i_3}\eta_{t-k-i_4})\right|, \\
  u_4 & := \sum_{i_1 = 0}^{\infty} \sum_{i_2 = 0}^{\infty} \sum_{i_3=0}^{\infty} \sum_{i_4 > \lfloor k/2 \rfloor} C\rho^{i_1+i_2+i_3+i_4} \left|\text{Cov}(\eta_{t+l-i_1}\eta_{t-i_2}, \eta_{t-k+r-i_3}\eta_{t-k-i_4})\right|, \\
  u_5 & := \sum_{i_1 = 0}^{\lfloor k/2\rfloor} \sum_{i_2 = 0}^{\lfloor k/2 \rfloor} \sum_{i_3=0}^{\lfloor k/2 \rfloor} \sum_{i_4 =0}^{\lfloor k/2 \rfloor} C\rho^{i_1+i_2+i_3+i_4} \left|\text{Cov}(\eta_{t+l-i_1}\eta_{t-i_2}, \eta_{t-k+r-i_3}\eta_{t-k-i_4})\right|, \\ 
  v_1 & : = \sum_{i_1 > \lfloor k /2 \rfloor}\sum_{i_3=0}^{\infty}\|\Cov(\bm{c}_{t+l}^{i_1}\otimes \bm{c}_{t}^{-l+i_1},\bm{c}_{t-k+r}^{i_3}\otimes\bm{c}_{t-k}^{-r+i_3})\|, \\
  v_2 & : = \sum_{i_1=0}^{\infty}\sum_{i_3>\lfloor k/2 \rfloor}\|\Cov(\bm{c}_{t+l}^{i_1}\otimes \bm{c}_{t}^{-l+i_1},\bm{c}_{t-k+r}^{i_3}\otimes\bm{c}_{t-k}^{-r+i_3})\|, \\
  v_3  &: = \sum_{i_1 = 0}^{\lfloor k/2\rfloor}\sum_{i_3 = 0}^{\lfloor k/2 \rfloor}\|\Cov(\bm{c}_{t+l}^{i_1}\otimes \bm{c}_{t}^{-l+i_1},\bm{c}_{t-k+r}^{i_3}\otimes\bm{c}_{t-k}^{-r+i_3})\|.
\end{align*}
\noindent Under \((\mathbf{A_3})\) and by Cauchy-Schwarz inequality, we also have 
\begin{align*}
 \left|\Cov(\eta_{t-i_1}\eta_{t+l-i_2}, \eta_{t-k-i_3}\eta_{t-k+r-i_4})\right|\leq  \mathbb{E}|\eta_t|^4 + (\mathbb{E}|\eta_{t}|^2)^2<\infty,
\end{align*}
and
\begin{align*}
  \|\Cov(\bm{c}_{t+l}^{i_1}\otimes \bm{c}_{t}^{-l+i_1},\bm{c}_{t-k+r}^{i_3}\otimes\bm{c}_{t-k}^{-r+i_3})\|\leq \kappa \rho^{i_1+i_3}. 
\end{align*}
\noindent Hence, there exists a family of strictly positive constants \((\kappa_i)_{1 \leq i \leq 4}\)  and $(\kappa'_{j})_{1\leq j \leq 2}$ such that

\begin{align}
\label{eq:3.14}
    u_i\leq \kappa_i\rho^{k/2} \ , 1\leq i\leq 4 \quad \ \  \text{and} \quad  v_j \leq \kappa'_j\rho^{k/2} \ , 1\leq j\leq 2 .
\end{align}
\noindent Using the Cauchy-Schwarz inequality and under Assumption \((\mathbf{A_3})\), we know that \(\|\eta_t \eta_{t^{\prime}}\|_{2+\nu} \) is upper bounded by some constant independent from $t$ and $t'$, for some \(\nu > 0\). Given that \((2+\nu)^{-1} + (2+\nu)^{-1} + \nu(2+\nu)^{-1} = 1\), \(0 \leq i_1, i_2, i_3, i_4 \leq k/2\) and \(1 \leq l, r \leq N\), by using Lemma~\ref{lem:2}, we can find two positive constants \(\kappa_5\) and \(\kappa_6\) such that
\begin{align}
\label{eq:3.15}
    u_5 \leq & \sum_{i_1=0}^{\lfloor k/2 \rfloor} \sum_{i_2 = 0}^{\lfloor k/2 \rfloor} \sum_{i_3=0}^{\lfloor k/2 \rfloor} \sum_{i_4=0}^{\lfloor k/2 \rfloor}\kappa_5 \rho^{i_1+i_2+i_3+i_4}\|\eta_{t+l-i_1}\eta_{t-i_2}\|_{2+\nu}\|\eta_{t-k+r-i_3}\eta_{t-k-i_4}\|_{2+\nu} \notag \\ & \times \Big(\alpha_{\eta}\left(\min\left\{k-r+l+i_3-i_1,k-r+i_3-i_2,,k+l-i_1+i_4,k+i_4-i_2\right\}\right)\Big)^{\frac{\nu}{2+\nu}} \notag \\ \leq &  \ \kappa_6\Big(\alpha_{\eta}\Big(\lfloor k/2\rfloor-r\Big)\Big)^{\frac{\nu}{2+\nu}}.
\end{align}
\noindent Furthermore, since by assumption the Markov chain \((\Delta_t)_{t \in \mathbb{Z}}\) is irreducible, aperiodic and is strictly stationary, with a finite state space, it follows from \citet[Theorem 3.1]{bradley2005basic} that the process \((\Delta_t)_{t \in \mathbb{Z}}\) is mixing (in the ergodic-theoretic sense) and satisfies \(\sum_{h=0}^{\infty} \alpha_{\Delta}(h)^{\nu/2+\nu} < \infty\) for some \(\nu > 0\), where \(\alpha_{\Delta}\) is defined as in Equation (\ref{eq:12}). Thus, using the Davidov inequality recalled in Lemma \ref{lem:2}, there exists a strictly positive constant $\kappa_7$ such that
\begin{align}
\label{eq:3.16}
   v_3 \leq &  \ \|\bm{c}_{t+l}^{i_1}\otimes\bm{c}_{t}^{-l+i_1}\|_{2+\nu} \|\bm{c}_{t-k+r}^{i_3'}\otimes\bm{c}_{t-k}^{-r+i_3'}\|_{2+\nu} \times \alpha_{\Delta}^{\nu/(2+\nu)}(\min \{k-l-1,\lfloor k/2 \rfloor\}) \notag \\  \leq & \  \kappa_7 \alpha_{\Delta}^{\nu/(2+\nu)}(\min \{k-l-1,\lfloor k/2 \rfloor\}).
\end{align}
\noindent In view of Equations (\ref{eq:3.14}), (\ref{eq:3.15}) and (\ref{eq:3.16}), there exists three positive constants $\kappa_8, \kappa_9$ and $\kappa_{10}$ such that for $k\geq 0$,
\begin{align*}
   \sum_{k=0}^{\infty} \|\bm{d}_k(l,r)(\theta_0)\| \leq & \ \kappa_8 \sum_{k=0}^{\infty}\rho^{k/2} + \kappa_9\sum_{k=0}^{\infty} \alpha^{\nu/2+\nu}_{\eta}(\lfloor k/2 \rfloor-r) + \kappa_{10}\sum_{k=0}^{\infty}\alpha_{\Delta}^{\nu/(2+\nu)}(\min \{k-l-1,\lfloor k/2 \rfloor\})<\infty.
\end{align*}
\noindent By following  similar arguments to the ones used earlier, it can be easily shown that
\begin{align*}
   \sum_{k=-\infty}^{0} \|\bm{d}_k(l,r)(\theta_0)\| < \infty,
\end{align*}
and therefore $\sum_{k=-\infty}^{\infty} \|\bm{d}_k(l,r)(\theta_0)\| < \infty$. Thus, by invoking the dominated convergence theorem, we arrive at
\begin{align*}
     \bm{I}_n^N(\theta_0)(l,r) = \frac{1}{n}\displaystyle\sum_{k = -n+1}^{n-1}\left(n-|k|\right)\text{Cov}(Z_t(\theta_0)(l),Z_{t-k}(\theta_0)(r))\xrightarrow[n\rightarrow\infty]\displaystyle\sum_{k = -\infty}^{\infty}\|\bm{d}_k(l,r)(\theta_0)\|.
\end{align*}
\noindent The conclusion follows. \qed

\begin{l1}\label{lem:5}
Under the assumptions of Theorem~\ref{thm:3}, the random vector $\sqrt{n}\bm{F}^{N,n}(\theta_0)$ has a limiting normal distribution with mean $0_{\nbR^{p^2 N}}$ and covariance matrix $\bm{I}^{N}(\theta_0)$ as $n\to \infty$.
\end{l1} 
\subsubsection*{Proof.}
\noindent Using the Cauchy product between two convergent series, we observe that for all \( l \in \mathbb{N} \), we have
\begin{align*}
\underline{X}_{t+l} \otimes \underline{X}_t = & \sum_{i=0}^{\infty}\sum_{k=0}^{i}  \left(\left(\prod_{j=0}^{k-1} \mathbf{A}(\Delta_{t+l-j}) \right) \underline{f}(\Delta_{t+l-k}) \eta_{t+l-k}\right)\otimes \left(\left( \prod_{n=0}^{i-k-1} \mathbf{A}(\Delta_{t-n}) \right)\underline{f}(\Delta_{t-(i-k)}) \eta_{t-(i-k)} \right) \\ = & \sum_{i=0}^{\infty}\sum_{k=0}^{i} \bm{c}_{t+l}^{k}\otimes\bm{c}_t^{i-k}\eta_{t+l-k}\eta_{t-i+k}
\end{align*}
Let $r$ be a positive integer. We introduce the following notations
\begin{align*}
    Q_{t,r}(\theta_0):= \sum_{i=0}^{r}\sum_{k=0}^{i}\Big((\bm{c}_{t+1}^k\otimes\bm{c}_t^{i-k})'\eta_{t+1-k}\eta_{t-i+k},\cdots,(\bm{c}_{t+N}^k\otimes\bm{c}_t^{i-k})'\eta_{t+N-k}\eta_{t-i+k}\Big)
\end{align*}
and
\begin{align*}
   R_{t,r}(\theta_0):= \sum_{i=r+1}^{\infty}\sum_{k=0}^{i}\Big((\bm{c}_{t+1}^k\otimes\bm{c}_t^{i-k})'\eta_{t+1-k}\eta_{t-i+k},\cdots,(\bm{c}_{t+N}^k\otimes\bm{c}_t^{i-k})'\eta_{t+N-k}\eta_{t-i+k}\Big)
\end{align*}
so that $Q_{t,r}(\theta_0) + R_{t,r}(\theta_0) = Z_t(\theta_0)$. 

\noindent Keeping the previous notations, we then have
\begin{align*}
    \sqrt{n}\bm{F}^{N,n}({\theta_0}) =  \frac{1}{\sqrt{n}}\displaystyle\sum_{t=1}^{n}\left(Q_{t,r}(\theta_0)-\mathbb{E}_{\theta_0}(Q_{t,r}(\theta_0))\right) + \frac{1}{\sqrt{n}}\displaystyle\sum_{t=1}^{n}\left(R_{t,r}(\theta_0)-\mathbb{E}_{\theta_0}(R_{t,r}(\theta_0))\right).
\end{align*} 
\noindent On the one hand, since the processes $(\Delta_t)_{t\in\nbZ}$ and $(\eta_t)_{t\in\nbZ}$ are assumed to be strongly mixing, it follows that the process $(Q_{t,r}(\theta_0))_{t\in\nbZ}$ is also strongly mixing, see \citet[Theorem 14.1]{davidson1994stochastic}. Furthermore, it may be proved by similar arguments to those developed in the proof of Lemma \ref{lem:4} that $\sum_{h=-\infty}^{\infty} \Cov(Q_{t,r}(\theta_0), Q_{t-h,r}(\theta_0))$ is an absolute convergent series. Therefore, applying  \citet[Theorem 1.7, p.367]{ibragimov1962some} for strongly mixing processes, we have
\begin{align}
\label{eq:3.17}
  n^{-1/2}\sum_{t=1}^{n}\{Q_{t,r} - \mathbb{E}_{\theta_0}(Q_{t,r})\} \xrightarrow[n\rightarrow\infty]{\mathcal{D}} \mathcal{N}(0,\bm{I}_r), \quad \quad \bm{I}_r := \sum_{h=-\infty}^{\infty} \Cov(Q_{t,r}(\theta_0), Q_{t-h,r}(\theta_0)).
\end{align}
\noindent On the other hand, using a similar approach to that established in \citet[Proof of Lemma 6]{mainassara2025estimating} or the argument given in \citet[Lemma 4]{francq1998estimating}, it can be shown that 
\begin{align}
\label{eq:3.18}
    \sup_{n\ge 1} \Var \left( n^{-1/2} \sum_{t=1}^{n} R_{t,r}(\theta_0) - \mathbb{E}_{\theta_0}(R_{t,r}(\theta_0)) \right) \xrightarrow[r \rightarrow \infty]{} 0.
\end{align}
\noindent Thus, considering Equations (\ref{eq:3.17}) and (\ref{eq:3.18}), by using  \citet[Corollary 7.7.1, p. 426]{anderson1971family}, we obtain
\begin{align*}
  \sqrt{n} \bm{F}^{N,n}(\theta_0) \xrightarrow[n\to\infty]{\mathcal{D}} \mathcal{N}(0, \bm{I})  
\end{align*}
where we show that $\bm{I}  = \lim_{r\rightarrow \infty} \bm{I}_r$, see \citet[Proof of Lemma 3]{francq1998estimating}. This completes the proof.\qed

\subsubsection*{Proof of Theorem \ref{thm:3}}
\noindent Similarly to \citet[Proof of Theorem 3]{mainassara2025estimating}, this proof is based on a Taylor series argument.

\noindent Noting that the functions $C_{k,0}(.)$, $k=1,\ldots,N$, defined in \eqref{def_theta_functions} are smooth, and since $\Theta$ is relatively compact, it follows that
\begin{align*}
    \sup_{n\ge 1}\sup_{\theta\in\Theta}\|\nabla^2\bm{\mathcal{F}}^{N,n}(\theta)\| < \infty
\end{align*}
where $ \bm{\mathcal{F}}^{N,n}(\cdot)$ is defined in \eqref{eq:1.8}. Moreover, since by Theorem \ref{thm:2}, $\hat{\theta}_n$ converges almost surely to $\theta_0$, using a Taylor expansion around $\theta_0$ and noting that by construction $\bm{\mathcal{F}}^{N,n}(\hat{\theta}_n) = 0$, we have
\begin{align*}
    0 = \sqrt{n}\bm{\mathcal{F}}^{N,n}(\theta_0) + \sqrt{n}\nabla \bm{\Psi}^{N}(\theta_n^{\star})(\hat{\theta}_n-\theta_0)
\end{align*}
\noindent where the parameter ${\theta_n^\star}$ lies on the segment in $\mathbb{R}^{{Kp+K^2}}$ with endpoints $\hat{\theta}_n$ and $\theta_0$.

\noindent Another Taylor expansion leads to 
\begin{align*}
    \|\nabla\bm{\mathcal{F}}^{N,n}(\theta_n^\star)-\nabla\bm{\mathcal{F}}^{N,n}(\theta_0)\| = & \ \|\nabla^2\bm{\mathcal{F}}^{N,n}(\theta_0)(\theta_n^\star-\theta_0)\| \\ 
    \leq & \ \sup_{n\ge 1}
    \sup_{\theta\in\Theta}\|\nabla^2\bm{\mathcal{F}}^{N,n}(\theta)\|\;\|\theta_n^\star-\theta_0\| \xrightarrow[n\rightarrow\infty]{} 0.
\end{align*}
By the ergodic theorem, we also have
\begin{align*}
    \nabla\bm{\mathcal{F}}^{N,n}(\theta_0) \xrightarrow[n\rightarrow\infty]{\mbox{a.s.}} \Xi^{N}(\theta_0).
\end{align*}
Thus, under the assumption that $J_{\bm{\Psi}^N}(\theta_0)$ is invertible, we obtain
\begin{align*}
  \sqrt{n}\left(\hat{\theta}_n-\theta_0\right) = & \ \left(\nabla\bm{\mathcal{F}}^{N,n}(\theta_0)\right)^{-1}\sqrt{n}\bm{\mathcal{F}}^{N,n}(\theta_0) + \mathrm{o}_\mathbb{P}(1) \\ = & \ \left(\nabla\bm{\mathcal{F}}^{N,n}(\theta_0)\right)^{-1} \sqrt{n} J_{\bm{\Psi}^N}(\theta_0)'\bm{F}^{N,n}(\theta_0) + \mathrm{o}_\mathbb{P}(1).
\end{align*}
Consequently, from Lemma \ref{lem:4}, Lemma \ref{lem:5}, and using Slutsky's theorem, we obtain as $n\rightarrow\infty$,
\begin{align*}
  \sqrt{n}(\hat{\theta}_n - \theta_0) \xrightarrow[n\rightarrow\infty]{\mathcal{D}} \mathcal{N}(0,\Xi^{-1}\bm{J}'\bm{I}\bm{J}'\Xi^{-1}).
\end{align*}
This completes the proof.
\qed
\subsection{Proof of Theorem \ref{thm:4}}
\label{proofthm:4}

\noindent The proofs of this theorem is quite technical. They are adaptations of the arguments used in \citet[Theorem 3]{mainassara2025estimating} when $p=1$, with the difference being that we are manipulating matrix expressions. In the following, we will use several lemmas to prove this theorem. However, it should be noted that among the lemmas presented, we will only prove those whose proof differs significantly from what was done in \cite{mainassara2025estimating}.

%

\noindent We then define
\begin{align*}
    \underline{\bm{\varphi}}_r^{\star} & := (\varphi_1,\ldots,\varphi_r) \in \mathbb{R}^{p^2N \times rp^2N}, & \underline{\bm{\varphi}}_r & := (\varphi_{r,1}, \ldots, \varphi_{r,r}) \in \mathbb{R}^{p^2N \times rp^2N}, \\
    \underline{\Upsilon}_{r,t} & := \left(\Upsilon_{t-1}', \ldots, \Upsilon_{t-r}'\right)' \in \mathbb{R}^{rp^2N}, & \underline{\hat{\Upsilon}}_{r,t} & := (\hat{\Upsilon}_{t-1}', \ldots, \hat{\Upsilon}_{t-r}')' \in \mathbb{R}^{rp^2N}.
\end{align*}
\noindent And introduce the following (real or empirical) covariance matrices
\begin{align*}
    \hat{\Sigma}_{\Upsilon,\underline{\Upsilon}_r} &= \frac{1}{n} \sum_{t=1}^{n} \Upsilon_t \underline{\Upsilon}_{r,t}' \in \mathbb{R}^{p^2N \times rp^2N}, & 
    \hat{\Sigma}_{\underline{\Upsilon}_r} &= \frac{1}{n} \sum_{t=1}^{n} \underline{\Upsilon}_{r,t} \underline{\Upsilon}_{r,t}' \in \mathbb{R}^{rp^2N \times rp^2N}, \\
    \hat{\Sigma}_{\hat{\Upsilon},\underline{\hat{\Upsilon}}_r} &= \frac{1}{n} \sum_{t=1}^{n} \hat{\Upsilon}_t \underline{\hat{\Upsilon}}_{r,t}' \in \mathbb{R}^{p^2N \times rp^2N}, &
    \hat{\Sigma}_{\underline{\hat{\Upsilon}}_r} &= \frac{1}{n} \sum_{t=1}^{n} \underline{\hat{\Upsilon}}_{r,t} \underline{\hat{\Upsilon}}_{r,t}' \in \mathbb{R}^{rp^2N \times rp^2N}, \\
    \Sigma_{\Upsilon_t,\underline{\Upsilon}_{r}} &= \mathbb{E}[\Upsilon_t \underline{\Upsilon'}_{r,t}] \in \mathbb{R}^{p^2N \times rp^2N}, & 
    \Sigma_{\Upsilon} &= \mathbb{E}[\Upsilon_t \Upsilon_t'] \in \mathbb{R}^{p^2N \times p^2N}, \\
    \Sigma_{\underline{\Upsilon}_r} &= \mathbb{E}[\underline{\Upsilon}_{r,t} \underline{\Upsilon'}_{r,t}] \in \mathbb{R}^{rp^2N \times rp^2N}, & 
    \hat{\Sigma}_{\hat{\Upsilon}} &= \frac{1}{n} \sum_{t=1}^{n} \hat\Upsilon_t \hat\Upsilon_t' \in \mathbb{R}^{p^2N \times p^2N}.
\end{align*}
\noindent Additionally, we define
\begin{align}
    \underline{\check{\bm{\varphi}}}_r &= \hat{\Sigma}_{\Upsilon,\underline{\Upsilon}_r} \hat{\Sigma}_{\underline{\Upsilon}_r}^{-1} \in \mathbb{R}^{p^2N \times rp^2N}, \quad  
    \hat{\Sigma}_{\hat{u}_r} = \frac{1}{n} \sum_{t=1}^{n} \left(\Upsilon_t - \check{\underline{\bm{\varphi}}}_r \underline{\Upsilon}_{r,t}\right)\left(\Upsilon_t - \check{\underline{\bm{\varphi}}}_r \underline{\Upsilon}_{r,t}\right)' \in \mathbb{R}^{p^2N \times p^2N},\label{Sigma_hat_u}
\end{align}
and
\begin{align*}
    \underline{\hat{\bm{\varphi}}}_r &= \hat{\Sigma}_{\hat{\Upsilon},\underline{\hat{\Upsilon}}_r} \hat{\Sigma}_{\underline{\hat{\Upsilon}}_r}^{-1} \in \mathbb{R}^{p^2N \times rp^2N}, \quad  
    \hat{\Sigma}_{\hat{u}_r} = \frac{1}{n} \sum_{t=1}^{n} \left(\hat{\Upsilon}_t - \hat{\underline{\bm{\varphi}}}_r \underline{\hat{\Upsilon}}_{r,t}\right)\left(\hat{\Upsilon}_t - \hat{\underline{\bm{\varphi}}}_r \underline{\hat{\Upsilon}}_{r,t}\right)' \in \mathbb{R}^{p^2N \times p^2N}.
\end{align*}
\noindent {In the following, for any matrix $M$ of arbitrary dimensions, we denote by $\|\cdot\|_\star$ the matrix norm induced by the Euclidean vector norm, that is},
\[
\|M\|_\star=\sup_{\|x\|_2\le 1}\|Mx\|_2=\rho^{1/2}(M'M),
\]
{where $\|\cdot\|_2$ denotes the Euclidean norm on vectors and $\rho(\cdot)$ is the spectral radius. This norm is chosen because it satisfies the useful inequality}
\begin{align}
\label{eq:5.7}
\|M\|_\star^2 \leq \sum_{i,j} m_{i,j}^2.
\end{align}

\noindent The choice of the norm is crucial for the following results to hold (with e.g. the Euclidean norm, this result is not valid).
We will now able to establish the various lemmas required to prove Theorem \ref{thm:4}.
\begin{l1}
\label{lem:6}
Under the assumptions of Theorem~\ref{thm:4}, $\sup_{r\geq1}\left\|\Sigma_{\Upsilon,\underline{\Upsilon}_r}\right\|_\star$, $\sup_{r\geq1}\left\|\Sigma_{\underline{\Upsilon}_r}\right\|_\star$ and $\sup_{r\geq1}\left\|\Sigma_{\underline{\Upsilon}_r}^{-1}\right\|_\star$ are finite.
\end{l1}
\noindent The proof is exactly the same that the proof of Lemma \ref{lem:110} from \cite{mainassara2025estimating} so its is omitted.
\begin{l1}\label{lem:110}
We assume that the condition $\|\eta_t\|_{8+4\nu}<\infty$ holds for some $\nu>0$ and that Assumption $(\mathbf{A_2})$ is satisfied. Then we have
\begin{align}
\label{eq:5.8}
\sup_{\ell,u, s\in\mathbb{N}}\sum_{h=-\infty}^{\infty}\left|\Cov\left(\Upsilon_1[w_1](m_1)\Upsilon_{1+s}[w_2](m_2), \Upsilon_{\ell+h}[w_1][m_1]\Upsilon_{1+u+h}[w_2](m_2)\right)\right| < \infty,
\end{align}
\begin{align}
\label{eq:5.9}
\sup_{s\in\mathbb{N}}\sum_{h=-\infty}^{\infty}\left|\Cov\left(\hat{\Upsilon}_1[w_1](m_1)\hat{\Upsilon}_{1+s}[w_2](m_2), \hat{\Upsilon}_{1+h}[w_1](m_1)\hat{\Upsilon}_{1+s+h}[w_2](m_2)\right)\right| < \infty
\end{align}
and
\begin{align}
\label{eq:5.10}
    \sup_{s\in\mathbb{N}}\sum_{h=-\infty}^{\infty}\left| \Cov\left(\Upsilon_1[w_1](m_1)\Upsilon_{1+s}[w_2](m_2), \hat{\Upsilon}_{1+h}[w_1](m_1)\hat{\Upsilon}_{1+s+h}[w_2](m_2)\right) \right| < \infty,
\end{align}
where $w_1, w_2 \in \{1, \ldots, N\}$, $m_1, m_2 \in \{1, \ldots, p^2\}$, and $\Upsilon_t[w](m)$ (resp. $\hat{\Upsilon}_t[w](m)$) denotes the $m$-th element of the $w$-th block of $\Upsilon_t$ (resp. $\hat{\Upsilon}_t$).
\end{l1}

\subsubsection*{Proof.}
\noindent Let \( h \in \mathbb{Z}\), $w_1,w_2\in\{1,\dots,N\}$ and \( m_1, m_2 \in \{1, \ldots, p^2\} \). we define:
{\small
\begin{align*}
\mathcal{G}_1: = & \ + \text{Cov}\left(\{\underline{X}_{1+w_1}\otimes\underline{X}_1\}(m_1)\{\underline{X}_{1+s+w_2}\otimes \underline{X}_{1+s}\}(m_2),\{\underline{X}_{\ell+h+w_1}\otimes\underline{X}_{\ell+h}\}(m_1)\{\underline{X}_{1+s+h+w_2}\otimes \underline{X}_{1+u+h}\}(m_2)\right) \\ 
& - \text{Cov}(\{\underline{X}_{1+w_1}\otimes\underline{X}_{1}\}(m_1)\{\underline{X}_{1+s+w_2}\otimes\underline{X}_{1+s}\}(m_2),\{\underline{X}_{\ell+h+w_1}\otimes \underline{X}_{\ell+h}\}(m_1))\mathbb{E}(\{\underline{X}_{1+u+h+w_2}\otimes \underline{X}_{1+u+h}\}(m_2))\\ 
& - \mathbb{E}(\underline{X}_{\ell+h+w_1}\otimes\underline{X}_{\ell+h})(m_1)\text{Cov}(\{\underline{X}_{1+w_1}\otimes\underline{X}_{1}\}(m_1)\{\underline{X}_{1+s+w_2}\otimes \underline{X}_{1+s}\}(m_2),\{\underline{X}_{1+u+h+w_2}\otimes\underline{X}_{1+u+h}\}(m_2)),\\ \\
\mathcal{G}_2 : = & \ -\text{Cov}(\{\underline{X}_{1+w_1}\otimes \underline{X}_{1}\}(m_1),\{\underline{X}_{\ell+h+w_1}\otimes \underline{X}_{\ell+h}\}(m_1)\{\underline{X}_{1+u+h+w_2}\otimes \underline{X}_{1+u+h}\}(m_2)\mathbb{E}(\{\underline{X}_{1+s+w_2}\otimes \underline{X}_{1+s}\}(m_2)))\\ 
& + \text{Cov}(\{\underline{X}_{1+w_1}\otimes\underline{X}_1\}(m_1)\mathbb{E}(\{\underline{X}_{1+s+w_2}\otimes\underline{X}_{1+s}\}(m_2))\ ,\{\underline{X}_{\ell+h+w_1}\otimes \underline{X}_{\ell+h}\}(m_1)\mathbb{E}(\{\underline{X}_{1+u+h+w_2}\otimes \underline{X}_{1+u+h}\}(m_2)))\\ 
& +\text{Cov}(\{\underline{X}_{1+w_1}\otimes\underline{X}_1\}(m_1)\mathbb{E}(\{\underline{X}_{1+s+w_2}\otimes\underline{X}_{1+s}\}(m_2)),\{\underline{X}_{1+u+h+w_2}\otimes \underline{X}_{1+u+h}\}(m_2)) \mathbb{E}(\{\underline{X}_{\ell+h+w_1}\otimes\underline{X}_{\ell+h}\}(m_1)),\\ \\
\mathcal{G}_3 : = & \ -\mathbb{E}(\underline{X}_{1+w_1}\otimes \underline{X}_1)(m_1)\text{Cov}(\{\underline{X}_{1+s+w_2}\otimes \underline{X}_{1+s}\}(m_2),\{\underline{X}_{\ell+h+w_1}\otimes \underline{X}_{\ell+h}\}(m_1)\{\underline{X}_{1+u+h+w_2}\otimes \underline{X}_{1+u+h}\}(m_2))\\ 
& + \mathbb{E}(\underline{X}_{1+w_1}\otimes\underline{X}_1)(m_1)\text{Cov}(\{\underline{X}_{1+s+w_2}\otimes \underline{X}_{1+s}\}(m_2),\underline{X}_{\ell+h+w_1}\otimes \underline{X}_{\ell+h}(m_1)\mathbb{E}(\{\underline{X}_{1+u+h+w_2}\otimes\underline{X}_{1+u+h}\}(m_2)))\\
& + \mathbb{E}(\underline{X}_{1+w_1}\otimes \underline{X}_{1})(m_1)\text{Cov}(\{\underline{X}_{1+s+w_2}\otimes \underline{X}_{1+s}\}(m_2),\{\underline{X}_{1+u+h+m_2}\otimes \underline{X}_{1+u+h}\}(m_2))\mathbb{E}(\{\underline{X}_{\ell+h+w_1}\otimes \underline{X}_{\ell+h}\}(m_1)).
\end{align*}
}
\noindent Using the definition of $\Upsilon_t$ given in Equation (\ref{eq:5.2}), we then have
\begin{align*}
\sum_{h=-\infty}^{\infty}\left|\Cov\left(\Upsilon_1[w_1](m_1)\Upsilon_{1+s}[w_2](m_2), \Upsilon_{\ell+h}[w_1][m_1]\Upsilon_{1+u+h}[w_2](m_2)\right)\right| \leq \displaystyle\sum_{h=-\infty}^{\infty}\sum_{j=1}^{3}|\mathcal{G}_j(s,h,\ell,u,m_1,m_2)|
\end{align*}
where
    $\mathcal{G}_j(s,h,\ell,u,m_1,m_2) : = \mathcal{G}_j$, $1\leq j\leq 3$.

\noindent In addition, similarly to \citet[Lemma 8]{mainassara2025estimating}, we may prove that
\begin{align*}
\sup_{s,\ell,u\in\mathbb{N}}|\mathcal{G}_j(s,h,\ell,u,m_1,m_2)|<\infty \quad \forall 1\leq j\leq 3. 
\end{align*}
\noindent This proves that
\begin{align*}
 \sup_{\ell,u,s\in\mathbb{N}}\sum_{h=-\infty}^{\infty}\left|\Cov\left(\Upsilon_1[w_1](m_1)\Upsilon_{1+s}[w_2](m_2), \Upsilon_{\ell+h}[w_1][m_1]\Upsilon_{1+u+h}[w_2](m_2)\right)\right| < \infty.   
\end{align*}
\noindent Proceeding in a similar manner, we also prove Inequalities (\ref{eq:5.9}) and (\ref{eq:5.10}). See \citet[Lemmas 8 and 9]{mainassara2025estimating} for a detailed proof of this results.
\begin{l1}\label{lem:8}
Under the assumptions of Theorem~\ref{thm:4}, the terms \(\sqrt{r}\|\hat{\Sigma}_{\underline{\Upsilon}_r}-\Sigma_{\underline{\Upsilon}_r}\|_\star\), \(\sqrt{r}\|\hat{\Sigma}_{\Upsilon}-\Sigma_{\Upsilon}\|_\star\) and \(\sqrt{r}\|\hat{\Sigma}_{\Upsilon,\underline{\Upsilon}_r}-\Sigma_{\Upsilon,\underline{\Upsilon}_r}\|_\star\) tend towards 0 in probability as \(n \rightarrow \infty\) when \(r = \mathrm{o}(n^{1/3})\).
\end{l1}
\subsubsection*{Proof.}
\noindent We will focus on proving the convergence for \(\sqrt{r}\|\hat{\Sigma}_{\underline{\Upsilon}_r}-\Sigma_{\underline{\Upsilon}_r}\|_\star\), as the others can be proved in a similar manner.
\noindent By applying Markov's inequality, we obtain
\begin{align}
\label{eq:31}
    \forall \ \epsilon > 0, \ \mathbb{P}\left(\sqrt{r}\|\hat{\Sigma}_{\underline{\Upsilon}_r} - \Sigma_{\underline{\Upsilon}_r}\|_\star > \epsilon\right) \leq \epsilon^{-2}\mathbb{E}\left(r\|\hat{\Sigma}_{\underline{\Upsilon}_r} - \Sigma_{\underline{\Upsilon}_r}\|_\star^2\right).
\end{align}

\medskip
\noindent Let \(1 \leq m_1, m_2 \leq p^2\), \(1 \leq w_1, w_2 \leq N\) and \(1 \leq r_1, r_2 \leq r\). The element located at the \((r_1 p^2N + w_1N+m_1)\)-th row and \((r_2 p^2N + w_2N+m_2)\)-th column of the matrix \( (\hat{\Sigma}_{\underline{\Upsilon}_r} - \Sigma_{\underline{\Upsilon}_r})\in \nbR^{rp^2N \times rp^2N}\) is given by
\[
n^{-1} \sum_{t=1}^{n} \Upsilon_{t-r_1}[w_1](m_1) \Upsilon_{t-r_2}[w_2](m_2) - \mathbb{E}\left(\Upsilon_{t-r_1}[w_1](m_1) \Upsilon_{t-r_2}[w_2](m_2)\right).
\]
where we recall that \(\Upsilon_t[w]\) denotes the \(w\)-th block of \(\Upsilon_t\). 

\noindent Next, we define for $t\in\mathbb{Z}$, $\mathcal{M}_t$ = $\mathcal{M}_t(m_1, m_2, r_1, r_2) := \Upsilon_{t - r_1}[w_1](m_1) \Upsilon_{t - r_2}[w_2](m_2)$. Using the norm inequality
in (\ref{eq:5.7}) and Equation (\ref{eq:5.8}), it follows by the stationarity of the process \((\Upsilon_t)_{t \in \mathbb{Z}}\) that
\begin{align}
\label{eq:32}
\mathbb{E}\left(r\|\hat{\Sigma}_{\underline{\Upsilon}_r} - \Sigma_{\underline{\Upsilon}_r}\|_\star^2\right) & \leq \sum_{m_1,m_2=1}^{p^2} \sum_{r_1,r_2=1}^{r}\sum_{w_1,w_2=1}^{N}\left\{\frac{1}{n^2}\mathbb{E}\left(r\left(\sum_{t=1}^n \mathcal{M}_t -\mathbb{E}(\mathcal{M}_t)\right)^2\right)\right\} \notag\\  & \leq \sum_{m_1,m_2=1}^{p^2} \sum_{r_1,r_2=1}^{r}\sum_{w_1,w_2=1}^{N}
     \frac{r}{n^2} \ \text{Cov}\left(\sum_{t=1}^n \mathcal{M}, \sum_{t=1}^n \mathcal{M}_t\right) \notag\\ & \leq \sum_{m_1,m_2=1}^{p^2} \sum_{r_1,r_2=1}^{r}\sum_{w_1,w_2=1}^{N}
      \frac{r}{n^2}\sum_{k=-n+1}^{n-1}(n-|k|)\text{Cov}(\mathcal{M}_t, \mathcal{M}_{t-k}) \notag\\ & \leq \sum_{m_1,m_2=1}^{p^2} \sum_{r_1,r_2=1}^{r}\sum_{w_1,w_2=1}^{N}
      \frac{r}{n}\sum_{k=-\infty}^{\infty}\left|\text{Cov}(\mathcal{M}_t, \mathcal{M}_{t-k})\right|\notag\\ & \leq  \frac{\mathcal{C}N^2p^4r^3}{n}\xrightarrow[n\rightarrow\infty]{r \ = \ \mathrm{o}(n^{1/3})}0\quad\mbox{thanks to \eqref{eq:5.8},}
\end{align}
where \(\mathcal{C}\) is a positive constant independent of \(m_1, m_2, r\) and \(n\). The proof is completed by applying Equations (\ref{eq:31}) and (\ref{eq:32}). 

\noindent Similarly, when $r = \mathrm{o}(n^{1/3})$, we also prove that:
\begin{align*}
  \sqrt{r}\|\hat{\Sigma}_{\Upsilon}-\Sigma_{\Upsilon}\|_\star = \mathrm{o}_{\mathbb{P}}(1) \ \text{and} \ \sqrt{r}\|\hat{\Sigma}_{\Upsilon,\underline{\Upsilon}_r}-\Sigma_{\Upsilon,\underline{\Upsilon}_r}\|_\star = \mathrm{o}_{\mathbb{P}}(1).
\end{align*}
This completes the proof. \qed

\begin{l1}\label{lem:9}
Let \(\hat{\Sigma}_{\Upsilon}\) be the matrix obtained by replacing \(\hat{\Upsilon}_t\) with \(\Upsilon_t\) in \(\hat{\Sigma}_{\hat{\Upsilon}}\). Under the assumptions of Theorem~\ref{thm:4}, the terms \(\sqrt{r}\|\hat{\Sigma}_{\underline{\hat{\Upsilon}}_r} - \Sigma_{\underline{\Upsilon}_r}\|_\star\), \(\sqrt{r}\|\hat{\Sigma}_{\hat{\Upsilon}} - \Sigma_{\Upsilon}\|_\star\) and \(\sqrt{r}\|\hat{\Sigma}_{\hat{\Upsilon}, \underline{\hat{\Upsilon}}_r} - \Sigma_{\Upsilon, \underline{\Upsilon}_r}\|_\star\) converge to zero in probability as \(n \rightarrow \infty\), provided that \(r = \mathrm{o}(n^{1/3})\).
\end{l1}

\subsubsection*{Proof.}
\noindent Let \(r = \mathrm{o}(n^{1/3})\). By applying the triangular inequality, we observe that
\begin{align}
\label{eq:444}
    \sqrt{r} \|\hat{\Sigma}_{\underline{\hat{\Upsilon}}_r} - \Sigma_{\underline{\Upsilon}_r}\|_\star \leq \sqrt{r} \|\hat{\Sigma}_{\underline{\hat{\Upsilon}}_r} - \hat{\Sigma}_{\underline{\Upsilon}_r}\|_\star + \sqrt{r} \|\hat{\Sigma}_{\underline{\Upsilon}_r} - \Sigma_{\underline{\Upsilon}_r}\|_\star.
\end{align}

\noindent Furthermore, by Markov's inequality, we have
\begin{align*}
    \forall \ \varepsilon > 0, \quad \mathbb{P}\left(\sqrt{r} \|\hat{\Sigma}_{\underline{\hat{\Upsilon}}_r} - \hat{\Sigma}_{\underline{\Upsilon}_r}\|_\star > \varepsilon\right) \leq \varepsilon^{-2} \mathbb{E}\left(r \|\hat{\Sigma}_{\underline{\hat{\Upsilon}}_r} - \hat{\Sigma}_{\underline{\Upsilon}_r}\|_\star^2\right).
\end{align*}
\noindent Let \(1 \leq m_1, m_2 \leq p^2\), $1\leq w_1,w_2\leq N$ and \(1 \leq r_1, r_2 \leq r\). The element located at the \((r_1 p^2N + w_1N+m_1)\)-th row and \((r_2 p^2N + w_2N+m_2)\)-th column of the matrix \( (\hat{\Sigma}_{\underline{\hat{\Upsilon}}_r} - \hat{\Sigma}_{\underline{\Upsilon}_r})\in \nbR^{rp^2N \times rp^2N}\) is given by
\[
n^{-1} \left(\sum_{t=1}^{n} \hat{\Upsilon}_{t-r_1}[w_1](m_1) \hat{\Upsilon}_{t-r_2}[w_2](m_2) - \Upsilon_{t-r_1}[w_1](m_1) \Upsilon_{t-r_2}[w_2](m_2)\right).
\]

\noindent For \(t \in \mathbb{Z}\), we define
\[
\mathcal{X}_t=\mathcal{X}_t(m_1, m_2, r_1, r_2,n) := \hat{\Upsilon}_{t-r_1}[w_1](m_1) \hat{\Upsilon}_{t-r_2}[w_2](m_2) - \Upsilon_{t-r_1}[w_1](m_1) \Upsilon_{t-r_2}[w_2](m_2).
\]
Since the processes \((\hat{\Upsilon}_t)_{t \in \mathbb{Z}}\) and \((\Upsilon_t)_{t \in \mathbb{Z}}\) are stationary, it follows that the process \((\mathcal{X}_t)_{t \in \mathbb{Z}}\) is also stationary. We thus have
\begin{align}
\label{eq:333}
 \mathbb{E}\left(r\|\hat{\Sigma}_{\underline{\hat{  {\Upsilon}}}_r}-\hat{\Sigma}_{\underline{{\Upsilon}}_r}\|_\star^{2}\right) \leq & \displaystyle\sum_{m_1,m_2=1}^{p^2}\sum_{r_1,r_2=1}^{r}\sum_{w_1,w_2=1}^{N}r\mathbb{E}\left(n^{-2}\displaystyle\left(\sum_{t=1}^n\mathcal{X}_t\right)^2\right) \notag\\ \leq &  \sum_{m_1,m_2=1}^{p^2}\sum_{r_1,r_2=1}^{r}\sum_{w_1,w_2=1}^{N}r\left(\left(\mathbb{E}(\mathcal{X}_1\right)^2 + n^{-2}\text{Var}\left(\displaystyle\sum_{t=1}^n\mathcal{X}_t\right)\right) \notag\\ \leq &\sum_{m_1,m_2=1}^{p^2}\sum_{r_1,r_2=1}^{r}\sum_{w_1,w_2=1}^{N}r\left(\left(\mathbb{E}(\mathcal{X}_1\right)^2 + n^{-2}\sum_{k=-n+1}^{n-1}(n-|k|)\text{Cov}(\mathcal{X}_t,\mathcal{X}_{t-k})\right) \notag\\ \leq & \sum_{m_1,m_2=1}^{p^2}\sum_{r_1,r_2=1}^{r}\sum_{w_1,w_2=1}^{N}r \left((\mathbb{E}\left(\mathcal{X}_1)\right)^2+n^{-1}\displaystyle\sum_{k=-\infty}^{\infty}\left|\text{Cov}(\mathcal{X}_t,\mathcal{X}_{t-k})\right|\right).
\end{align}
Let \(1\leq m\leq p^2\) and $1\leq w\leq N$. By stationary, we have on the one hand
\begin{align}
\label{eq:33}
\|\hat{\Upsilon}_t[w](m)\|_2 = \left\|\{\underline{X}_{t+w} \otimes \underline{X}_t\}(m) - \hat{C}'_{w,0}(m)\right\|_2 \leq \left\|\{\underline{X}_{t+w} \otimes \underline{X}_t\}\right\|_2 + \left\|\hat{C}_{w,0}\right\|_2 \leq 2 \|\underline{X}_0\|_4^2 < \infty,
\end{align}
and on the other hand
\begin{align}
\label{eq:34}
\|\Upsilon_t[w](m)\|_2^2 = \left\|\{\underline{X}_{t+w} \otimes \underline{X}_t\}(m) - \mathbb{E}(\underline{X}_{t+w} \otimes \underline{X}_t)(m)\right\|_2^2 = \Var(\{\underline{X}_{t+w} \otimes \underline{X}_t\}(m)) \leq 2 \|\underline{X}_0\|_4^2 < \infty.
\end{align}
Thanks to the limit obtained in Lemma \ref{lem:4}, it follows that the sequence $\{(\hat{C}_{w,0} - C_{w,0})(m)\}_{n \in \mathbb{N}}$ satisfies
\begin{align*}
    \left\|n\left(\hat{C}_{w,0} - C_{w,0}\right)(m)\right\|_2 := \text{diag}^{(m)}\left(\text{diag}^{[w]}(\Var(\sqrt{n} \bm{F}^{N,n}(\theta_0)))\right) \xrightarrow[n \to \infty]{} \text{diag}^{(m)}\left(\text{diag}^{[w]}(I^N(\theta_0))\right) < \infty,
\end{align*}
where \(\text{diag}^{[w]}(\cdot)\) denotes the \(w\)-th diagonal block of a matrix, and \(\text{diag}^{(m)}(\cdot)\) refers to the \(m\)-th diagonal element of a matrix.
Hence we then have
\begin{align}
\label{eq:35}
\|\{\hat{C}_{w,0} - C_{w,0}\}(m)\|_2 \underset{n \to \infty}{\sim} n^{-1/2} \mathcal{C}(m,w).
\end{align}
where $\mathcal{C}(m,w)$ is a positive constant depends only on $w$ and $m$.

\noindent By applying the Cauchy-Schwarz inequality, it follows from Equations (\ref{eq:33}), (\ref{eq:34}) and (\ref{eq:35}) that
\begin{align}
\label{eq:36}
\left|\mathbb{E}(\mathcal{X}_1)\right|& \leq \|\hat{\Upsilon}_{1 - r_1}[w_1](m_1) - \Upsilon_{1 - r_1}[w_1](m_1)\|_2\|\hat{\Upsilon}_{1 - r_2}[w_2](m_2)\|_2 \notag\\ &\hspace*{0.25 cm} + \|\Upsilon_{1 - r_1}[w_1](m_1)\|_2 \|\hat{\Upsilon}_{1 - r_2}[w_2](m_2) - \Upsilon_{1 - r_2}[w_2](m_2)\|_2\notag\\
&\leq \|\{\hat{C}_{w_1,0} - C_{w_1,0}\}(m_1)\|_2\|\hat{\Upsilon}_{1 - r_2}[w_2](m_2)\|_2\notag 
+ \|\hat{\Upsilon}_{1 - r_1}[w_1](m_1)\|_2\|\{\hat{C}_{w_2,0} - C_{w_2,0}\}(m_2)\|_2\notag\\ &  = \mathrm{O}(n^{-1/2}),
\end{align}
so that we obtain
\begin{align}
\label{eq:63}
    (\mathbb{E}(\mathcal{X}_1))^2 = \mathrm{O}(n^{-1}).
\end{align}
Finally, using \eqref{eq:5.10} and Equation (\ref{eq:63}), we arrive at
\begin{align*}
\mathbb{E}\left(r\|\hat{\Sigma}_{\underline{\hat{{\Upsilon}}}_r}-\hat{\Sigma}_{\underline{{\Upsilon}}_r}\|_\star^{2}\right) \leq \mathcal{K}p^4N^2 \frac{r^3}{n}\xrightarrow[n\rightarrow\infty]{r \ = \ \mathrm{o}(n^{1/3})}0.
\end{align*}
where \(\mathcal{K}\) denotes a positive constant independent of \(n\).

\noindent Thus, we have shown that \(\sqrt{r} \|\hat{\Sigma}_{\underline{\hat{\Upsilon}}_r} - \hat{\Sigma}_{\underline{\Upsilon}_r}\|_\star\) converges to zero in probability as \(n \rightarrow \infty\), provided that \(r = \mathrm{o}(n^{1/3})\). 

\noindent By consequently we have using Equation (\ref{eq:444}) and the previous Lemma \ref{lem:8}, when $r=\mathrm{o}(n^{1/3})$:
\begin{align*}
   \sqrt{r}\|\hat{\Sigma}_{\underline{\hat{\Upsilon}}_r} - \Sigma_{\underline{\Upsilon}_r}\|_\star = \mathrm{o}_{\mathbb{P}}(1).
\end{align*}
Similarly, we also prove that \(\sqrt{r}\|\hat{\Sigma}_{\hat{\Upsilon}} - \Sigma_{\Upsilon}\|_\star\) and \(\sqrt{r}\|\hat{\Sigma}_{\hat{\Upsilon}, \underline{\hat{\Upsilon}}_r} - \Sigma_{\Upsilon, \underline{\Upsilon}_r}\|_\star\) converges to zero in probability as \(n \rightarrow \infty\), provided that \(r = \mathrm{o}(n^{1/3})\). This completes the proof.\qed

\begin{l1}
\label{lem:10}
Under the assumptions of Theorem~\ref{thm:4}, we have
    $\sqrt{r} \left\|\hat{\Sigma}_{\underline{{\hat{\Upsilon}}}_r}^{-1}-\Sigma_{\underline{{\Upsilon}}_r}^{-1}\right\|_\star = \circ_{\mathbb{P}}(1)$
as $n\rightarrow\infty$ when $r = \mathrm{o}(n^{1/3})$ and $r\rightarrow\infty$.
\end{l1}
\noindent The proof is exactly the same that the proof of Lemma 13 from \cite{mainassara2025estimating} so its is omitted. \qed

\begin{l1}
\label{lem:11}
Under the assumptions of Theorem~\ref{thm:4}, we have
    $\sqrt{r}\|\underline{\varphi}_r^{\star} - \underline{\varphi}_r\|_\star$ converges to $0$ as $n\rightarrow\infty$ when $r = \mathrm{o}(n^{1/3})$ and $r\rightarrow\infty$.
\end{l1}

\subsubsection*{Proof.}
 \noindent Let $r=r(n)\rightarrow\infty$ and $r=\mathrm{o}(n^{1/3})$. Considering the regression defined by the Equation (\ref{eq:5.4}) and (\ref{eq:22}), we have 
 \begin{align*}
     \sum_{i=1}^r\varphi_i\Upsilon_{t-i}+\left(u_t + \sum_{i=r+1}^{\infty} \varphi_i\Upsilon_{t-i}\right) = \Upsilon_t = \sum_{i=1}^{r}\varphi_{r,i}\Upsilon_{t-i}+u_{r,t}.
 \end{align*}
 \noindent Consequently, we have
 \begin{align*}
   \underline{\bm{\varphi}}^{\star}_r\underline{\Upsilon}_{r,t}+\left\{\sum_{i=r+1}^{\infty}\varphi_i\Upsilon_{t-i}+u_t\right\} = \underline{\bm{\varphi}}_r\underline{\Upsilon}_{r,t}+u_{r,t}
 \end{align*}
where, we recall that $\underline{\bm{\varphi}}^{\star}_r: = (\varphi_1,\ldots,\varphi_r)$ and $\underline{\bm{\varphi}}_r: = (\varphi_{r,1}, \ldots, \varphi_{r,r})$.

\noindent Using additionally the orthogonality conditions given in Equations~(\ref{eq:5.4}), we obtain
\begin{align*}
  \left(\underline{\bm{\varphi}}_r^{\star} - \underline{\bm{\varphi}}_r\right)\mathbb{E}(\underline{\Upsilon}_{r,t}\underline{\Upsilon}'_{r,t}) & =  \mathbb{E}\left(\left\{-\sum_{i=r+1}^{\infty} \varphi_i\Upsilon_{t-i} + u_t + u_{r,t}\right\}\underline{\Upsilon}'_{r,t}\right) \\ 
  & =  \mathbb{E}\left(\left\{-\sum_{i=r+1}^{\infty} \varphi_i\Upsilon_{t-i}\right\}\underline{\Upsilon}'_{r,t}\right).
\end{align*}
\noindent In other words, we have
\begin{align*}
\underline{\bm{\varphi}}_r^{\star} - \underline{\bm{\varphi}}_r = -\mathbb{E}\left(\left\{\sum_{i=r+1}^{\infty} \varphi_i \Upsilon_{t-i}\right\}\underline{\Upsilon}'_{r,t}\right)\Sigma_{\underline{\Upsilon}_r}^{-1},
\end{align*}
where we recall that $\Sigma_{\underline{\Upsilon}_r} := \mathbb{E}(\underline{\Upsilon}_{r,t}\underline{\Upsilon}'_{r,t})$ is in particular invertible thanks to Lemma \ref{lem:6}.


\noindent Furthermore, using the norm defined in Equation~(\ref{eq:5.7}), Assumption $(\mathbf{A_5})$, Minkowski's inequality, Hölder's inequality, the assumption ${\|\bm{\varphi}_i\|} = \mathrm{o}(i^{-2})$ and Equation~(\ref{eq:34}), we arrive at
\begin{align*}
  \sqrt{r}\|\underline{\bm{\varphi}}_r^{\star}-\underline{\bm{\varphi}}_r\|_{\star}&  = \sqrt{r}\left\|\mathbb{E}\left(\sum_{i=r+1}^{\infty}\varphi_i\Upsilon_{t-i}\underline{\Upsilon}_{r,t}'\right)\Sigma_{\underline{\Upsilon}_r}^{-1}\right\|_\star \\ & \leq \sqrt{r}\left\|\mathbb{E}\left(\sum_{i=r+1}^{\infty}\varphi_i\Upsilon_{t-i}\underline{\Upsilon}_{r,t}'\right)\right\|\left\|\Sigma_{\underline{\Upsilon}_r}^{-1}\right\|_\star\\ & \leq \sqrt{r}\sum_{i=1}^{\infty}\|\varphi_{r+i}\|_\star\left\|\mathbb{E}\left(\Upsilon_{t-r-i}\underline{\Upsilon}'_{r,t}\right)\right\|_{\star}\left(\sup_{s\geq 1}\left\|\Sigma_{\underline{\Upsilon}_s}^{-1}\right\|_\star\right)\\
  & \leq \sqrt{r}\left\{\sum_{i=1}^{\infty}\|\varphi_{r+i}\|_\star\left\{\sum_{m_1,m_2 = 1}^{N}\sum_{j=1}^r\left\|\Upsilon_{t-r-i}[m_1]\right\|_2^2\left\|\Upsilon'_{t-j}[m_2]\right\|_2^2\right\}\left(\sup_{s\geq 1}\left\|\Sigma_{\underline{\Upsilon}_s}^{-1}\right\|_{\star}\right)\right\}\\ & \leq \mathcal{C} N^2r^{3/2}\sum_{i=1}^{\infty} \mathrm{o}(r+i)^{-2}\left(\sup_{s\geq 1}\left\|\Sigma_{\underline{\Upsilon}_s}^{-1}\right\|_\star\right) \xrightarrow[r\rightarrow \infty]{} 0
\end{align*}
where $\mathcal{C}$ denotes the positive constant independent of $r$. Hence, the proof is concluded. \qed


\begin{l1}
\label{lem:12}
Under the assumptions of Theorem~\ref{thm:4}, we have
\begin{align*}
 \sqrt{r}\left\|\hat{\underline{\bm{\varphi}}_r}-\underline{\bm{\varphi}}_r\right\| = \circ_{\mathbb{P}}(1)
\end{align*}
as $n\rightarrow\infty$ when $r=\circ(n^{1/3})$.
\end{l1}
\noindent The proof is exactly the same that the proof of Lemma 15 from \cite{mainassara2025estimating} so its is omitted. \qed

\subsubsection*{Proof of Theorem ~\ref{thm:4}.}
\noindent We assume that the assumptions of Theorem~\ref{thm:4} are satisfied and recall the expressions for the matrices $\hat{\bm{I}}^{SP}$ and $\bm{I}$, defined as follows
\begin{align*}
  \hat{\bm{I}}^{SP} = \hat{\bm{\varphi}}(1)^{-1}\hat{\Sigma}_{\hat{u}_r}\hat{\bm{\varphi}}'(1)^{-1}, \quad\quad \bm{I} = \bm{\varphi}(1)^{-1}\Sigma_u\bm{\varphi}'(1)^{-1}.
\end{align*}
\noindent Given the expressions for $\hat{\bm{I}}^{SP}$ and $\bm{I}$, it is therefore sufficient to demonstrate that $\hat{\Sigma}_{\hat{u}_r}$ converges in probability to $\Sigma_u$ and that $\hat{\bm{\varphi}}(1)$ converges in probability to $\bm{\varphi}(1)$ as $r=r(n)\rightarrow\infty$ and $r=\mathrm{o}(n^{1/3})$.

\noindent Denoting by $0_{p^2N}$ (respectively recall that $I_{p^2N }$) the null matrix (respectively the identity matrix) of dimension $p^2N$, we introduce the following notation 
\begin{align*}
    \mathcal{Q}_i : = \left( \begin{array}{ccccc}
0_{p^2N} & \dots & I_{p^2N } & \dots & 0_{p^2N } 
\end{array} \right)' \in \mathbb{R}^{p^2N \times rp^2N}, \quad 1\leq i\leq r.
\end{align*}
where $I_{p^2N}$ is located in the $i$-th block of $\mathcal{Q}$.

\noindent Thus, we arrive at
\begin{align*}
    \hat{\varphi}_{r,i}-\varphi_{r,i} = \left(\underline{\hat{\bm{\varphi}}}_r-\underline{\bm{\varphi}}_r\right)\mathcal{Q}_i \qquad \text{and} \quad \varphi_i-\varphi_{r,i} = \left(\underline{\bm{\varphi}}_r^{\star}-\underline{\bm{\varphi}}_r\right)\mathcal{Q}_i \qquad \forall \ 1\leq i\leq r.
\end{align*}
\noindent Using  Lemmas~\ref{lem:11}, ~\ref{lem:12} and (\ref{eq:5.7}), we obtain
\begin{align*}
    \left\|\hat{\bm{\varphi}}(1)-\bm{\varphi}(1)\right\|_\star \leq & \ \left\|\displaystyle\sum_{i=1}^{r}\left(\hat{\varphi}_{i,r}-\varphi_{r,i}\right)\right\|_\star+\left\|\displaystyle\sum_{i=1}^{r}\left(\varphi_{r,i}-\varphi_i\right)\right\|_\star + \left\|\sum_{i=r+1}^{\infty}\varphi_i\right\|_\star \\ \leq & \ 
    \left\|\left(\underline{\hat{\bm{\varphi}}}_r-\underline{\bm{\varphi}}_r\right)\sum_{i=1}^{r}\mathcal{Q}_i\right\|_\star+\left\|\left(\underline{\bm{\varphi}}_r^{\star}-\underline{\bm{\varphi}}_r\right)\sum_{i=1}^{r}\mathcal{Q}_i\right\|_\star+ \displaystyle\sum_{i=r+1}^{\infty}\left\|\varphi_i\right\|_\star
    \\ \leq & \ \left\|\sum_{i=1}^{r}\mathcal{Q}_i\right\|_\star\left\{\left\|\underline{\hat{\bm{\varphi}}}_r-\underline{\bm{\varphi}}_r\right\|_\star + \left\|\left(\underline{\bm{\varphi}}_r^{\star}-\underline{\bm{\varphi}}_r\right)\right\|_\star\right\}+ \sum_{i=r+1}^{\infty}\left\|\varphi_i\right\|_\star \\ \leq & \ p\sqrt{N} \sqrt{r} \left\{\left\|\underline{\hat{\bm{\varphi}}}_r-\underline{\bm{\varphi}}_r\right\|_\star + \left\|\left(\underline{\bm{\varphi}}_r^{\star}-\underline{\bm{\varphi}}_r\right)\right\|_\star\right\}+\displaystyle\sum_{i=r+1}^{\infty} \left\|\varphi_i\right\|_\star =\circ_{\mathbb{P}}(1).
\end{align*}

\noindent We can note from Definition \eqref{Sigma_hat_u} that
\begin{align*}
   \hat{\Sigma}_{\hat{u}_r} = \hat{\Sigma}_{\hat{\Upsilon}}-\underline{\hat{\bm{\varphi}}}_{r}\hat{\Sigma}_{\hat{\Upsilon},\underline{\hat{\Upsilon}}_r}^{'},
\end{align*}
\noindent and
\begin{align*}
    \Sigma_{u} = & \ \mathbb{E}(u_tu_t^{'}) = \mathbb{E}\left(u_t\Upsilon_t^{'}\right)=\mathbb{E}\left\{\left(\Upsilon_t-\displaystyle\sum_{i=1}^{\infty}\varphi_i\Upsilon_{t-i}\right)\Upsilon_t^{'}\right\} \\
    = & \ \Sigma_{\Upsilon}-\displaystyle\sum_{i=1}^{\infty}\varphi_i\mathbb{E}\left(\Upsilon_{t-i}\Upsilon_t^{'}\right) = \Sigma_{\Upsilon}-\underline{\bm{\varphi}}_r^{\star}\Sigma_{\Upsilon,\underline{\Upsilon}_r}^{'}-\displaystyle\sum_{i=r+1}^{\infty}\varphi_i\mathbb{E}\left(\Upsilon_{t-i}\Upsilon_{t}^{'}\right).
\end{align*}
\noindent Consequently we also obtain 
\begin{align*}
    \left\|\hat{\Sigma}_{\hat{u}_r}-\Sigma_{u}\right\|_\star = &\left\|\hat{\Sigma}_{\hat{\Upsilon}}-\Sigma_{\Upsilon}-\left(\underline{\hat{\bm{\varphi}}}_r-\underline{\bm{\varphi}}_r^{\star}\right)\hat{\Sigma}_{\hat{\Upsilon},\underline{\hat{\Upsilon}}_{r}}^{'}-\underline{\bm{\varphi}}_r^{\star}\left(\hat{\Sigma}_{\hat{\Upsilon},\underline{\hat{\Upsilon}}_{r}}^{'}-\Sigma_{\Upsilon,\underline{\Upsilon}_r}^{'}\right)+\sum_{i=r+1}^{\infty}\varphi_i\mathbb{E}\left(\Upsilon_{t-i}\Upsilon_t^{'}\right)\right\|_\star \\ 
    \leq & \left\|\hat{\Sigma}_{\hat{\Upsilon}}-\Sigma_{\Upsilon}\right\|_\star+ \left\|\left(\underline{\hat{\bm{\varphi}}}_r-\underline{\bm{\varphi}}_r^{\star}\right)\left(\hat{\Sigma}_{\hat{\Upsilon},\underline{\hat{\Upsilon}}_{r}}^{'}-\Sigma_{\Upsilon,\underline{\Upsilon}_r}^{'}\right)\right\|_\star + \left\|\left(\underline{\hat{\bm{\varphi}}}_r-\underline{\bm{\varphi}}_r^{\star}\right)\Sigma_{\Upsilon,\underline{\Upsilon}_r}^{'}\right\|_\star\\ & + \left\|\underline{\bm{\varphi}}_r^{\star}\left(\hat{\Sigma}_{\hat{\Upsilon},\underline{\hat{\Upsilon}}_r}^{'}-\Sigma_{\Upsilon,\underline{\Upsilon}_r}^{'}\right)\right\|_\star + \left\|\sum_{i=r+1}^{\infty}\varphi_{i}\mathbb{E}\left(\Upsilon_{t-i}\Upsilon_t^{'}\right)\right\|_\star \\ 
    \leq & \left\|\hat{\Sigma}_{\hat{\Upsilon}}-\Sigma_{\Upsilon}\right\|_\star+\left\|\underline{\hat{\bm{\varphi}}}_r-\underline{\bm{\varphi}}_r^{\star}\right\|_\star\left\|\hat{\Sigma}_{\hat{\Upsilon},\underline{\hat{\Upsilon}}_{r}}^{'}-\Sigma_{\Upsilon,\underline{\Upsilon}_r}^{'}\right\|_\star+\left\|\underline{\hat{\bm{\varphi}}}_r-\underline{\bm{\varphi}}_r^{\star}\right\|_\star \left(\sup_{s\geq 1}\left\|\Sigma_{\Upsilon,\underline{\Upsilon}_s}^{'}\right\|_\star\right)\\ 
    + & \left\|\underline{\bm{\varphi}}_r^{\star}\right\|_\star\left\|\hat{\Sigma}_{\hat{\Upsilon},\underline{\hat{\Upsilon}}_r}^{'}-\Sigma_{\Upsilon,\underline{\Upsilon}_r}^{'}\right\|_\star + \displaystyle\sum_{i=r+1}^{\infty}\left\|\varphi_i\right\|_\star\left\|\mathbb{E}\left(\Upsilon_{t-i}\Upsilon_{t}^{'}\right)\right\|_\star.
\end{align*}
\noindent According to Lemma~\ref{lem:8}, we have established that $\hat{\Sigma}_{\hat{\Upsilon}} - \Sigma_{\Upsilon} = \mathrm{o}_{\mathbb{P}}(r^{-1/2})$. Furthermore, Lemmas~\ref{lem:11} and~\ref{lem:12} show that $\left\|\underline{\hat{\bm{\varphi}}}_r - \underline{\bm{\varphi}}_r^{\star}\right\|_\star = \mathrm{o}_{\mathbb{P}}(1)$. Additionally, Lemma~\ref{lem:10} indicates that $\left\|\hat{\Sigma}_{\hat{\Upsilon}, \underline{\hat{\Upsilon}}_r}^{'} - \Sigma_{\Upsilon, \underline{\Upsilon}_r}^{'}\right\|_\star = \mathrm{o}_{\mathbb{P}}(r^{-1/2})$. According to Lemma~\ref{lem:6}, we also have $\sup_{s \geq 1} \left\|\Sigma_{\Upsilon, \underline{\Upsilon}_s}^{'}\right\|_\star = \mathrm{O}(1)$. Finally, from Equation~(\ref{eq:5.7}), we deduce that $\left\|\underline{\bm{\varphi}}_r^{\star}\right\|_\star \leq \left(\sum_{i=1}^{\infty} \text{Tr}(\varphi_i \varphi_i^{'})\right)^{1/2} < \infty$ and $\left\|\mathbb{E}\left(\Upsilon_{t-i}\Upsilon_{t}^{'}\right)\right\|_\star = \mathrm{O}(1)$. Together, these results allow us to conclude the proof. \qed
\subsection{Proof of Theorem \ref{thm:5}} \label{proofthm:5}
\noindent
The proof of this theorem follows directly by applying Lemma~\ref{lem:1} together with Theorem~\ref{thm:1}. Indeed, using the independence between the processes \((\eta_t)_{t \in \mathbb{Z}}\) and \((\Delta_t)_{t \in \mathbb{Z}}\), along with the fact that \((\eta_t)_{t \in \mathbb{Z}}\) is a  white noise, and by exploiting standard properties of the tensor product for matrices and vectors, we obtain for all \(k \geq 0\),
\begin{align*}
  \mathbb{E}(\text{vec}\{\underline{Y}_0\underline{Y}_k'\}): = \mathbb{E}(\underline{Y}_k\otimes\underline{Y}_0) & =  \mathbb{E}(\{\underline{X}_k+\underline{m}(\Delta_k)\}\otimes\{\underline{X}_0+\underline{m}(\Delta_0)\})\\ &= \mathbb{E}(\underline{X}_k\otimes\underline{X}_0)+\mathbb{E}(\underline{m}(\Delta_k)\otimes\underline{m}(\Delta_0))\\ &  = \mathbb{E}(\underline{X}_k\otimes\underline{X}_0) + \mathbb{E}(\{\underline{m}(\Delta_k)\otimes I_p\})\underline{m}(\Delta_0))\notag\\
  & = C_{k,0} + (\bm{1}'\otimes I_{p^2})Q_{\underline{m}\otimes I_p}\Pi_{\underline{m}},
\end{align*}
\noindent where the $C_{k,0}, k\in\nbN\setminus\{0\}$ are defined in Theorem \ref{thm:1} (see Equation (\ref{eq:2})).

\noindent This completes the proof.

\qed

{
\noindent {\bf Acknowledgements.} \\

We sincerely thank the anonymous referees and Editor-in-chief for helpful remarks.
}


\begin{thebibliography}{}

\bibitem[Anderson, 1971]{anderson1971family}
Anderson, T.~W. (1971).
\newblock {\em The statistical analysis of time series}.
\newblock John Wiley \& Sons, Inc., New York-London-Sydney.

\bibitem[Andrews, 1991]{A91}
Andrews, D. W.~K. (1991).
\newblock Heteroskedasticity and autocorrelation consistent covariance matrix
  estimation.
\newblock {\em Econometrica}, 59(3):817--858.

\bibitem[Baum and Petrie, 1966]{baum1966statistical}
Baum, L.~E. and Petrie, T. (1966).
\newblock Statistical inference for probabilistic functions of finite state
  {M}arkov chains.
\newblock {\em Ann. Math. Statist.}, 37:1554--1563.

\bibitem[Bauwens et~al., 2017]{BauwensCarpantierDufays2017}
Bauwens, L., Carpantier, J.-F., and Dufays, A. (2017).
\newblock Autoregressive moving average infinite hidden markov-switching
  models.
\newblock {\em Journal of Business \& Economic Statistics}, 35(2):162--182.

\bibitem[Berk, 1974]{berk1974consistent}
Berk, K.~N. (1974).
\newblock Consistent autoregressive spectral estimates.
\newblock {\em Ann. Statist.}, 2:489--502.
\newblock Collection of articles dedicated to Jerzy Neyman on his 80th
  birthday.

\bibitem[Billio et~al., 1999]{billio1999bayesian}
Billio, M., Monfort, A., and Robert, C.~P. (1999).
\newblock Bayesian estimation of switching {ARMA} models.
\newblock {\em J. Econometrics}, 93(2):229--255.

\bibitem[Bollerslev, 1986]{bollerslev1986generalized}
Bollerslev, T. (1986).
\newblock Generalized autoregressive conditional heteroskedasticity.
\newblock {\em J. Econometrics}, 31(3):307--327.

\bibitem[Boubacar~Mainassara, 2011]{BM2011}
Boubacar~Mainassara, Y. (2011).
\newblock Multivariate portmanteau test for structural {VARMA} models with
  uncorrelated but non-independent error terms.
\newblock {\em J. Statist. Plann. Inference}, 141(8):2961--2975.

\bibitem[Boubacar~Ma{\"i}nassara, 2012]{yac_jtsa}
Boubacar~Ma{\"i}nassara, Y. (2012).
\newblock Selection of weak {VARMA} models by modified {A}kaike's information
  criteria.
\newblock {\em J. Time Series Anal.}, 33(1):121--130.

\bibitem[Boubacar~Ma\"inassara et~al., 2026]{mainassara2025estimating}
Boubacar~Ma\"inassara, Y., Bra, A., and Rabehasaina, L. (2026).
\newblock {E}stimating weak {M}arkov-switching {AR}(1) models.
\newblock {\em Electron. J. Stat.}, 20(1):2254--2313.

\bibitem[Boubacar~Mainassara et~al., 2012]{BMCF2012}
Boubacar~Mainassara, Y., Carbon, M., and Francq, C. (2012).
\newblock Computing and estimating information matrices of weak {ARMA} models.
\newblock {\em Comput. Statist. Data Anal.}, 56(2):345--361.

\bibitem[Boubacar~Ma{\"i}nassara et~al., 2021]{BMES2021}
Boubacar~Ma{\"i}nassara, Y., Esstafa, Y., and Saussereau, B. (2021).
\newblock Estimating {FARIMA} models with uncorrelated but non-independent
  error terms.
\newblock {\em Stat. Inference Stoch. Process.}, 24:549--608.

\bibitem[Boubacar~Ma\"inassara et~al., 2023]{BMES2023}
Boubacar~Ma\"inassara, Y., Esstafa, Y., and Saussereau, B. (2023).
\newblock Diagnostic checking in {FARIMA} models with uncorrelated but
  non-independent error terms.
\newblock {\em Electron. J. Stat.}, 17(1):1160--1239.

\bibitem[Boubacar~Mainassara and Francq, 2011]{BMF2011}
Boubacar~Mainassara, Y. and Francq, C. (2011).
\newblock Estimating structural {VARMA} models with uncorrelated but
  non-independent error terms.
\newblock {\em J. Multivariate Anal.}, 102(3):496--505.

\bibitem[Boubacar~Ma\"inassara et~al., 2022]{BMKS2022}
Boubacar~Ma\"inassara, Y., Kadmiri, O., and Saussereau, B. (2022).
\newblock Portmanteau test for the asymmetric power {GARCH} model when the
  power is unknown.
\newblock {\em Statist. Papers}, 63(3):755--793.

\bibitem[Boubacar~Ma{\"i}nassara and Kokonendji, 2016]{BMK2016}
Boubacar~Ma{\"i}nassara, Y. and Kokonendji, C.~C. (2016).
\newblock Modified {S}chwarz and {H}annan-{Q}uinn information criteria for weak
  {VARMA} models.
\newblock {\em Stat. Inference Stoch. Process.}, 19(2):199--217.

\bibitem[Boubacar~Ma{\"i}nassara and Rabehasaina, 2020]{boubacar2020estimation}
Boubacar~Ma{\"i}nassara, Y. and Rabehasaina, L. (2020).
\newblock Estimation of weak {ARMA} models with regime changes.
\newblock {\em Stat. Inference Stoch. Process.}, 23(1):1--52.

\bibitem[Bougerol and Picard, 1992]{bougerol1992strict}
Bougerol, P. and Picard, N. (1992).
\newblock Strict stationarity of generalized autoregressive processes.
\newblock {\em Ann. Probab.}, 20(4):1714--1730.

\bibitem[Bradley, 2005]{bradley2005basic}
Bradley, R.~C. (2005).
\newblock Basic properties of strong mixing conditions. {A} survey and some
  open questions.
\newblock {\em Probab. Surv.}, 2:107--144.
\newblock Update of, and a supplement to, the 1986 original.

\bibitem[Brandt, 1986]{brandt1986stochastic}
Brandt, A. (1986).
\newblock The stochastic equation {$Y_{n+1}=A_nY_n+B_n$} with stationary
  coefficients.
\newblock {\em Adv. in Appl. Probab.}, 18(1):211--220.

\bibitem[Brockwell and Davis, 1991]{brockwell1991time}
Brockwell, P.~J. and Davis, R.~A. (1991).
\newblock {\em Time series: theory and methods}.
\newblock Springer Series in Statistics. Springer-Verlag, New York, second
  edition.

\bibitem[Cai, 1994]{Cai1994}
Cai, J.~Y. (1994).
\newblock A markov model of switching-regime arch.
\newblock {\em Journal of Business \& Economic Statistics}, 12:155--163.

\bibitem[Cavicchioli, 2014]{Cavi2014}
Cavicchioli, M. (2014).
\newblock Analysis of the likelihood function for {M}arkov-switching
  {VAR}({CH}) models.
\newblock {\em J. Time Series Anal.}, 35(6):624--639.

\bibitem[Cavicchioli, 2016]{Cavi2016}
Cavicchioli, M. (2016).
\newblock Statistical analysis of mixture vector autoregressive models.
\newblock {\em Scand. J. Stat.}, 43(4):1192--1213.

\bibitem[Chan and Yau, 2024]{CY2024}
Chan, K.~W. and Yau, C.~Y. (2024).
\newblock Asymptotically constant risk estimator of the time-average variance
  constant.
\newblock {\em Biometrika}, 111(3):825--842.

\bibitem[Chib, 1996]{chib1996calculating}
Chib, S. (1996).
\newblock Calculating posterior distributions and modal estimates in {M}arkov
  mixture models.
\newblock {\em J. Econometrics}, 75(1):79--97.
\newblock Annals of econometrics: Bayes, Bernoullis, and Basel (1993).

\bibitem[Chiba, 2023]{Chiba2023}
Chiba, M. (2023).
\newblock Robust and efficient specification tests in markov‑switching
  autoregressive models.
\newblock {\em Statistical Inference for Stochastic Processes}, 26(1):99--137.

\bibitem[Davidson, 1994]{davidson1994stochastic}
Davidson, J. (1994).
\newblock {\em Stochastic limit theory}.
\newblock Advanced Texts in Econometrics. The Clarendon Press, Oxford
  University Press, New York.
\newblock An introduction for econometricians.

\bibitem[Davydov, 1968]{davydov1968convergence}
Davydov, Y.~A. (1968).
\newblock Convergence of distributions generated by stationary stochastic
  processes.
\newblock {\em Theory of Probability \& Its Applications}, 13(4):691--696.

\bibitem[den Haan and Levin, 1997]{den1996practitioner}
den Haan, W.~J. and Levin, A.~T. (1997).
\newblock A practitioner's guide to robust covariance matrix estimation.
\newblock In {\em Robust inference}, volume~15 of {\em Handbook of Statist.},
  pages 299--342. North-Holland, Amsterdam.

\bibitem[Douc et~al., 2004]{douc2004asymptotic}
Douc, R., Moulines, {\'E}., and Ryd{\'e}n, T. (2004).
\newblock Asymptotic properties of the maximum likelihood estimator in
  autoregressive models with {M}arkov regime.
\newblock {\em Ann. Statist.}, 32(5):2254--2304.

\bibitem[Engle, 1982]{engle1982autoregressive}
Engle, R.~F. (1982).
\newblock Autoregressive conditional heteroscedasticity with estimates of the
  variance of {U}nited {K}ingdom inflation.
\newblock {\em Econometrica}, 50(4):987--1007.

\bibitem[Escanciano and Velasco, 2006]{EscVel2006}
Escanciano, J.~C. and Velasco, C. (2006).
\newblock Testing the martingale difference hypothesis using integrated
  regression functions.
\newblock {\em Computational Statistics \& Data Analysis}, 51(4):2278--2294.
\newblock Nonlinear Modelling and Financial Econometrics.

\bibitem[Francq and Gautier, 2004a]{francq2004estimation}
Francq, C. and Gautier, A. (2004a).
\newblock Estimation of time-varying {ARMA} models with {M}arkovian changes in
  regime.
\newblock {\em Statist. Probab. Lett.}, 70(4):243--251.

\bibitem[Francq and Gautier, 2004b]{FG2004jtsa}
Francq, C. and Gautier, A. (2004b).
\newblock Large sample properties of parameter least squares estimates for
  time-varying {ARMA} models.
\newblock {\em J. Time Ser. Anal.}, 25(5):765--783.

\bibitem[Francq and Roussignol, 1997]{francq1997white}
Francq, C. and Roussignol, M. (1997).
\newblock On white noises driven by hidden {M}arkov chains.
\newblock {\em J. Time Ser. Anal.}, 18(6):553--578.

\bibitem[Francq and Roussignol, 1998]{francq1998ergodicity}
Francq, C. and Roussignol, M. (1998).
\newblock Ergodicity of autoregressive processes with {M}arkov-switching and
  consistency of the maximum-likelihood estimator.
\newblock {\em Statistics}, 32(2):151--173.

\bibitem[Francq et~al., 2005]{FRZ2005}
Francq, C., Roy, R., and Zako{\"i}an, J.-M. (2005).
\newblock Diagnostic checking in {ARMA} models with uncorrelated errors.
\newblock {\em J. Amer. Statist. Assoc.}, 100(470):532--544.

\bibitem[Francq and Zako{\"{\i}}an, 1998]{francq1998estimating}
Francq, C. and Zako{\"{\i}}an, J.-M. (1998).
\newblock Estimating linear representations of nonlinear processes.
\newblock {\em J. Statist. Plann. Inference}, 68(1):145--165.

\bibitem[Francq and Zako{\"{\i}}an, 2001]{francq2001stationarity}
Francq, C. and Zako{\"{\i}}an, J.-M. (2001).
\newblock Stationarityof multivariate markov-switching arma models.
\newblock {\em J. Econometrics}, 102(2):339--364.

\bibitem[Francq and Zako{\"{\i}}an, 2005]{fz05}
Francq, C. and Zako{\"{\i}}an, J.-M. (2005).
\newblock Recent results for linear time series models with non independent
  innovations.
\newblock In {\em Statistical modeling and analysis for complex data problems},
  volume~1 of {\em GERAD 25th Anniv. Ser.}, pages 241--265. Springer, New York.

\bibitem[Francq and Zako{\"i}an, 2019]{francq2019garch}
Francq, C. and Zako{\"i}an, J.-M. (2019).
\newblock {\em {GARCH} Models: Structure, Statistical Inference and Financial
  Applications}.
\newblock Wiley.

\bibitem[Hamilton, 1988]{hamilton1988rational}
Hamilton, J.~D. (1988).
\newblock Rational-expectations econometric analysis of changes in regime: an
  investigation of the term structure of interest rates.
\newblock {\em J. Econom. Dynam. Control}, 12(2-3):385--423.
\newblock Economic time series with random walk and other nonstationary
  components.

\bibitem[Hamilton, 1989]{hamilton1989new}
Hamilton, J.~D. (1989).
\newblock A new approach to the economic analysis of nonstationary time series
  and the business cycle.
\newblock {\em Econometrica}, 57(2):357--384.

\bibitem[Hamilton, 1990]{Hamilton1990}
Hamilton, J.~D. (1990).
\newblock Analysis of time series subject to changes in regime.
\newblock {\em J. Econometrics}, 45(1-2):39--70.

\bibitem[Hamilton and Susmel, 1994]{HS1994autoregressive}
Hamilton, J.~D. and Susmel, R. (1994).
\newblock Autoregressive conditional heteroskedasticity and changes in regime.
\newblock {\em Journal of econometrics}, 64(1):307--333.

\bibitem[Ibragimov, 1962]{ibragimov1962some}
Ibragimov, I.~A. (1962).
\newblock Some limit theorems for stationary processes.
\newblock {\em Teor. Verojatnost. i Primenen.}, 7:361--392.

\bibitem[Kamgaing et~al., 2009]{KamgaingOmbaoDavis2009}
Kamgaing, J.~T., Ombao, H., and Davis, R.~A. (2009).
\newblock Autoregressive processes with data-driven regime switching.
\newblock {\em Journal of Time Series Analysis}, 30(5):505--533.

\bibitem[Kim, 1994]{Kim1994}
Kim, C.-J. (1994).
\newblock Dynamic linear models with {M}arkov-switching.
\newblock {\em J. Econometrics}, 60(1-2):1--22.

\bibitem[Krolzig, 1997]{Krolzig1997}
Krolzig, H.-M. (1997).
\newblock {\em Markov-switching vector autoregressions}, volume 454 of {\em
  Lecture Notes in Economics and Mathematical Systems}.
\newblock Springer-Verlag, Berlin.
\newblock Modelling, statistical inference, and application to business cycle
  analysis.

\bibitem[Leroux, 1992]{leroux1992maximum}
Leroux, B.~G. (1992).
\newblock Maximum-likelihood estimation for hidden {M}arkov models.
\newblock {\em Stochastic Process. Appl.}, 40(1):127--143.

\bibitem[Liu and Chan, 2025]{LC2025}
Liu, X. and Chan, K.~W. (2025).
\newblock Positive-definite converging kernel estimation of long-run variance.
\newblock {\em J. Bus. Econom. Statist.}, 43(4):835--849.

\bibitem[McCulloch and Tsay, 1994]{mcculloch1994bayesian}
McCulloch, R.~E. and Tsay, R.~S. (1994).
\newblock Bayesian inference of trend- and difference-stationarity.
\newblock {\em Econometric Theory}, 10(3-4):596--608.

\bibitem[Newey and West, 1987]{newey}
Newey, W.~K. and West, K.~D. (1987).
\newblock A simple, positive semidefinite, heteroskedasticity and
  autocorrelation consistent covariance matrix.
\newblock {\em Econometrica}, 55(3):703--708.

\bibitem[Petrie, 1969]{petrie1969probabilistic}
Petrie, T. (1969).
\newblock Probabilistic functions of finite state {M}arkov chains.
\newblock {\em Ann. Math. Statist.}, 40:97--115.

\bibitem[Politis, 2011]{P2011}
Politis, D.~N. (2011).
\newblock Higher-order accurate, positivesemidefinite estimation of
  large-sample covarianceand spectral density matrices.
\newblock {\em Econometric Theory}, 27(4):703–744.

\bibitem[Politis and Romano, 1995]{PDR1995}
Politis, D.~N. and Romano, J.~P. (1995).
\newblock Bias-corrected nonparametric spectral estimation.
\newblock {\em J. Time Ser. Anal.}, 16(1):67--103.

\bibitem[Romano and Thombs, 1996]{romano1996inference}
Romano, J.~P. and Thombs, L.~A. (1996).
\newblock Inference for autocorrelations under weak assumptions.
\newblock {\em J. Amer. Statist. Assoc.}, 91(434):590--600.

\bibitem[Vats and Flegal, 2022]{VF2022}
Vats, D. and Flegal, J.~M. (2022).
\newblock Lugsail lag windows for estimating time-average covariance matrices.
\newblock {\em Biometrika}, 109(3):735--750.

\bibitem[Xie et~al., 2008]{xie2008general}
Xie, Y., Yu, J., and Ranneby, B. (2008).
\newblock A general autoregressive model with {M}arkov switching: estimation
  and consistency.
\newblock {\em Math. Methods Statist.}, 17(3):228--240.

\end{thebibliography}

\end{document}